\documentclass[final,3p,times,sort&compress]{elsarticle}
\usepackage{mwe}
\usepackage{microtype}
\usepackage{tikz}
\usetikzlibrary{external}
\usetikzlibrary{shapes.geometric}
\usetikzlibrary{decorations.pathmorphing,patterns}
\usepackage{structmech}
\usepackage{float}
\usepackage{hyperref}
\usepackage{caption}
\usepackage{subcaption}
\usepackage{amsmath,dsfont}
\DeclareMathOperator*{\argmax}{arg\,max}
\DeclareMathOperator*{\argmin}{arg\,min}

\DeclareMathOperator*{\arglocmin}{arg\,loc\,min}
\usepackage{comment}
\usepackage[lined]{algorithm2e}
\SetKwComment{Comment}{/* }{ */}
\usepackage{amsthm}
\usepackage{amssymb}
\usepackage{stmaryrd}
\usepackage{mathtools}
\usepackage{tabularx}

\usepackage{expl3}
\ExplSyntaxOn
\newcommand\latinabbrev[1]{
  \peek_meaning:NTF . {% Same as \@ifnextchar
    #1\@}%
  { \peek_catcode:NTF a {% Check whether next char has same catcode as \'a, i.e., is a letter
      #1.\@ }%
    {#1.\@}}}
\ExplSyntaxOff
\usepackage{mathrsfs}

\usepackage{amsmath,amssymb,amsthm,color,xfrac,mathrsfs}
\usepackage{bbm}
\usepackage{makecell}
\usepackage{epsfig,amsfonts}
\usepackage{graphicx,epsfig,epstopdf} 
\usepackage{array} 
\usepackage{color,soul}
\soulregister\citep{7}
\soulregister\citep{7}
\soulregister\ref{7}
\soulregister\pageref{7}
\sethlcolor{yellow}
\usepackage{appendix}
     
\usepackage{pmat} 
\usepackage{indentfirst} 
\usepackage{enumerate} 
\usepackage{lineno}
 \usepackage{placeins}
 \usepackage{silence}
\usepackage{booktabs,ctable,multirow,longtable} 
 \usepackage{rotating}
\usepackage{hyphenat}

\usetikzlibrary{backgrounds}           % For on background layer

\usepackage{ulem}

\usepackage{pifont}

\definecolor{myBlue}{rgb}{0.1294, 0.3608, 0.6863}
\definecolor{myRed}{rgb}{0.7176, 0.2078, 0.1765}

\usepackage{changepage}

\NewDocumentCommand{\evalat}{sO{\big}mm}{%
  \IfBooleanTF{#1}
   {\mleft. #3 \mright|_{#4}}
   {#3#2|_{#4}}%
}

\DeclareMathAlphabet{\mathpzc}{OT1}{pzc}{m}{it}
\DeclareMathAlphabet{\mathcalligra}{OT1}{calligra}{m}{it}

\AtBeginDocument{

}

\newcolumntype{P}[1]{>{\centering\arraybackslash}m{#1}}

\graphicspath{{./Images/}}
\newtheorem{remark}{Remark}

\journal{}
\makeatletter
\def\ps@pprintTitle{%
  \let\@oddhead\@empty
  \let\@evenhead\@empty
  \let\@oddfoot\@empty
  \let\@evenfoot\@oddfoot
}
\makeatother

\newcommand{\jump}[1]{\llbracket #1 \rrbracket}

\begin{document}
\begin{frontmatter}

\title{Variational phase-field modeling of cohesive fracture with flexibly tunable strength surface}
\author[1]{F. Vicentini\corref{cor1}}
\author[1]{J. Heinzmann}
\author[1]{P. Carrara}
\author[1]{L. De Lorenzis}
%\ead{ldelorenzis@ethz.ch}
%\ead{corrado.maurini@sorbonne-universite.fr}

%\ead{l.delorenzis@tu-braunschweig.de}
\address[1]{Computational Mechanics Group, Eidgen\"{o}ssische Technische Hochschule Z\"{u}rich, Tannenstrasse 3, 8092 Z\"{u}rich, Switzerland}
%\date{}
%\maketitle

%--------------------------------------------------------------------------

\begin{abstract}
Variational phase-field models of brittle fracture are powerful tools for studying Griffith-type crack propagation in complex scenarios. However, as approximations of Griffith’s theory—which does not incorporate a strength criterion—these models lack flexibility in prescribing material-specific strength surfaces. Consequently, they struggle to accurately capture crack nucleation under multiaxial stress conditions.
In this paper, inspired by \citep{alessi2014gradient}, we propose a variational phase-field model that approximates cohesive fracture. The model accommodates an arbitrary (convex) strength surface, independent of the regularization length scale, and allows for flexible tuning of the cohesive response.
Our formulation results in sharp cohesive cracks and naturally enforces a sharp non-interpenetration condition, thereby eliminating the need for additional energy decomposition strategies. It inherently satisfies stress softening and produces "crack-like" residual stresses by construction. To ensure strain hardening, the ratio of the regularization length to the material’s cohesive length must be sufficiently small; however, if crack nucleation is desired, this ratio must also be large enough to make the homogeneous damaged state unstable.
We investigate the model in one and three dimensions, establishing first- and second-order stability results. The theoretical findings are validated through numerical simulations using the finite element method, employing standard discretization and solution techniques.
\end{abstract}
%--------------------------------------------------------------------------

%\begin{keyword}
%Alternate-minimization \sep 
%Line-search \sep 
%Multi-axial\sep 
%Dimensional-analysis
%\sep Star-convex
%Energy decomposition \sep Fracture \sep Multi-axial  \sep Phase-field \sep Star-convex 
%\end{keyword}

\end{frontmatter}

%\linenumbers

%===================================================================================================
\section{Introduction}

The two fundamental processes that must be described for accurate modeling of fracture in brittle and quasi-brittle materials are the nucleation of new cracks, i.e. the first appearance of discontinuities within an otherwise continuous body, and the propagation of pre-existing cracks \citep{belytschko}. For many quasi-brittle materials of practical interest such as concrete, rocks and composites, proper modeling requires in addition the description of a third process, i.e. the appearance of diffuse damage, which may possibly precede nucleation \citep{bazant_planas}.

``Failure" for a material point subjected to a uniaxial tensile stress is postulated to occur when the stress attains a critical value, denoted as the \textit{tensile strength} and considered as a material property. Under these conditions, at the material point a new crack nucleates (possibly after some diffuse damage) along the direction orthogonal to the applied stress. If the uniaxial stress is compressive, the notion of failure is less clear and usually associated to diffuse damage due to crushing, again under a critical value of compressive stress known as the \textit{compressive strength}. If a material point undergoes a multiaxial stress state, ``failure" is postulated to occur when the stress tensor at the point reaches the so-called \textit{strength surface}—a critical surface in the stress space (described implicitly by $\mathcal{F}(\boldsymbol{\sigma}) = 0$, with $\boldsymbol{\sigma}$ as the stress tensor) that separates the admissible stress states from those that, regardless of the exact ``failure mechanism" (cracking, crushing or combinations thereof), cannot possibly be attained. These surfaces are typically asymmetric, reflecting a markedly different behavior in tension and compression. Various models (also denoted as  ``failure criteria") have been proposed to represent strength surfaces, including the Rankine, Mohr–Coulomb and Drucker–Prager criteria. The shape of the strength surface is considered a material property and constitutes a fundamental ingredient in the description of material failure \cite{bazant_planas, mroz}. 
%In this paper, our focus lies on multiaxial stress states including the special case of uniaxial tension. 
As hinted to above, the attainment of the strength surface at a material point may coincide with nucleation of a new crack, defined as the transition from homogeneous to localized damage \citep{de2022nucleation}, but may also lead to the development of diffuse damage. 

To study propagation of existing cracks, the historically first approach was Griffith's brittle fracture theory, which gave birth to the field of fracture mechanics in 1921. Griffith's theory is unable to predict the nucleation of a new crack; it postulates the existence of a crack and -- under some simplifying assumptions, including the restriction to idealized geometries and a planar crack path -- it can predict whether this crack will or not propagate based on the comparison between energy release rate and fracture toughness. According to Griffith, the energy cost associated to fracture is independent of the magnitude of the displacement jump across the crack and proportional to the area of the crack. For decades after Griffith's seminal contribution, the modeling of failure in brittle and quasi-brittle materials evolved along two separate theoretical paths: strength-based criteria, which predict crack nucleation (or initiation of damage) when a critical strength surface is reached, and fracture mechanics, which describes crack propagation based on energy release rates. These approaches remained conceptually disconnected until the introduction of cohesive zone models \citep{dugdale1960yielding, barenblatt1962mathematical} provided a bridge between them by postulating a traction–separation relationship within a finite process zone ahead of the crack tip (or, equivalently, an energy cost of fracture depending on the magnitude of the displacement jump). This allows for a continuous transition from intact material to fully developed cracks and unifies the treatment of crack initiation and propagation. For quasi-brittle materials, cohesive approaches gained prominence following the work of Hillerborg et al. \cite{Hillerborg}.

In the variational reformulation of Griffith's theory \cite{francfort1998revisiting}, the evolution of the displacement field and of the crack set within a body is dictated by a sequence of minimizers of a total energy, given by the sum of the elastic bulk energy and of the energy cost associated to fracture. As postulated by Griffith, this cost is proportional to the area of the crack set, an assumption that prevents the correct prediction of crack nucleation also within the variational framework. To incorporate the notion of cohesive zone model into a variational formulation, the fracture cost can be modified to depend on the magnitude of the displacement jump \cite{bourdin2008variational}. The resulting model associates finite values to critical stresses and provides sufficient freedom to prescribe the strength surface $\mathcal{F}(\boldsymbol{\sigma})=0$ \cite{charlotte2006initiation}. Variational formulations for both brittle and cohesive fracture remove the restriction to simple geometries and planar crack paths, thus solving - at least in theory - the important issue of crack path selection. However, they treat cracks as discontinuities, leading to a complex free-discontinuity problem that is challenging to solve numerically. 

This motivates the regularization carried out (for brittle fracture) in \cite{bourdin2000numerical} and now known as \textit{variational phase-field model of brittle fracture}; the sharp crack of the original model is replaced by a diffusive approximation over a regularization length scale $\ell$. Mathematically, the link between the regularized and the original (sharp-crack) model is provided by the notion of \(\Gamma\)-convergence, i.e., global minimizers of the regularized problem tend to those of the sharp-crack model for $\ell$ tending to zero \cite{marigo2016overview}. On the other hand, the regularized model can also be viewed as a special family of gradient damage models, whereby the phase-field variable can be interpreted as a damage variable \cite{Pham2010}. Due to the finite regularization length, the phase-field approach is able to predict the nucleation of a new crack which, within the variational phase-field framework, is identified with the localization of the phase-field variable. In turn, adopting local minimization as a criterion to select the stable states during a quasi-static evolution, localization is associated to the loss of stability of the solution corresponding to an almost uniform damage value. The associated nucleation load is the one at which the current solution branch stops being a local minimum of the energy, and crucially depends on $\ell$. This definition of nucleation leads to the interpretation of the length scale as a material parameter to be calibrated based on the tensile strength of the material, as proposed in \cite{TanLiBou17}. This approach proved successful for quantitative predictions of fracture under predominant tension (mode I), see e.g. \citep{amor2009regularized, pham2011gradient,TanLiBou17,nguyen_choice_2016,PhamRavi2017,WU2017658}. 

When it comes to fracture under multiaxial stress states, calibration of the tensile strength is no longer sufficient to correctly reproduce experimental data. While in standard isotropic phase-field models \citep{bourdin2000numerical} the nucleation threshold is symmetric in tension and compression,
more complex variational phase-field models have been developed to avoid crack interpenetration in compression and to obtain an asymmetric strength surface. The basic idea of these models is the decomposition of the strain energy density into active and inactive parts \citep{amor2009regularized,MieWelHof10, freddi2010regularized}, a concept known also in earlier literature on gradient damage \citep{mazars1986description,carol2001formulation,peerlings1996gradient}. Unfortunately, none of these models provides sufficient flexibility to match a desired shape of the strength surface. A first attempt to propose a variational phase-field model based on a more flexible energy decomposition to recover a strength surface of the Drucker-Prager type is made in \cite{de2022nucleation}. However, while achieving the goal of a greater flexibility with the strength surface, this model introduces spurious shear stresses at fully developed cracks \citep{vicentini2024energy}. Indeed, as extensively discussed in \citep{vicentini2024energy}, the obtainment of the desired strength surface  is only one of the important requirements of a variational phase-field model; additionally, the underlying local damage model must satisfy the strain-hardening and the stress-softening conditions, and the residual stress at fully developed damage must be ``crack-like", i.e. purely volumetric and depend on the negative trace of the strain tensor (see \citep{chambolle2018approximation} and earlier studies on damage models, e.g. \citep{RICHARD20101203}). In \citep{vicentini2024energy}, the authors propose a phase-field model based on a new energy decomposition specifically designed to meet all the aforementioned requirements. The model delivers a strength surface of ``star-convex" shape, which allows for the independent calibration of tensile and compressive strength while ensuring ``crack-like" residual stresses. A star-convex shape is not a realistic description of experimentally observed strength surfaces; indeed, the model in \citep{vicentini2024energy} is to be intended as a means of preventing spurious damage under compression in multiaxial settings (e.g. in the plate with hole under compression where points under tensile and compressive stress states compete for nucleation), and not as a model leading to a realistic strength surface. At the same time, this model is the only successful result of many alternative attempts based on the energy decomposition approach (see Appendix B in \citep{vicentini2024energy}), which seems to indicate the need to depart from this approach if a realistic strength surface is desired. 

The limitation of phase-field models of brittle fracture in accurately capturing the shape of the strength surface is not surprising, since all these models (despite adding some more complexity) at their core approximate Griffith’s theory, which inherently lacks a strength criterion. On the other hand, the study of phase-field regularizations of cohesive models and of the opportunities they offer for flexibly shaping the strength surface remains largely unexplored, despite being a promising path \citep{marigo2023modelling}. One of the first attempts in this direction is the use of gradient damage models with more complex degradation functions, which exhibit an asymptotically cohesive behavior (see \citep{lorentz2011convergence, Lorentz2012, zolesi2024stability} and the later generalization by \cite{WU201820}). A key advantage of these models is their ability to predict a tensile strength independent of the regularization length, which makes them particularly suitable for crack nucleation analyses. However, in the multiaxial case, these models offer limited control over the shape of the strength surface \cite{zolesi2024stability}. Another approach, proposed in the mathematical community, consists of a family of gradient damage models that \(\Gamma\)-converge to a sharp cohesive model \cite{conti2024phase}. However, these models require modifications to make them numerically implementable \cite{freddi2017numerical}. Moreover, due to their similarity in structure to phase-field models of brittle fracture, the extent to which they may enable a more flexible control of the shape of the strength surface remains open to question. 

In the past years, the above difficulties motivated several authors to propose non-variational phase-field models, in which the damage criterion does not stem from an energy minimization principle \cite{Abrari,miehe2015phase,zhang2017modification,feng2022phase,kumar2020revisiting,lopez2025classical} as in damage models of the pre-phase-field generation~\citep{lemaitre1990mechanics, mazars1986description, COMI20016427}. Abandoning the variational structure grants more flexibility to retrieve experimental strength surfaces under multiaxial loading. However, it also gives up many theoretical and practical advantages. The variational approach allows the use of the tools of calculus of variations to study the existence of solutions and $\Gamma$-convergence. It also enables quasi-static equilibrium states and evolution processes to be interpreted, as in many other areas of physics, as corresponding to the minimization or incremental minimization of a clearly defined energy. This also provides a clear distinction between admissible and stable, therefore observable, solutions. Moreover, abandoning the variational structure breaks the connection with the Drucker-Ilyushin postulate \cite{marigo1989constitutive}; as a result, non-variational models may lack thermodynamic consistency. Finally, from the computational standpoint a variational structure guarantees a symmetric tangent stiffness matrix, enables the utilization of optimization algorithms, and provides bounds to the error of the finite element discretization which may be conveniently used for mesh adaptation \cite{ORTIZ1999419}.

In this paper, we propose a new variational phase-field model of cohesive fracture which enables a flexibly tunable strength surface, while preserving all the other desired properties mentioned earlier. An inspiration for this model is the variational phase-field model of fracture coupled with plasticity in \cite{alessi2014gradient} which is shown to lead to the nucleation of cohesive cracks and whose $\Gamma$-convergence to a cohesive model is studied in \citep{dal2016fracture}. In the model proposed in this paper, the plastic strain is replaced by a more general notion of eigenstrain, reminiscent of the structured deformation studied in \citep{del1993structured}; this notion had already found a role in variational phase-field fracture models in the context of energy decompositions \citep{freddi2010regularized, freddi2011variational}, see also \citep{de2022nucleation}. 
The structural similarity between the proposed model and a plasticity model enables us to quite straightforwardly select a priori the desired shape of the strength surface \cite{marigo2019micromechanical}. In a parallel study \citep{Maggiorelli25} we also show that, in the discrete setting, the proposed model $\Gamma$-converges to a sharp cohesive fracture model. 

The paper is structured as follows. In Section~\ref{sec:general}, we introduce the fundamental assumptions and the general formulation of the novel variational phase-field model of cohesive fracture. In Section~\ref{sec:1D}, we specialize the formulation to the one-dimensional case, in order to highlight the main features of the model, in particular the homogeneous solution, its stability, and the cohesive behavior of the localized solution. We then move on to the three-dimensional case, where we focus on the study of the shape of the strength surface $\mathcal{F}(\boldsymbol{\sigma})=0$. Section~\ref{sec:numerics} presents numerical implementation aspects and provides illustrative examples that validate the analytical findings. The main conclusions are summarized in Section~\ref{sec:conclusions}.

Regarding the notation, we denote by $d$ the dimension of the problem; thus, $d=1$ for the one-dimensional case, and $d=3$ for the three-dimensional, plane-strain and plane-stress cases. Vectors and second-order tensors are written in bold, e.g. $\boldsymbol{u}$ and $\boldsymbol{\sigma}$ represent the displacement vector and the stress tensor, respectively. We refer to the set of symmetric second-order tensors as $\mathbb{M}_s^d$. Fourth-order tensors are written using blackboard bold notation, e.g. $\mathbb{C}$ represents the elasticity tensor. Their components are denoted with subscripts, e.g. $u_i$, $\sigma_{ij}$, $\mathbb{C}_{ijkl}$, where $1 \leq i, j, k, l \leq d$. The inner product of vectors and tensors is denoted using the dot symbol. Therefore, we have $\boldsymbol{u} \cdot \boldsymbol{v} = u_i\, v_i$, $\boldsymbol{\varepsilon} \cdot \boldsymbol{\varepsilon} = \varepsilon_{ij}\, \varepsilon_{ij}$, and $\mathbb{C} \boldsymbol{\varepsilon} \cdot \boldsymbol{\varepsilon} = \mathbb{C}_{ijkl}\, \varepsilon_{ij}\, \varepsilon_{kl}$, where we used Einstein summation convention. The Euclidean norm is denoted by $\Vert \cdot \Vert$, e.g., $\Vert \boldsymbol{u} \Vert = \sqrt{\boldsymbol{u} \cdot \boldsymbol{u}}$. The symbols $\nabla$ and $\nabla_{\text{sym}}$ denote the spatial gradient and its symmetric part, respectively. The Laplace operator $\Delta$ is defined as $\Delta:=\nabla\cdot \nabla$. The symbol $\otimes_{\text{sym}}$ denotes the symmetric tensor product. Accordingly, given the vectors $\boldsymbol{u}$ and $\boldsymbol{n}$, we have $\boldsymbol{u} \otimes_{\text{sym}} \boldsymbol{n} = \frac{1}{2} \left( \boldsymbol{u} \otimes \boldsymbol{n} + \boldsymbol{n} \otimes \boldsymbol{u} \right)$, with the symbol $\otimes$ denoting the tensor product. The symbol $\langle \cdot \rangle_{-}$ represents the negative Macaulay bracket, that is, $\langle \cdot \rangle_- := \frac{1}{2}(\cdot - |\cdot|)$. 

For the orthogonal decomposition into volumetric and deviatoric components of a second-order tensor $\boldsymbol{\sigma}$, we use the notation:
\begin{equation}
    \boldsymbol{\sigma} = \boldsymbol{\sigma}_{\text{vol}} + \boldsymbol{\sigma}_{\text{dev}}, \quad
    \boldsymbol{\sigma}_{\text{vol}} = \frac{\text{tr}(\boldsymbol{\sigma})}{d}\,\boldsymbol{I}, \quad
    \boldsymbol{\sigma}_{\text{dev}} = \boldsymbol{\sigma} - \frac{\text{tr}(\boldsymbol{\sigma})}{d}\,\boldsymbol{I}, \quad
    \boldsymbol{\sigma}_{\text{vol}} \cdot \boldsymbol{\sigma}_{\text{dev}} = 0,
\end{equation}
where $\boldsymbol{I}$ is the second-order identity tensor.

For an isotropic elastic material with Young's modulus $E$ and Poisson's ratio $\nu$, we define the first Lamé parameter $\lambda$, the bulk modulus $\kappa$ and the shear modulus $\mu$ as:
\begin{equation}
    \lambda:= \frac{E\,\nu}{(1+\nu)\,(1-(d-1)\nu)},\quad \kappa := \frac{E}{d\,(1-(d-1)\,\nu)}, \quad \mu := \frac{E}{2\,(1+\nu)}.
\end{equation}
The fourth-order elasticity tensor for isotropic materials in terms of the Lamé parameters reads $\mathbb{C}_{ijkl}=\lambda\,\delta_{ij}\,\delta_{kl}+\mu\,(\delta_{ik}\,\delta_{jl}+\delta_{il}\,\delta_{jk})$.

Given the vector field $\boldsymbol{z}$ and the functional $\mathcal{E}(\boldsymbol{z})$, the expression $\mathcal{E}'(\boldsymbol{z})[\boldsymbol{\delta}]$ denotes the Gâteaux derivative of $\mathcal{E}$ at $\boldsymbol{z}$ along the direction $\boldsymbol{\delta}$:
\begin{equation}
    \mathcal{E}'(\boldsymbol{z})[\boldsymbol{\delta} ] :=\frac{d}{dh}\,\mathcal{E}(\boldsymbol{z}+h\boldsymbol{\delta})\Bigg\vert_{h=0}.
\end{equation}
For functionals, we refer to differentiability and directional differentiability as Fréchet and Gâteaux differentiability, respectively \cite{2007introductory}.

\section{General formulation}
\label{sec:general}

In this section, we illustrate the general formulation of our variational phase-field model of cohesive fracture. A central ingredient of the proposed model is the eigenstrain variable \(\boldsymbol{\eta}\), which plays a role similar to that of the plastic strain in \cite{alessi2014gradient}. This variable is crucial in generating cohesive cracks, as we will show that \(\boldsymbol{\eta}\) is directly linked to the displacement jump occurring at the crack. However, unlike the plastic strain, \(\boldsymbol{\eta}\) retains a reversible nature.
%, leading to a \textit{non-linear elastic} behavior.

In the framework of eigenstrain theory, a similar non-linear prestrain is used to capture the effects of mechanical microdefects \cite{mura2013micromechanics}, which are otherwise often modeled using plasticity-based approaches \cite{marigo2019micromechanical}. A similar concept is also employed within the theory of structured deformations \citep{del1993structured} and was already connected to phase-field modeling of brittle fracture through the work of \citep{freddi2010regularized,freddi2011variational}, see also \citep{de2022nucleation}, to formulate different types of energy decomposition.
%account for residual elastic energies, i.e., the energy attained for a fully developed damage state  \cite{de2022nucleation}. Here, we extend this approach to define the elastic energy for any value of the damage variable.

\subsection{Total energy density}
\label{subsec:tot_en_dens}

Consider a homogeneous body that occupies the domain $\Omega\subset \mathbb{R}^d$. The \textit{state of the volume element} at a point \( \boldsymbol{x} \in \Omega\) is represented by the strain tensor $\boldsymbol{\varepsilon}$, the eigenstrain \( \boldsymbol{\eta} \), the local damage \( \alpha \), and its gradient \( \nabla\alpha \).  The damage variable \(\alpha\) is a scalar internal variable that expresses the level of degradation at a given point \(x\). It is bounded between $0$ and $1$ which represent the intact and the fully damaged states, respectively.
\par The first energetic quantity that we introduce is the \textit{elastic energy density} $\psi$ defined as
\begin{equation}
    \psi(\boldsymbol{\varepsilon}, \boldsymbol{\eta}, \alpha) := \psi_e(\boldsymbol{\varepsilon}-\boldsymbol{\eta})+ \pi(\boldsymbol{\eta},\alpha)
    \label{eq:strain_en_d}
\end{equation}  
with
\begin{equation}
\psi_e(\boldsymbol{\varepsilon}-\boldsymbol{\eta}) :=\frac{1}{2} \mathbb{C}\left(\boldsymbol{\varepsilon} - \boldsymbol{\eta}\right)\cdot \left(\boldsymbol{\varepsilon} - \boldsymbol{\eta}\right),
\end{equation}  
where we adopted the kinematic assumption that \( \boldsymbol{\varepsilon} = \boldsymbol{\varepsilon}_e + \boldsymbol{\eta} \), with  $\boldsymbol{\varepsilon}_e $ as the linear elastic strain, \( \mathbb{C} \) is the fourth-order elasticity tensor, and \( \pi \) is the \textit{eigenstrain potential}.
By differentiation, we obtain the stress tensor $\boldsymbol{\sigma}$
\begin{equation}
    \boldsymbol{\sigma}(\boldsymbol{\varepsilon},\boldsymbol{\eta}):=\frac{\partial \psi(\boldsymbol{\varepsilon},\boldsymbol{\eta},\alpha)}{\partial \boldsymbol{\varepsilon}}= \frac{\partial \psi_e(\boldsymbol{\varepsilon}-\boldsymbol{\eta})}{\partial \boldsymbol{\varepsilon}}= \mathbb{C}\,\left(\boldsymbol{\varepsilon}-\boldsymbol{\eta}\right).
\end{equation}
The second-order tensors $\boldsymbol{\varepsilon}$, $\boldsymbol{\eta}$, and $\boldsymbol{\sigma}$ are all symmetric, hence they are elements of $\mathbb{M}_s^d$. 
\par The eigenstrain potential \(\pi\) governs the elastic non-linearities. The onset of these non-linearities is determined by specifying the non-empty closed set \(\mathcal{S}(\alpha) \subset \mathbb{M}^d_s\), which the stress tensor \(\boldsymbol{\sigma}\) is constrained to lie in. We refer to this set as the (stress) \textit{elastic domain}.  When the stress $\boldsymbol{\sigma}$ is on the boundary of $\mathcal{S}(\alpha)$, $\boldsymbol{\eta}$ may take non-zero values and the material exhibits non-linear elastic behavior.
We make the following key assumptions:  

\begin{enumerate}
    \item \( \mathcal{S}(\alpha) \) is convex;
    \item \( \mathcal{S}(\alpha) \) is axisymmetric with respect to the direction of purely volumetric stress;
    \item \(\mathcal{S}(\alpha)\) is bounded in all \(\boldsymbol{\sigma}\) directions in \(\mathbb{M}_s^d\), except in the negative direction of purely volumetric stresses, i.e., it is unbounded for \(\boldsymbol{\sigma}_{\text{dev}}=\mathbf{0},\text{tr}(\boldsymbol{\sigma})<0\);
    \item \( \mathcal{S}(\alpha) \) scales homothetically with respect to \(\alpha\), i.e., \(\mathcal{S}(\alpha)=a(\alpha)\,\mathcal{S}_0\),  
\end{enumerate}  
where $\mathcal{S}_0$ is the \textit{initial elastic domain} and \(a(\alpha)\) is the \textit{degradation function}, which must satisfy the conditions:  

\begin{equation}
a(0) = 1, \hspace{3mm} a(1) = 0, \hspace{3mm} a'(1)=0, \hspace{3mm}a'(\alpha) \leq 0,
\hspace{3mm}a''(\alpha)\geq 0\quad \text{for } 0 \leq \alpha < 1.
\end{equation}

%Without loss of generality, 
The second assumption simplifies our subsequent formulation as, for \( d = 3 \), it will be sufficient to provide a two-dimensional description of \( \mathcal{S}(\alpha) \) in the \( \text{tr}(\boldsymbol{\sigma}) - \Vert\boldsymbol{\sigma}_{\text{dev}}\Vert \) plane.
 As we will see later, the third assumption automatically implies a non-interpenetration condition. Finally, the fourth assumption along with $a'(\alpha) \leq 0$ inherently ensures the \textit{stress-softening} condition, i.e., for all \(\beta>\alpha\), \(\mathcal{S}(\beta)\subset \mathcal{S}(\alpha)\) \cite{Pham2010}. The initial elastic domain $\mathcal{S}_0$, assumed to be a material property, is a non-empty closed convex set defined through a convex function $\mathcal{F}:\mathbb{M}_s^d\rightarrow \mathbb{R}$ as follows
\begin{equation}
    \mathcal{S}_0:=\left\{\boldsymbol{\sigma}\in \mathbb{M}_s^d:\mathcal{F}(\boldsymbol{\sigma})\leq 0\right\}.
    \label{eq:elastic_dom}
\end{equation}
We denote the surface $\mathcal{F}(\boldsymbol{\sigma})=0$ or, equivalently, the boundary $\partial\mathcal{S}_0$ of the initial elastic domain as the \textit{strength surface} specific of the material. Additionally we assume that $\boldsymbol{0}\in \mathcal{S}_0$. 
\par To each convex set, we can associate a convex \textit{support function} \cite{boyd2004convex}. Let us define $\pi_0$ as the support function of the initial elastic domain 
\begin{equation}
\pi_0(\boldsymbol{\eta}):=\sup_{\boldsymbol{\sigma}\in \mathcal{S}_0}\boldsymbol{\sigma}\cdot \boldsymbol{\eta}\in\overline{\mathbb{R}}
\label{eq:support_fun}
\end{equation}
\noindent where $\overline{\mathbb{R}}= \mathbb{R} \cup \{-\infty,\infty\}$ is the extended real line. To highlight its connection with the initial elastic domain, whose boundary is the strength surface, in the following we denote $\pi_0(\boldsymbol{\eta})$ as \textit{strength potential}. By construction, $\pi_0(\boldsymbol{\eta})$ is convex and \textit{positively} homogeneous of degree 1, i.e.,
\begin{equation}
    \pi_0(\lambda\,\boldsymbol{\eta})=\lambda\,\pi_0(\boldsymbol{\eta}),\hspace{3mm}\forall \lambda>0,\hspace{3mm}\forall \boldsymbol{\eta}\in \mathbb{M}_s^d.
\end{equation}
Additionally,  $\pi_0(\boldsymbol{\eta})$ is nonnegative since $\boldsymbol{0}\in\mathcal{S}_0$ and it vanishes at $\boldsymbol{\eta}=\boldsymbol{0}$. The support function $\pi_0(\boldsymbol{\eta})$ may not be differentiable, but it is always \textit{directionally} differentiable \cite{rockafellar1997convex}. Regarding it as a functional, we also say that it may not be Fréchet differentiable, but it is always Gâteaux differentiable.

Similarly, for an arbitrary admissible value of the damage variable $\alpha$, we define the eigenstrain potential $\pi(\boldsymbol{\eta}, \alpha)$ as the support function of the convex set $\mathcal{S}(\alpha)$. Based on our assumptions about the elastic domain, we can express it as
\begin{equation}
\pi(\boldsymbol{\eta}, \alpha):=
\begin{cases}
a(\alpha)\pi_0(\boldsymbol{\eta}) & \text{if } \text{tr}(\boldsymbol{\eta}) \geq 0, \\
+\infty & \text{otherwise},
\end{cases}
\label{eq:barrier}
\end{equation}
where 
%the set of $\boldsymbol{\eta}\in \mathbb{M}_s^d$ such that $\text{tr}(\boldsymbol{\eta}) \geq 0$ represents the so-called \textit{barrier cone} of $\mathcal{S}_0$. 
the constraint $\text{tr}(\boldsymbol{\eta}) \geq 0$ in order to obtain a finite potential  directly stems from the assumption of an unbounded stress elastic domain exclusively in the negative direction of purely volumetric stresses.

%\begin{equation}
%    \pi(\boldsymbol{\eta},\alpha):=\begin{cases}
%        a(\alpha)\pi_0(\boldsymbol{\eta}),&\hspace{3mm}\text{if }\boldsymbol{\eta}\in b(\mathcal{S}_0)\\
%        +\infty &\hspace{3mm}\text{otherwise},
%    \end{cases}
%    \label{eq:support_h}
%\end{equation}
%where $b(\mathcal{S}_0)$ is the \textit{barrier cone} of $\mathcal{S}_0$, i.e., the cone on which $\pi_0(\boldsymbol{\eta})<+\infty$. Specifically, beacuse of our assumption of having a bounded domain except along the negative volumetric stress direction, we have that
%\begin{equation}
%    b(\mathcal{S}_0)=\left\{\boldsymbol{\eta}: \text{tr}(\boldsymbol{\eta})\geq 0\right\}.
%    \label{eq:barrier}
%\end{equation}

The second energetic quantity that we introduce is the \textit{fracture energy density} \( d_{\ell} \). For this quantity, we adopt the same definition used in the context of the phase-field formulation for brittle fracture \cite{marigo2016overview}:  

\begin{equation}
d_{\ell}(\alpha, \nabla\alpha) = \frac{G_{\text{c}}}{c_w} \left( \frac{w(\alpha)}{\ell} + \ell \, \nabla\alpha\cdot \nabla\alpha \right), \hspace{3mm} \text{with } c_w = 4 \int_0^1 \sqrt{w(\beta)} \, d\beta.
\label{eq:Am_To}
\end{equation}
Here, \( \ell \) is the \textit{regularization length}, \( G_\text{c} \) is the \textit{fracture toughness}, and \( w(\alpha) \) is the \textit{dissipation function}, which must satisfy
\begin{equation}
w(0) = 0, \hspace{3mm} w(1) = 1, \hspace{3mm} w'(\alpha) \geq 0 \quad \text{for } 0 \leq \alpha \leq 1.    
\end{equation}
The functional expression in (\ref{eq:Am_To}) with a specific choice for $a(\alpha)$ and $w(\alpha)$ was originally introduced to regularize the Mumford-Shah image segmentation problem in \cite{ambrosio1990approximation}, applied to regularize the variational fracture problem in \cite{bourdin2000numerical} and generalized to a wider family of gradient damage models in \cite{Pham2010}. In most of this paper, we adopt for the degradation function \( a(\alpha) \) and the dissipation function \( w(\alpha) \) the choices of the so-called \texttt{AT2} model \cite{ambrosio1990approximation,Pham2010,ambati2015review}, as follows
\begin{equation}
a(\alpha) = (1 - \alpha)^2, \hspace{3mm} w(\alpha) = \alpha^2, \hspace{3mm} c_w = 2.
\label{eq:AT2}
\end{equation}

The sum of the elastic energy density and the fracture energy density yields the state function  

\begin{equation}
    W_\ell(\boldsymbol{\varepsilon}, \boldsymbol{\eta}, \alpha, \nabla\alpha) := \psi(\boldsymbol{\varepsilon}, \boldsymbol{\eta}, \alpha) + d_\ell(\alpha, \nabla\alpha),
\end{equation}  
denoted as \textit{total energy density}. For a given strain \(\boldsymbol{\varepsilon}\), we require the total energy density to be convex to ensure a well-posed problem. To this end, we impose the \textit{strain-hardening} condition:  

\begin{equation}
    W_{\ell}''(\boldsymbol{\varepsilon},\boldsymbol{\eta},\alpha, \boldsymbol{0}) [\boldsymbol{0},\boldsymbol{\zeta}, \beta,\boldsymbol{0}][\boldsymbol{0},\boldsymbol{\zeta}, \beta,\boldsymbol{0}]\geq 0,
    \label{eq:strain_hardening}
\end{equation}
where $W_{\ell}''(\boldsymbol{\varepsilon},\boldsymbol{\eta},\alpha, \boldsymbol{0}) [\boldsymbol{0},\boldsymbol{\zeta}, \beta,\boldsymbol{0}][\boldsymbol{0},\boldsymbol{\zeta}, \beta,\boldsymbol{0}]$ denotes the second Gâteaux derivative of $W_\ell$ at $(\boldsymbol{\varepsilon},\boldsymbol{\eta},\alpha, \boldsymbol{0})$ evaluated in the direction of the admissible perturbation $(\boldsymbol{0},\boldsymbol{\zeta}, \beta,\boldsymbol{0})$. 
%\begin{equation}
%    W_{\ell}(\boldsymbol{\varepsilon},\boldsymbol{\eta},\alpha, \boldsymbol{0}) \text{ is convex w.r.t. } (\boldsymbol{\eta},\alpha)\, \forall \boldsymbol{\varepsilon}.
%    \label{eq:strain_hardening}
%\end{equation}
This condition guarantees that, for a given strain value, the solution is unique in terms of eigenstrain and damage.

\subsection{Total energy functional}
\par Our next goal is to formulate the variational problem to describe the evolution of the state variables: the displacement $\boldsymbol{u}$, the eigenstrain $\boldsymbol{\eta}$ and the damage variable $\alpha$. We assume to be in the quasi-static regime parametrized by the (pseudo-)time $t$ and we consider the time-discrete setting. For the sake of simplicity, we assume zero body forces and surface tractions, and Dirichlet boundary conditions on the part $\partial_D\Omega$ of the boundary of $\Omega$ (which we denote as $\partial\Omega$) where a displacement $\boldsymbol{U}_t$ is prescribed. By integrating the total energy density $W_{\ell}$, we obtain the \textit{total energy functional}
\begin{equation}
    \mathcal{E}_{\ell}(\boldsymbol{u},\boldsymbol{\eta},\alpha):=\int_{\Omega} W_{\ell}(\boldsymbol{\varepsilon}(\boldsymbol{u}(\boldsymbol{x})),\boldsymbol{\eta}(\boldsymbol{x}),\alpha(\boldsymbol{x}),\nabla\alpha(\boldsymbol{x}))\,d\boldsymbol{x}
    \label{eq:total_energy}
\end{equation}
which is the sum of the \textit{elastic energy functional} $\Psi(\boldsymbol{u}, \boldsymbol{\eta},\alpha)$ and of the \textit{fracture energy functional} $\mathcal{D}_{\ell}(\alpha)$
\begin{equation}
    \mathcal{E}_{\ell}(\boldsymbol{u},\boldsymbol{\eta}, \alpha) = \Psi(\boldsymbol{u},\boldsymbol{\eta}, \alpha)+\mathcal{D}_{\ell}(\alpha),
\end{equation}
with 
\begin{equation}
    \Psi(\boldsymbol{u},\boldsymbol{\eta},\alpha):=\int_{\Omega} \psi(\boldsymbol{\varepsilon}(\boldsymbol{u}(\boldsymbol{x})),\boldsymbol{\eta}(\boldsymbol{x}),\alpha(\boldsymbol{x}))\,d\boldsymbol{x}
\hspace{3mm}\text{and}\hspace{3mm}\mathcal{D}_{\ell}(\alpha):= \int_{\Omega} d_{\ell}(\alpha(\boldsymbol{x}),\nabla\alpha(\boldsymbol{x}))\,d\boldsymbol{x}.
\label{eq:energy_quantities}
\end{equation}

\begin{figure}
\centering
\centering
    \includegraphics[scale=1]{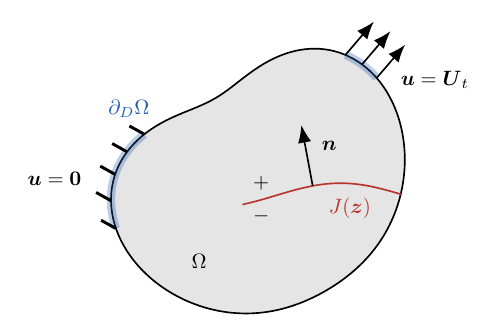}

\caption{The problem domain $\Omega$ with the Dirichlet boundary $\partial_D\Omega$ highlighted in blue and the jump set $J(\boldsymbol{z})$ shown in red. At a generic point on the jump set we depict the unit normal vector $\boldsymbol{n}$, which defines the positive side ``$+$'' and the negative side ``$-$'' in the neighborhood of the jump set.}
\label{fig:prob_domain}
\end{figure}

We introduce the vector field collecting the state variables, $\boldsymbol{z}(\boldsymbol{x}) = (\boldsymbol{u}(\boldsymbol{x}), \boldsymbol{\eta}(\boldsymbol{x}),\alpha(\boldsymbol{x}))$, which we also refer to as the \textit{state field}. We assume that $\boldsymbol{z}$ is piecewise smooth, with a singular part localized on the \textit{jump set} $J(\boldsymbol{z})$, which we assume to consist of a finite number of smooth, non-intersecting subsets of $\Omega\setminus\partial_D\Omega$ of codimension 1 (i.e. curves for $d=2$ or surfaces for $d=3$). Thus, to every point on $J(\boldsymbol{z})$ we can associate a positively oriented unit normal vector $\boldsymbol{n}$ (Figure~\ref{fig:prob_domain}). 

Despite its singularity on $J(\boldsymbol{z})$, the state field $\boldsymbol{z}$ must be sufficiently regular to ensure that the total energy functional remains finite. To guarantee this, we follow the framework outlined in \cite{alessi2015gradient}:

\begin{itemize}
    \item We require that the displacement field $\boldsymbol{u}$ be continuously differentiable on $\Omega \setminus J(\boldsymbol{z})$, while it may admit jumps across $J(\boldsymbol{z})$. The displacement jump is denoted as $\jump{\boldsymbol{u}(\boldsymbol{x})} = \boldsymbol{u}^+(\boldsymbol{x}) - \boldsymbol{u}^-(\boldsymbol{x})$, where $\boldsymbol{u}^+(\boldsymbol{x})$ is the limit of $\boldsymbol{u}$ approaching $\boldsymbol{x}$ from the direction pointed at by the normal $\boldsymbol{n}$ and $\boldsymbol{u}^-(\boldsymbol{x})$ is the limit of $\boldsymbol{u}$ approaching $\boldsymbol{x}$ from the opposite direction (Figure~\ref{fig:prob_domain}). Accordingly, the strain tensor $\boldsymbol{\varepsilon}(\boldsymbol{u})$ is decomposed into a regular component $\boldsymbol{\varepsilon}_R(\boldsymbol{u})$ and a singular component $\boldsymbol{\varepsilon}_S(\boldsymbol{u})$. With a slight abuse of notation, we write
    \begin{equation}
        \boldsymbol{\varepsilon}(\boldsymbol{u}) = \boldsymbol{\varepsilon}_R(\boldsymbol{u}) + \boldsymbol{\varepsilon}_S(\boldsymbol{u}),
        \quad \text{with} \quad
        \boldsymbol{\varepsilon}_R(\boldsymbol{u}) = \nabla_{\text{sym}} \boldsymbol{u}, 
        \quad \text{and} \quad
        \boldsymbol{\varepsilon}_S(\boldsymbol{u})  = \left(\jump{\boldsymbol{u}} \otimes_{\text{sym}} \boldsymbol{n}\right) \, \delta_{J(\boldsymbol{z})},
        \label{eq:strain_jump}
    \end{equation}
    where $\delta_{J(\boldsymbol{z})}$ denotes the Dirac surface measure concentrated on the jump set $J(\boldsymbol{z})$ \cite{alessi2015gradient}. At time $t$, a displacement field $\boldsymbol{u}$ is regarded as admissible if it satisfies the stated regularity properties and it additionally fulfills the boundary condition $\boldsymbol{u} = \boldsymbol{U}_t$ on $\partial_D\Omega$.
    
    \item Similarly, the field $\boldsymbol{\eta}$ is decomposed into a regular part $\boldsymbol{\eta}_R$ and a singular part $\boldsymbol{\eta}_S$, such that $\boldsymbol{\eta} = \boldsymbol{\eta}_R + \boldsymbol{\eta}_S$. To ensure finite energy, we impose the condition that $\boldsymbol{\varepsilon} - \boldsymbol{\eta} \in L^2(\Omega)$, i.e., this difference must be square-integrable. As a consequence, we obtain $\boldsymbol{\eta}_S = \boldsymbol{\varepsilon}_S$, hence we can write
    \begin{equation}
        \boldsymbol{\eta} = \boldsymbol{\eta}_R + \left(\jump{\boldsymbol{u}} \otimes_{\text{sym}} \boldsymbol{n}\right) \, \delta_{J(\boldsymbol{z})}.
        \label{eq:eta_jump}
    \end{equation}
    Recall also that from (\ref{eq:barrier}), in order to obtain a finite energy, it must be $\text{tr}(\boldsymbol{\eta}) \geq 0$.
    
    \item The damage field $\alpha$ takes values in $[0,1]$ and its gradient must be square-integrable, implying that $\alpha$ belongs to the Sobolev space $H^1(\Omega)$. Consequently, \( \alpha \) remains continuous, although its gradient may exhibit discontinuities on \( J(\boldsymbol{z}) \). Thus, in general, for $\beta\in H^1(\Omega)$ with $\beta$ vanishing on $\partial\Omega$, we have  
\begin{equation}
    -\int_{\Omega} \nabla \alpha \cdot \nabla \beta \, d\boldsymbol{x} = \int_{\Omega \setminus J(\boldsymbol{z})} \Delta \alpha \, \beta \, d\boldsymbol{x} + \int_{J(\boldsymbol{z})} \jump{\nabla \alpha} \cdot \boldsymbol{n} \,\beta\,\delta_{J(\boldsymbol{z})}.
\end{equation}
Moreover, the damage variable must satisfy the \textit{irreversibility} condition. In the time-discrete setting, this simply requires that at the current time step \( \alpha \geq \alpha_p \), where \( \alpha_p \) denotes the damage from the previous step.
\end{itemize}

A more rigorous definition of the functional space for $\boldsymbol{z}$ is beyond the scope of this work. If the state field $\boldsymbol{z}$ satisfies the conditions specified above, we say that $\boldsymbol{z}$ is \textit{admissible}. Concisely, we denote the admissibility condition as $\boldsymbol{z} \in \mathcal{Z}_t$.

\subsection{The time-discrete evolution problem}
We now have all the ingredients to formulate a time-discrete variational principle. The evolution of the displacement, eigenstrain, and damage at time $t$ is determined by solving the minimization problem:
\begin{equation}
    (\boldsymbol{u},\boldsymbol{\eta},\alpha) = \arglocmin_{(\boldsymbol{v}, \boldsymbol{\zeta},\beta)\in\mathcal{Z}_t} \mathcal{E}_{\ell}(\boldsymbol{v},\boldsymbol{\zeta},\beta).
    \label{eq:locmin}
\end{equation}
Equation \eqref{eq:locmin} requires that $(\boldsymbol{u},\boldsymbol{\eta}, \alpha)$ satisfy the \textit{local stability condition}
\begin{equation}
\begin{gathered}
    \forall (\boldsymbol{v},\boldsymbol{\zeta},\beta) \in \mathcal{Z}_t, \quad \exists \Bar{h} > 0 \text{ such that } \forall h \in [0,\Bar{h}], \\
    \mathcal{E}_{\ell}(\boldsymbol{u} + h (\boldsymbol{v} - \boldsymbol{u}), \boldsymbol{\eta} + h (\boldsymbol{\zeta} - \boldsymbol{\eta}), \alpha + h (\beta - \alpha)) \geq \mathcal{E}_{\ell}(\boldsymbol{u},\boldsymbol{\eta},\alpha)
\end{gathered}
\end{equation}
and a necessary condition for local stability is the \textit{first-order stability condition}
\begin{equation}
    \mathcal{E}'_{\ell}(\boldsymbol{u},\boldsymbol{\eta},\alpha)[\boldsymbol{v}-\boldsymbol{u},\boldsymbol{\zeta}-\boldsymbol{\eta},\beta-\alpha]\geq 0,\hspace{3mm}\forall(\boldsymbol{v},\boldsymbol{\zeta},\beta)\in\mathcal{Z}_t.
    \label{eq:first_opt}
\end{equation}
By standard arguments of calculus of variation, the first-order optimality condition (\ref{eq:first_opt}) with respect to $\boldsymbol{u}$ is equivalent to the \textit{equilibrium equation} and boundary conditions
\begin{equation}
    \text{div}\,\boldsymbol{\sigma}(\boldsymbol{\varepsilon},\boldsymbol{\eta})=\boldsymbol{0}\hspace{3mm}\text{in }\Omega,\hspace{5mm}\boldsymbol{\sigma}(\boldsymbol{\varepsilon},\boldsymbol{\eta})\,\boldsymbol{m}=\boldsymbol{0}\hspace{3mm}\text{on }\partial\Omega\setminus\partial_D\Omega,
    \label{eq:equilibrium}
\end{equation}
where $\boldsymbol{m}$ is the outer unit normal to the boundary $\partial\Omega$. From the optimality with respect to $\boldsymbol{\eta}$ we obtain the \textit{eigenstrain evolution criterion} in the $\boldsymbol{\zeta}$ direction\footnote{Due to condition~(\ref{eq:barrier}), in order to ensure finite energy, the eigenstrain tensor must have a non-negative trace. Therefore, we distinguish between the case \(\text{tr}(\boldsymbol{\eta}) = 0\), in which only perturbations \(\boldsymbol{\zeta} \in \mathbb{M}_s^d\) with \(\text{tr}(\boldsymbol{\zeta}) > 0\) are admissible, and the case \(\text{tr}(\boldsymbol{\eta}) > 0\), in which any \(\boldsymbol{\zeta} \in \mathbb{M}_s^d\) is allowed.} 
\begin{equation}
    \boldsymbol{\sigma}(\boldsymbol{\varepsilon},\boldsymbol{\eta})\cdot \boldsymbol{\zeta}\leq \pi'(\boldsymbol{\eta},\alpha)[\boldsymbol{\zeta},0]\hspace{3mm}\text{in }\Omega
    \label{eq:nl_law} 
\end{equation}
 For the special case in which $\pi(\boldsymbol{\eta},\alpha)$ is Fréchet differentiable, we obtain the Karush-Kuhn-Tucker (KKT) conditions
\begin{equation}
    \boldsymbol{\sigma}(\boldsymbol{\varepsilon},\boldsymbol{\eta})\leq \frac{\partial\pi(\boldsymbol{\eta},\alpha)}{\partial \boldsymbol{\eta}}, \hspace{3mm}\text{tr}(\boldsymbol{\eta}) \ge 0, \hspace{3mm} \left[ \boldsymbol{\sigma}(\boldsymbol{\varepsilon},\boldsymbol{\eta})- \frac{\partial\pi(\boldsymbol{\eta},\alpha)}{\partial \boldsymbol{\eta}}\right]\text{tr}(\boldsymbol{\eta})=\mathbf{0} \hspace{3mm}\text{in }\Omega.
    \label{eq:nl_law_Fréchet} 
\end{equation}
Finally, the optimality with respect to $\alpha$ delivers the KKT conditions on the regular part of the domain $\Omega \setminus J(\boldsymbol{z})$
\begin{equation}
        -\frac{\partial\pi(\boldsymbol{\eta},\alpha)}{\partial \alpha}  \leq  \frac{G_{\text{c}}}{c_w} \left(\frac{w'(\alpha)}{\ell}- 2\ell\Delta\alpha\right), \hspace{2mm} \alpha \geq \alpha_{p}, \hspace{2mm} \left[\frac{\partial\pi(\boldsymbol{\eta},\alpha)}{\partial \alpha}+\frac{G_{\text{c}}}{c_w}\left(\frac{w'(\alpha)}{\ell} - 2\ell\Delta\alpha\right)\right](\alpha -\alpha_{p})=0\hspace{2mm}\text{in}\ \Omega\setminus J(\boldsymbol{z}),
    \label{eq:KKT_damage1}
\end{equation}
on the side of the jump set $J(\boldsymbol{z})$ with positive unit normal $\boldsymbol{n}$
\begin{equation}
    -\frac{\partial\pi(\jump{\boldsymbol{u}}\otimes_{\text{sym}}\boldsymbol{n},\alpha)}{\partial \alpha}\leq -2\,\frac{G_{\text{c}}}{c_w}\,\ell\,\jump{\nabla\alpha}\cdot \boldsymbol{n},\hspace{3mm}\alpha\geq \alpha_p,\hspace{3mm}\left[\frac{\partial\pi(\jump{\boldsymbol{u}}\otimes_{\text{sym}}\boldsymbol{n},\alpha)}{\partial \alpha} -2\frac{G_{\text{c}}}{c_w}\,\ell\,\jump{\nabla\alpha}\cdot \boldsymbol{n}\right](\alpha-\alpha_p)=0\hspace{3mm}\text{on }J(\boldsymbol{z}),
    \label{eq:KKT_damage2}
\end{equation}
and on the domain boundary $\partial\Omega$
\begin{equation}
\nabla\alpha\cdot\boldsymbol{m}\geq0, \hspace{3mm} \alpha \geq \alpha_{p}, \hspace{3mm} (\nabla\alpha\cdot \boldsymbol{m})(\alpha -\alpha_{p})=0\hspace{3mm}\text{on}\ \partial\Omega.
\label{eq:KKT_damage3}
\end{equation}
We denote these KKT conditions as \textit{damage criterion}, \textit{irreversibility} and \textit{loading-unloading condition}, respectively. 

\begin{remark}
The eigenstrain evolution criterion (\ref{eq:nl_law}) (or \ref{eq:nl_law_Fréchet}) delivers a $\mathrm{stress} \,\, \mathrm{criterion}$ that depends only on local quantities, unlike the equilibrium equation (\ref{eq:equilibrium}) and the KKT conditions (\ref{eq:KKT_damage1}--\ref{eq:KKT_damage3}). Moreover, since the total energy functional \( \mathcal{E}_{\ell} \) is \textit{Fréchet differentiable} with respect to \( \boldsymbol{u} \) and \( \alpha \), perturbation directions can be omitted in (\ref{eq:equilibrium}) and (\ref{eq:KKT_damage1}--\ref{eq:KKT_damage3}). In contrast, \( \mathcal{E}_{\ell} \) is generally only \textit{Gâteaux differentiable} in \( \boldsymbol{\eta} \), so directional dependence must be retained in (\ref{eq:nl_law}), which in turn affects the shape of \( \mathcal{S}(\alpha) \), namely its width along different stress directions, as we will see in Section \ref{sec:3D}.
\end{remark}

The choice to express the evolution problem in the variational form (\ref{eq:locmin}) has important implications. The first, which we highlight, is the automatic satisfaction of the \textit{non-interpenetration} condition. Indeed, according to (\ref{eq:barrier}), $\pi(\boldsymbol{\eta},\alpha)=+\infty$ when $\text{tr}(\boldsymbol{\eta})<0$, and since the evolution problem is governed by the minimization of total energy, this implies that $\text{tr}(\boldsymbol{\eta}(\boldsymbol{x}))\geq 0$ for any $\boldsymbol{x}\in\Omega$. In particular, on $J(\boldsymbol{z})$, we have $\text{tr}(\boldsymbol{\eta}_S)\geq 0$. Since $\boldsymbol{\eta}_S$ depends explicitly on the displacement jump $\jump{\boldsymbol{u}}$ as in (\ref{eq:eta_jump}), we obtain  
    \begin{equation}
        \text{tr}(\boldsymbol{\eta}_S)=\left(\jump{\boldsymbol{u}}\cdot \boldsymbol{n}\right)\,\delta_{J(\boldsymbol{z})}\geq 0,
    \end{equation}
    leading to  
    \begin{equation}
        \jump{\boldsymbol{u}}\cdot \boldsymbol{n}\geq 0\quad \text{on } J(\boldsymbol{z}).
        \label{eq:non_int}
    \end{equation}
   which represents the non-interpenetration condition.
    
Another important consequence is the \textit{normality rule}. As shown in \ref{append:hill}, a direct consequence of the variational formulation is that
\begin{equation}
    \boldsymbol{\sigma}(\boldsymbol{\varepsilon},\boldsymbol{\eta}) \in \mathcal{S}(\alpha), \quad (\boldsymbol{\sigma}(\boldsymbol{\varepsilon},\boldsymbol{\eta}) - \boldsymbol{\sigma}^*) \cdot \boldsymbol{\eta} \geq 0, \quad \forall\, \boldsymbol{\sigma}^* \in \mathcal{S}(\alpha),
    \label{eq:normality_rule}
\end{equation}
which is analogous to \textit{Hill's principle} in plasticity. Consequently, for a given stress tensor $\boldsymbol{\sigma}$, the orientation of the eigenstrain tensor $\boldsymbol{\eta}$ follows the normality rule. When the boundary of $\mathcal{S}(\alpha)$ is smooth, $\boldsymbol{\eta}$ is oriented as the gradient of the strength surface function $\mathcal{F}(\boldsymbol{\sigma})$. When the boundary has angular points, $\boldsymbol{\eta}$ belongs to the \textit{cone of outer normals} to $\mathcal{S}$ at the boundary point $\boldsymbol{\sigma}$~\cite{simo1997computational}.

An additional implication, useful later, follows from (\ref{eq:gat_der}) in \ref{append:hill}, where we have $\pi'(\boldsymbol{\eta},\alpha)[\boldsymbol{\eta},0] = -\pi'(\boldsymbol{\eta},\alpha)[-\boldsymbol{\eta},0] = \pi(\boldsymbol{\eta},\alpha)$.  
Thus, the eigenstrain evolution criterion (\ref{eq:nl_law}) implies that
\begin{equation}
    \boldsymbol{\sigma}(\boldsymbol{\varepsilon},\boldsymbol{\eta})\in\mathcal{S}(\alpha), \quad \text{for}\quad \text{tr}(\boldsymbol{\eta}) > 0, \quad \boldsymbol{\sigma}(\boldsymbol{\varepsilon},\boldsymbol{\eta})\cdot \boldsymbol{\eta} = \pi(\boldsymbol{\eta},\alpha).
    \label{eq:eta_ev}
\end{equation}

Furthermore, thanks to \eqref{eq:nl_law}, it can be shown (see \ref{app:strain_hard}) that a sufficient condition to satisfy the strain-hardening requirement \eqref{eq:strain_hardening} when using the \texttt{AT2} model is
\begin{equation}
    \ell \leq \frac{1}{4}\,\ell_{\text{ch}} \quad \text{with} \quad \ell_{\text{ch}} := \min_{\boldsymbol{\sigma}_\text{c} \in \partial \mathcal{S}_0} \frac{G_{\text{c}}}{\mathbb{S}\,\boldsymbol{\sigma}_\text{c} \cdot \boldsymbol{\sigma}_\text{c}},
    \label{eq:suff_SH}
\end{equation}
where $\ell_{\text{ch}}$ denotes the material's \textit{characteristic cohesive length}, $\mathbb{S}$ is the elasticity compliance tensor, and the second-order tensor $\boldsymbol{\sigma}_\text{c}$ is defined in \ref{app:strain_hard} and is considered a material property.

%\begin{remark}
%    If, for given $\boldsymbol{\eta}$ and $\alpha$, the non-linear elastic potential $\pi(\boldsymbol{\eta}, \alpha)$ is Fréchet differentiable with respect to $\boldsymbol{\eta}$, then its Gâteaux derivative is a linear operator. That is, there exists $\boldsymbol{\sigma}_\text{c}(\boldsymbol{\eta}, \alpha) \in \mathbb{M}_s^d$ such that
%    \[
%    \pi'(\boldsymbol{\eta}, \alpha)(\boldsymbol{\zeta}, 0) = \boldsymbol{\sigma}_\text{c}(\boldsymbol{\eta}, \alpha) \cdot \boldsymbol{\zeta}.
 %   \]
 %   Under this condition, the eigenstrain evolution law (\ref{eq:nl_law}) becomes
  %  \begin{equation}
  %      \boldsymbol{\sigma} \leq \boldsymbol{\sigma}_\text{c}(\boldsymbol{\eta}, \alpha),
  %  \end{equation}
   % therefore, in this specific case, the criterion for the growth of $\boldsymbol{\eta}$ results to be independent on the perturbation direction.
%end{remark}

\section{One-dimensional phase-field model of cohesive fracture}
\label{sec:1D}

So far, we have introduced the main energetic quantities and derived the governing equations for the evolution problem. Next, we aim to specialize the formulation to the one-dimensional case, i.e., \(d=1\). This will enable us to analytically demonstrate the cohesive behavior of the proposed model in a simplified context. Subsequently, we will study the three-dimensional case, \(d=3\), and, by proposing specific expressions for the eigenstrain potential \(\pi\), we will illustrate its role in nucleation of a cohesive crack by analyzing the resulting nucleation domains.
\subsection{Energetic quantities and evolution problem in the one-dimensional setting}
In the one-dimensional setting, the displacement \( u \), the strain \( \varepsilon \), and the eigenstrain \( \eta \) are scalar quantities, and the elastic energy density reads
\begin{equation}
    \psi(\varepsilon, \eta, \alpha) = \psi_e(\varepsilon-\eta) + \pi(\eta, \alpha),
    \label{eq:strain_en_d_1D}
\end{equation}
with
\begin{equation}
    \psi_e(\varepsilon-\eta) = \frac{1}{2}\, E\, (\varepsilon - \eta)^2,
\end{equation}
where \( E \) is the material's \textit{elastic modulus}. The stress
\begin{equation}
\label{stress1D}
    \sigma(\varepsilon, \eta) = \frac{\partial \psi(\varepsilon, \eta, \alpha)}{\partial \varepsilon} = \frac{\partial \psi_e(\varepsilon-\eta)}{\partial \varepsilon} = E\, (\varepsilon - \eta).
\end{equation}
is also a scalar quantity. The four assumptions on \( \mathcal{S} \) (Section \ref{subsec:tot_en_dens}) lead to the initial domain
\begin{equation}
    \mathcal{S}_0 = \left\{ \sigma \in \mathbb{R} : \sigma \leq \sigma_{\text{c}} \right\}.
\end{equation}
with $\sigma_{\text{c}}>0$ as the scalar counterpart of $\boldsymbol{\sigma}_\text{c}$ in (\ref{eq:suff_SH}).
Thus, the eigenstrain potential becomes
\begin{equation}
    \pi(\alpha, \eta) =
    \begin{cases}
        a(\alpha)\, \sigma_{\text{c}}\, \eta &\hspace{3mm} \text{if } \eta \geq 0, \\
        +\infty &\hspace{3mm} \text{otherwise}.
    \end{cases}
    \label{eq:nlpot}
\end{equation}

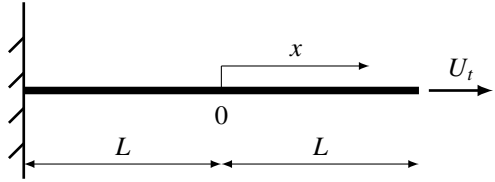
\begin{figure}[H]
        \centering
            \begin{tikzpicture}[scale = 0.65]
		        
		        %Beam
                \draw[line width=1 mm] (-4,0) -- (4,0);
		        
		        %Reference system
		        %horizontal line
		        \draw [-latex] (0,0.5) -- (3, 0.5) node[midway, above,scale=1.]{$x$};
		        %vertical line
		        \draw (0,0) -- (0,0.5);
		        %origin
		        \node[scale = 1.] (origin) at (0,-0.5) {$0$};
		        
		        %Length measure
		        %left
		        \draw [latex-latex] (-4,-1.5) -- (0, -1.5) node[midway, above, scale = 1.]{$L$};
		        %right
		        \draw [latex-latex] (0,-1.5) -- (4, -1.5) node[midway, above, scale = 1.]{$L$};
		        
		        %Prescribed displacement
			    \draw[-latex] [ line width = 1] (4.2, 0) -- (5.5, 0)
			    node[midway, above, scale = 1.]{$U_t$};
		        
		        %Clamp
		        %vertical
		        \draw[line width = 1] (-4,-1.8) -- (-4,1.8);
		        %inclined
		        \foreach \nn in {-1.08, -0.36, 0.36, 1.08} {
			    \draw[line width = 1
			    ](-4,\nn) -- (-4.3,\nn - 0.3);
			    };
		        
		\end{tikzpicture}
        \caption{One-dimensional setting: clamped bar under tension.}
        \label{fig:bar}
    \end{figure}

The one-dimensional domain corresponds to the bar \( \Omega = (-L, L) \) illustrated in Figure~\ref{fig:bar}. The bar is clamped at the left end, while a displacement \( U_t \) is imposed at the right end. Hence, the Dirichlet boundary conditions correspond to
\begin{equation}
    u(-L) = 0, \hspace{3mm} u(L) = U_t.
\end{equation}

Given the state field \( \boldsymbol{z} = (u, \eta, \alpha) \), the one-dimensional jump set \( J(\boldsymbol{z}) \) consists of a finite set of disjoint points. For simplicity, and without loss of generality, we assume that the jump set contains at most one point \( \in (-L, L) \). If this singular point exists, we assume that it is located at $x=0$. Consequently, in the absence of a crack, we have \( J(\boldsymbol{z}) = \emptyset \), while in presence of a crack \( J(\boldsymbol{z}) = \{0\} \). Under this assumption, Equations (\ref{eq:strain_jump}) and (\ref{eq:eta_jump}) simplify to
\begin{equation}
    \varepsilon(u) = u' + \jump{u} \, \delta_{0}, \hspace{3mm} \eta = \eta_R + \jump{u} \, \delta_{0},
    \label{eq:1Djump}
\end{equation}
where \( u'\) and \( \eta_R \) are the regular components of the strain and of the eigenstrain, respectively, and \( \delta_{0} \) denotes the Dirac delta at \( 0 \).

Using (\ref{eq:1Djump}), the total energy functional is given by
\begin{equation}
\begin{aligned}
        \mathcal{E}_{\ell}(u,\eta,\alpha) &= \int_{-L}^{L} \left\{ \frac{1}{2} E (\varepsilon(u) - \eta)^2 + a(\alpha) \sigma_{\text{c}} \eta + \frac{G_{\text{c}}}{c_w} \left( \frac{w(\alpha)}{\ell} + \ell \alpha'^2 \right) \right\} dx \\
        &= \int_{(-L,L) \setminus \{0\}} \left\{ \frac{1}{2} E (u' - \eta)^2 + a(\alpha) \sigma_{\text{c}} \eta + \frac{G_{\text{c}}}{c_w} \left( \frac{w(\alpha)}{\ell} + \ell \alpha'^2 \right) \right\} dx + a(\alpha(0)) \sigma_{\text{c}} \jump{u(0)}
        \label{eq:total_en_1D}
\end{aligned}
\end{equation}
if \( \eta(x) \geq 0 \) for all \( x \in [-L, L] \), otherwise, \( \mathcal{E}_{\ell}(u,\eta,\alpha) = +\infty \). As in the general formulation, since the total energy equals \( +\infty \) for \( \eta < 0 \) and the variational principle requires the minimization of the total energy, solutions with \( \eta < 0 \) are excluded. This implies that \( \eta(x) \geq 0 \) for all \( x \in [-L, L] \). From (\ref{eq:1Djump}) at $x=0$, we obtain
\begin{equation}
    \jump{u(0)} \geq 0,
\end{equation}
which represents the one-dimensional version of the non-interpenetration condition.

\par The one-dimensional counterpart of the equilibrium equation \eqref{eq:equilibrium}  
\begin{equation}
    \frac{\partial}{\partial x}\sigma(\varepsilon(u(x)),\eta(x))=0\hspace{3mm}\text{in }(-L,L)
    \label{eq:equilibrium1D}
\end{equation}
requires the stress \(\sigma\) to be constant throughout \( (-L, L) \).
The one-dimensional version of the eigenstrain evolution criterion (\ref{eq:nl_law_Fréchet}) reads
\begin{equation}
\sigma(\varepsilon,\eta) \leq a(\alpha)\,\sigma_{\text{c}},\hspace{3mm}\eta\geq 0,\hspace{3mm}(\sigma(\varepsilon,\eta)-a(\alpha)\,\sigma_{\text{c}})\,\eta=0\hspace{3mm}\text{in }(-L,L).
\label{eq:nl_law_1D}
\end{equation}
Along with (\ref{stress1D}), (\ref{eq:nl_law_1D}) can be equivalently rewritten as
\begin{equation}
\eta = \begin{cases}
    0&\hspace{3mm}\text{if }\varepsilon\leq a(\alpha)\,\frac{\sigma_{\text{c}}}{E},\\
    \varepsilon - a(\alpha)\,\frac{\sigma_{\text{c}}}{E}&\hspace{3mm}\text{otherwise}
\end{cases}
    \label{eq:eta_criterion}
\end{equation}
and, accordingly, 
\begin{equation}
    \sigma(\varepsilon,\eta)=\begin{cases}
        E\,\varepsilon \hspace{3mm}&\text{if }\eta=0,\\
        a(\alpha)\,\sigma_{\text{c}}\hspace{3mm}&\text{if }\eta>0
    \end{cases}\hspace{3mm}\text{in }(-L,L).
    \label{eq:stress_opt}
\end{equation}
From \eqref{eq:eta_criterion} and \eqref{eq:stress_opt} we can observe that, for a fully-damaged material (\(\alpha=1\)), the stress vanishes in tension (\(\sigma = 0 \ \text{for}\  \varepsilon \geq 0\)), while the material behaves as linear elastic in compression (\(\varepsilon < 0\)). Consequently, the elastic energy density \(\psi\) at \(\alpha=1\) is zero for \(\varepsilon \geq 0\) and quadratic in \(\varepsilon\) for \(\varepsilon < 0\).

The damage KKT conditions (\ref{eq:KKT_damage1}--\ref{eq:KKT_damage3}) read 
\begin{equation}
    -a'(\alpha)\,\sigma_{\text{c}}\,\eta\leq \frac{G_{\text{c}}}{c_w}\,\left(\frac{w'(\alpha)}{\ell}-2\,\ell\alpha''\right),\hspace{3mm}\alpha\geq \alpha_p,\hspace{3mm}\left[a'(\alpha)\,\sigma_{\text{c}}\,\eta+\frac{G_{\text{c}}}{c_w}\left(\frac{w'(\alpha)}{\ell}-2\,\ell\,\alpha''\right)\right](\alpha-\alpha_p)=0\hspace{3mm}\text{in }(-L,L)\setminus\{0\},
    \label{eq:kkt_1D}
\end{equation}
\begin{equation}
    -a'(\alpha)\,\sigma_{\text{c}}\,\jump{u}\leq -2\,\frac{G_{\text{c}}}{c_w}\,\ell\,\jump{\alpha'},\hspace{3mm}\alpha\geq \alpha_p,\hspace{3mm}\left(a'(\alpha)\,\sigma_{\text{c}}\,\jump{u}-2\,\frac{G_{\text{c}}}{c_w}\,\ell\,\jump{\alpha'}\right)(\alpha-\alpha_p)=0\hspace{3mm}\text{at }x=0,
    \label{eq:kkt_1D_J}
\end{equation}
\begin{equation}
        \alpha'\geq 0,\hspace{3mm}\alpha\geq \alpha_p,\hspace{3mm}\alpha'\,(\alpha-\alpha_p)=0\hspace{3mm}\text{at }x=\pm L.
    \label{eq:kkt_1D_BC}
\end{equation}
%These KKT conditions represent the one-dimensional damage criterion, irreversibility and loading-unloading condition, respectively.

\subsection{The effect of the eigenstrain on the elastic energy density}

\begin{figure}
    \centering
    \begin{subfigure}[t]{0.49\textwidth}
        \centering
        \includegraphics[scale=1]{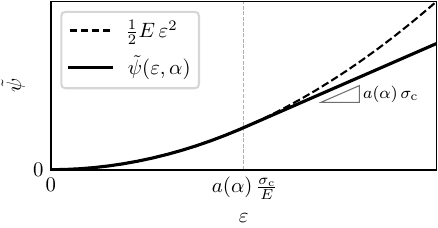}
        \caption{}
        \label{fig:relaxation_left}
    \end{subfigure}
    \begin{subfigure}[t]{0.49\textwidth}
        \centering
        \includegraphics[scale=1]{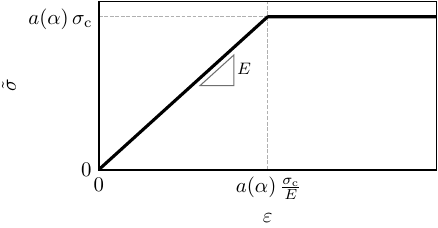}
        \caption{}
        \label{fig:relaxation_right}
    \end{subfigure}
    \caption{Effective elastic energy $\tilde{\psi}$ as a function of the strain $\varepsilon$ (solid line) for a given value of the damage $\alpha$. At the threshold strain $a(\alpha)\,\frac{\sigma_{\text{c}}}{E}$, the curve transitions from a quadratic to a linear behavior with slope $a(\alpha)\,\sigma_{\text{c}}$, reflecting the typical relaxation of the elastic energy density required for sharp cohesive fracture (see Sections 2.5 and 4.2 in~\cite{bourdin2008variational}) (a). Effective stress $\tilde{\sigma}$ as a function of the strain $\varepsilon$ for a given value of the damage $\alpha$ (b).}
    \label{fig:relaxation_double}
\end{figure}

In this section, we exploit the simplified setting of the one-dimensional problem to highlight the fundamental role of the eigenstrain $\eta$ on the elastic energy density. To illustrate this, we define the \textit{effective elastic energy density} \(\tilde{\psi}\) as  
\begin{equation}
\label{eff_en}
    \tilde{\psi}(\varepsilon,\alpha):=\min_{\eta\geq0}\psi(\varepsilon,\eta,\alpha)
\end{equation}  
which describes the mechanical response of the material for a given level of damage, accounting for the presence of eigenstrains. Exploiting (\ref{eq:eta_criterion}) (which is equivalent to the minimization in (\ref{eff_en})), we obtain  
\begin{equation}
    \tilde{\psi}(\varepsilon,\alpha)=\begin{cases}
        \frac{1}{2} E \varepsilon^2 &\hspace{3mm}\text{if } \varepsilon\leq a(\alpha)\,\frac{\sigma_{\text{c}}}{E},\\
        a(\alpha) \sigma_{\text{c}} \left[\varepsilon - \frac{1}{2} a(\alpha) \frac{\sigma_{\text{c}}}{E}\right] &\hspace{3mm}\text{otherwise},
    \end{cases}
    \label{eq:relax}
\end{equation}  
see Figure \ref{fig:relaxation_double}. From this relation, we observe that the effective elastic energy density follows the usual quadratic form for strain values \(\varepsilon\) below the critical threshold \(a(\alpha)\,\frac{\sigma_{\text{c}}}{E}\). Beyond this point, the function transitions to a linear dependence on \(\varepsilon\). We also define the \textit{effective stress} as the partial derivative of the effective elastic energy $\tilde{\psi}$ with respect to the strain $\varepsilon$, that is
\begin{equation}
    \tilde{\sigma}(\varepsilon,\alpha):=\frac{\partial\tilde{\psi}(\varepsilon,\alpha)}{\partial\varepsilon}=\begin{cases}
        E\,\varepsilon&\hspace{3mm}\text{if } \varepsilon\leq a(\alpha)\,\frac{\sigma_{\text{c}}}{E},\\ a(\alpha)\,\sigma_{\text{c}}&\hspace{3mm}\text{otherwise}.
    \end{cases}
    \label{eq:eff_stress}
\end{equation}

\begin{remark}  
\label{remark:effective}
    In the context of the variational sharp cohesive model, a relaxation of the elastic energy density analogous to the one in (\ref{eq:relax}) is required to obtain a lower-semicontinuous energy. This property is fundamental to ensure a well-posed minimization problem (see Sections 2.5 and 4.2 in~\cite{bourdin2008variational}).  
\end{remark}  

Exploiting the effective elastic energy density \(\tilde{\psi}\), we can write the effective total energy functional \(\tilde{\mathcal{E}}_\ell\), which depends only on the displacement \(u\) and the damage \(\alpha\), as
\begin{equation}
\tilde{\mathcal{E}}_\ell(u,\alpha)
:= \min_{\eta\geq0}\mathcal{E}_\ell(u,\eta,\alpha)=\int_{-L}^{L} \left\{
\tilde{\psi}\big(\varepsilon(u), \alpha\big)
+ \frac{G_c}{c_w} \left(
\frac{w(\alpha)}{\ell} + \ell {\alpha'}^2
\right)
\right\} \, dx.
\label{eq:effective_energy}
\end{equation}
In terms of displacement and damage, finding the pair \((u,\alpha)\) that locally minimizes the effective energy \(\tilde{\mathcal{E}}_\ell\) is equivalent to solving problem~(\ref{eq:locmin}).
The expression of the effective energy in~\eqref{eq:effective_energy} resembles that used in brittle phase-field fracture models \cite{marigo2016overview}, with the key difference that the effective elastic energy density \(\tilde{\psi}\) now introduces a non-linear constitutive law at fixed damage. The effect of the eigenstrain \(\eta\) is precisely to introduce this constitutive non-linearity, which physically represents the non-linear behavior in the process zone due to cohesive forces. In this sense, the eigenstrain can be regarded as a measure of the non-linearity within the cohesive zone. Consistently, the displacement jump corresponds to the integral of the singular part of \(\eta\). The energetic contribution of the cohesive non-linearity is represented by \(\pi\). In this work, \(\pi\) is interpreted as part of the elastic energy density \(\psi\) and not as a dissipative term, as expressed in \eqref{eq:strain_en_d} and \eqref{eq:strain_en_d_1D}. As highlighted in Remark~\ref{remark:effective}, this has a mathematical justification in the fact that \(\tilde{\psi}\) corresponds to the elastic energy density required to make the sharp cohesive model well-posed. With reference to \eqref{eq:strain_en_d}, \eqref{eq:strain_en_d_1D}, and \eqref{eq:relax}, \(\psi_e\) governs the quadratic part of \(\tilde{\psi}\), which provides the initial linear elastic behavior, while \(\pi\) governs the linear part of \(\tilde{\psi}\) necessary to achieve the limit stress \(\sigma_\text{c}\).

\begin{remark}
In plasticity, the formalism used here to derive the eigenstrains and the nonlinear elastic energy density $\tilde{\psi}$ in \eqref{eq:relax} originates from the \textit{deformation theory of plasticity} first proposed by Hencky \cite{hencky1924theorie}, which interprets elasto-plastic materials as nonlinear hyperelastic materials (see, e.g., \cite{budiansky1959reassessment, Kachanov1971}). Despite the formal similarity, we emphasize that in the present work the eigenstrain represents a reversible elastic strain, not a plastic one.
\end{remark}

Now that the energetic quantities and the governing equations of the problem have been formulated in the one-dimensional setting, in the following we aim to analyze two possible solutions obtained from first-order stability, namely the homogeneous solution and the solution localized about the point \( x=0 \), as well as to study the second-order stability of the homogeneous solution.
%In particular, we will demonstrate that, unlike in the classical phase-field approach for brittle fracture, the homogeneous solution is always unstable in this context. 

\subsection{The homogeneous solution} \label{sec:hom_sol}
The homogeneous solution is characterized by a damage value which is constant over the entire domain \( [-L, L] \), i.e., \( \alpha'(x) = 0 \). This implies that the strain \( \varepsilon \) and the eigenstrain \( \eta \) are both constant, leading to an empty jump set, i.e., \( J(\boldsymbol{z}) = \emptyset \). Consider the case where the initial values of displacement \( u \), eigenstrain \( \eta \), and damage  \( \alpha \) are zero, i.e., \( \boldsymbol{z} = \boldsymbol{0} \). The applied displacement \( U_t \) starts from zero and increases monotonically with time \( t \). In particular, for the displacement, the solution takes the form \( u(x) = \frac{U_t}{2L} x \), and the strain reads \( \varepsilon = u' = \frac{U_t}{2L} \).

\begin{figure}
    \centering
    \includegraphics[scale=1]{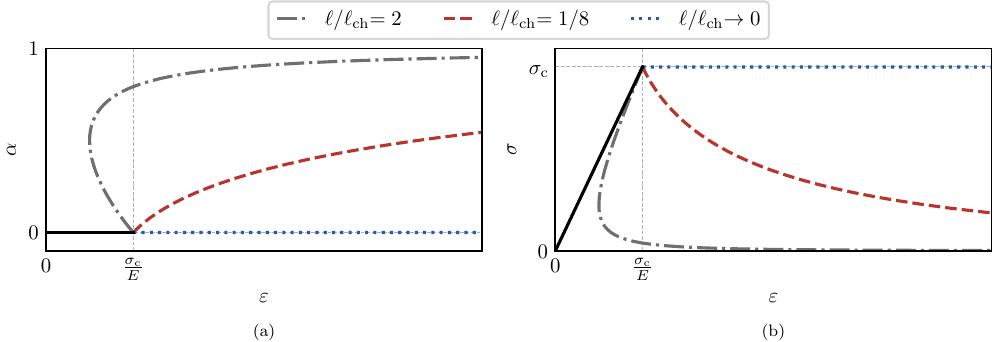}
    \caption{Homogeneous solution: Evolution of the damage $\alpha$ (a) and the stress $\sigma$ (b) as a function of $U_t/2L$. In both cases, the elastic regime response is represented by the solid black line. For the damaging regime, responses for different values of the length ratio $\ell/\ell_{\text{ch}}$ are presented. When the ratio $\ell/\ell_{\text{ch}} > 1/4$, the strain-hardening condition (\ref{eq:strain_hardening_Irwin}) is not satisfied, resulting in a snap-back (dash-dotted gray line).
} 
    \label{fig:hom_sol}
\end{figure}

For prescribed displacements \( U_t \leq \frac{\sigma_{\text{c}}}{E}\,2L \) (or \(\varepsilon\leq\frac{\sigma_{\text{c}}}{E}\)), the values \( \eta = 0 \) and \( \alpha = 0 \) satisfy both inequality~(\ref{eq:eta_criterion}) and the damage criterion~(\ref{eq:kkt_1D}). As a result, the material remains undamaged. According to equation~(\ref{eq:stress_opt}), the stress is given by \( \sigma = E \, \varepsilon \), indicating that the system is in the \textit{linear elastic regime} during this phase (Figure~\ref{fig:hom_sol}).

For prescribed displacements \( U_t > \frac{\sigma_{\text{c}}}{E} 2L \) (or \(\varepsilon > \frac{\sigma_{\text{c}}}{E}\)), the eigenstrain evolution criterion (\ref{eq:nl_law_1D}, \ref{eq:eta_criterion}) requires that
\begin{equation}
    \eta = \varepsilon - (1 - \alpha)^2 \frac{\sigma_{\text{c}}}{E}>0,
\end{equation}
where we have assumed to use the \texttt{AT2} model (as we do in the remainder of this section), see (\ref{eq:AT2}). Introducing this relation into the damage criterion in (\ref{eq:kkt_1D})$_1$ (where we use the equality since, under loading, the irreversibility constraint is not active), we obtain the expression of \( \varepsilon \) as a function of the damage variable
\begin{equation}
    \varepsilon = (1 - \alpha)^2 \frac{\sigma_{\text{c}}}{E} + \frac{1}{2}\,\frac{\alpha}{1 - \alpha} \frac{G_{\text{c}}}{\sigma_{\text{c}} \ell} =:\bar{\varepsilon}_{\ell}(\alpha).
    \label{eq:eps_law}
\end{equation}
Thus, when \( U_t > \frac{\sigma_{\text{c}}}{E} 2L \) (or \(\varepsilon > \frac{\sigma}{E}\)) the damage in the bar evolves according to the inverse law
\begin{equation}
    \alpha = \bar{\varepsilon}_{\ell}^{-1}\left(\varepsilon\right)
\end{equation}
and the material is in the \textit{damaging regime} (Figure~\ref{fig:hom_sol}). Using equation~(\ref{eq:nl_law_1D})$_1$ (written as equality since, during loading, the interpenetration constraint is not active), we obtain the stress as a function of $\varepsilon$
\begin{equation}
    \sigma = \sigma_{\text{c}}\,\left(1 - \bar{\varepsilon}_{\ell}^{-1}\left(\varepsilon\right)\right)^2 =:\bar{\sigma}_{\ell}\left(\varepsilon\right)
\end{equation}
whose trend is shown in Figure~\ref{fig:hom_sol}. Moreover, by combining equations~(\ref{eq:nl_law_1D})$_1$ and~(\ref{eq:eps_law}), with some simple manipulations we find that
\begin{equation}
    \lim_{\ell \rightarrow 0} \bar{\sigma}_{\ell}\left(\varepsilon\right)= \sigma_{\text{c}}. 
\end{equation}
The fact that, for vanishing regularization length, the stress remains constant and equal to \( \sigma_{\text{c}} \) (Figure~\ref{fig:hom_sol}) is a common feature of phase-field models exhibiting asymptotically cohesive behavior \cite{lorentz2011convergence, zolesi2024stability, freddi2017numerical}.

Given \( \bar{\varepsilon}_{\ell}(\alpha) \) in (\ref{eq:eps_law}), the strain-hardening condition can be expressed simply as \( \frac{d\bar{\varepsilon}_{\ell}(\alpha)}{d\alpha} \geq 0 \) for $\alpha\in[0,1)$, from which we derive the condition
\begin{equation}
    \ell \leq \frac{1}{4} \ell_{\text{ch}} \quad \text{with} \quad \ell_{\text{ch}} = \frac{G_{\text{c}} E}{\sigma_{\text{c}}^2},
    \label{eq:strain_hardening_Irwin}
\end{equation}
where \( \ell_{\text{ch}} \) represents the material’s \textit{Irwin length}. The Irwin length corresponds to the characteristic cohesive length of the material, already defined in (\ref{eq:suff_SH}), for the special case of the one-dimensional setting. When the strain-hardening condition is not satisfied, the local solution for $\eta$ and $\alpha$ at a given strain $\varepsilon$ is no longer unique, which leads to a snap-back in the damaging response, see Figure \ref{fig:hom_sol}. This loss of uniqueness gives rise to both numerical and theoretical difficulties \cite{zolesi2024stability}.
\par
\begin{figure}[H]
    \centering
    \includegraphics[scale=1, trim=0 0.55cm 0 0.3cm, clip]{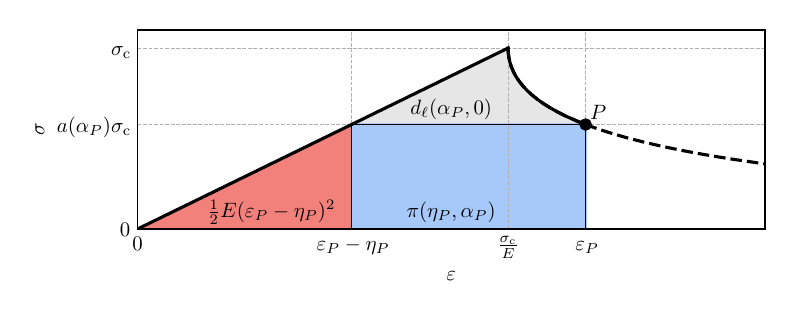}
    \caption{Energetic contributions to the total energy density $W_{\ell}$ at point $P$ in the stress-strain diagram corresponding to the local state $(\varepsilon_P,\eta_P,\alpha_P)$. The red triangular area represents the linear elastic energy density $\frac{1}{2}E(\varepsilon_P - \eta_P)^2$. The light blue rectangular area corresponds to the non-linear elastic energy density $\pi(\eta_P, \alpha_P)$. The remaining gray area denotes the local dissipation density $d_{\ell}(\alpha_P, 0)$.
} 
    \label{fig:areas}
\end{figure}

 Assuming that the bar has been loaded up to a state characterized by $(\varepsilon_P, \eta_P, \alpha_P)$, the total energy density $W_{\ell}$ everywhere in the bar is given by the sum of three contributions
\begin{equation}
    W_{\ell}(\varepsilon_P, \eta_P, \alpha_P,0) = \frac{1}{2}E(\varepsilon_P - \eta_P)^2 + \pi(\eta_P,\alpha_P)+ d_\ell(\alpha_P, 0)
\end{equation}
which are individually represented in Figure~\ref{fig:areas}. Of all three contributions, the only one that is dissipated is $d_{\ell}$, due to the irreversibility condition.

\par Let us now examine what happens during unloading, see the illustration in Table~\ref{tab:unloading}. Suppose that, following the softening curve $\bar{\sigma}_{\ell}\left(\varepsilon\right)$ during loading, we reach state B, which is characterized by a certain value of the eigenstrain $\eta$ and the damage variable $\alpha$. If unloading begins from this point, the eigenstrain $\eta$ starts to decrease; however, this does not result in a reduction of the damage, as $\alpha$ is constrained by the irreversibility condition. Therefore, in this non-linear elastic unloading phase (B$\rightarrow$C), the stress remains constant at the value $a(\alpha)\,\sigma_{\text{c}}$. Once the eigenstrain has decreased to zero, it cannot decrease any further, and unloading proceeds in a linear elastic fashion (C$\rightarrow$D). Upon reloading, the material first follows a linear elastic path again (D$\rightarrow$E) until the stress reaches the value $a(\alpha)\,\sigma_{\text{c}}$. Beyond this point, as loading continues, both the damage and the stress remain constant until the softening curve $\bar{\sigma}_{\ell}\left(\varepsilon\right)$ is encountered once more (E$\rightarrow$F). From there, further loading resumes along the softening curve, with a continued increase in both $\eta$ and $\alpha$.

Note that the above behavior, corresponding to the homogeneous solution and obtained from the first-order stability conditions of the total energy functional, is not observable in practice unless it can be proven to satisfy also second-order stability. The related analysis is carried out in the next section. 

\begin{table}
\centering
\begin{tabular}{|>{\centering\arraybackslash}m{4cm}|
                >{\centering\arraybackslash}m{6cm}|
                >{\centering\arraybackslash}m{4cm}|}
\hline
\textbf{\parbox[c][1cm][c]{4cm}{\centering Phase}} & \textbf{\parbox[c][1cm][c]{6cm}{\centering Stress-strain response}} & \textbf{\parbox[c][1cm][c]{4cm}{\centering Rates of internal variables}} \\
\hline
Initial linear elastic loading &
\vspace{0.5mm}\hspace{1mm}\includegraphics[scale=1, trim=0 0.5cm 0 0.2cm, clip]{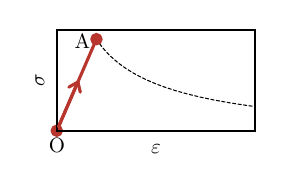}\hspace{1mm}\vspace{0.5mm} &
\begin{equation*}
    \begin{aligned}
        &\dot{\eta}=0\\
        &\dot{\alpha}=0
    \end{aligned}
\end{equation*} \\
\hline
Damaging loading &
\vspace{0.5mm}\hspace{1mm}\includegraphics[scale =1, trim=0 0.5cm 0 0.2cm, clip]{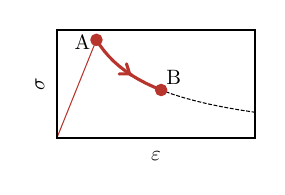}\hspace{1mm}\vspace{0.5mm} &
\begin{equation*}
    \begin{aligned}
        &\dot{\eta}>0\\
        &\dot{\alpha}>0
    \end{aligned}
\end{equation*} \\
\hline
Non-linear elastic unloading &
\vspace{0.5mm}\hspace{1mm}\includegraphics[scale =1, trim=0 0.5cm 0 0.2cm, clip]{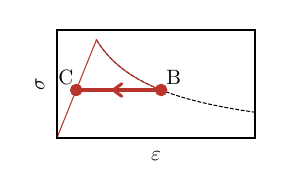}\hspace{1mm}\vspace{0.5mm} &
\begin{equation*}
    \begin{aligned}
        &\dot{\eta}<0\\
        &\dot{\alpha}=0
    \end{aligned}
\end{equation*} \\
\hline
Linear elastic unloading &
\vspace{0.5mm}\hspace{1mm}\includegraphics[scale =1, trim=0 0.5cm 0 0.2cm, clip]{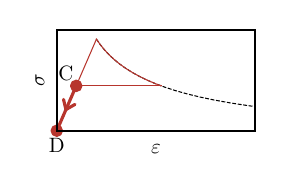}\hspace{1mm}\vspace{0.5mm} &
\begin{equation*}
    \begin{aligned}
        &\dot{\eta}=0\\
        &\dot{\alpha}=0
    \end{aligned}
\end{equation*} \\
\hline
Linear elastic reloading &
\vspace{0.5mm}\hspace{1mm}\includegraphics[scale =1, trim=0 0.5cm 0 0.2cm, clip]{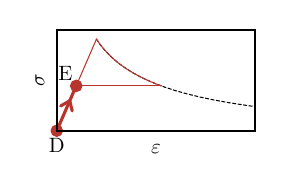}\hspace{1mm}\vspace{0.5mm} &
\begin{equation*}
    \begin{aligned}
        &\dot{\eta}=0\\
        &\dot{\alpha}=0
    \end{aligned}
\end{equation*} \\
\hline
Non-linear elastic reloading &
\vspace{0.5mm}\hspace{1mm}\includegraphics[scale =1, trim=0 0.5cm 0 0.2cm, clip]{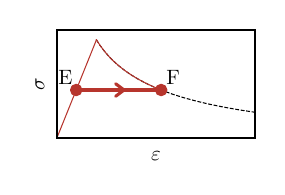}\hspace{1mm}\vspace{0.5mm} &
\begin{equation*}
    \begin{aligned}
        &\dot{\eta}>0\\
        &\dot{\alpha}=0
    \end{aligned}
\end{equation*} \\
\hline
Damaging reloading &
\vspace{0.5mm}\hspace{1mm}\includegraphics[scale =1, trim=0 0.5cm 0 0.2cm, clip]{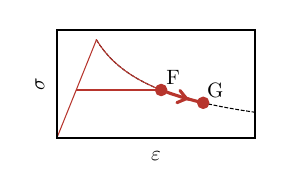}\hspace{1mm}\vspace{0.5mm} &
\begin{equation*}
    \begin{aligned}
        &\dot{\eta}>0\\
        &\dot{\alpha}>0
    \end{aligned}
\end{equation*} \\
\hline
\end{tabular}
\caption{Homogeneous solution: example of a loading–unloading history and corresponding evolution of stress and strain. The third column reports the sign of the rates of the eigenstrain $\dot{\eta}$ and of the damage variable $\dot{\alpha}$ (with a little abuse of notation since we are in the time-discrete setting).}
\label{tab:unloading}
\end{table}

\subsection{The instability of the homogeneous solution}

\par At this point, we aim to show that the obtained homogeneous solution $(u,\eta,\alpha)$, where
\begin{equation}
    u = (x+L)\,\varepsilon = (x+L)\,\bar{\varepsilon}_{\ell}(\alpha), \quad 
    \eta = \bar{\varepsilon}_{\ell}(\alpha) - (1-\alpha)^2 \frac{\sigma_{\text{c}}}{E} \quad \text{and} \quad \alpha'(x) = 0,
    \label{eq:hom_sol_full}
\end{equation}
valid for the \texttt{AT2} model and shown in Section \ref{sec:hom_sol} to be \textit{admissible}, i.e. to satisfy the first-order stability condition~\eqref{eq:first_opt}, is \textit{unstable}. 
\par To this end, it is sufficient to verify that there exists at least one admissible perturbation $(v,\zeta,\beta)$ such that the second Gâteaux derivative of the energy evaluated at $(u,\eta,\alpha)$ along $(v,\zeta,\beta)$ is negative, that is
\begin{equation}
    \exists(v,\zeta,\beta)\quad \text{such that} \quad 
    (u+v,\eta+\zeta,\alpha+\beta)\in \mathcal{Z}_t\quad\text{and}\quad \mathcal{E}_{\ell}''(u,\eta,\alpha)(v,\zeta,\beta) <0,
    \label{eq:second_stability}
\end{equation}
where the second Gâteaux derivative reads
\begin{equation}
    \mathcal{E}''_{\ell}(u,\eta,\alpha)[v,\zeta,\beta][v,\zeta,\beta] = 
    \int_{-L}^{L} \left[
    E\,(v' - \zeta)^2 
    - 4\,(1-\alpha)\,\sigma_{\text{c}}\,\zeta\,\beta 
    + 2\,\sigma_{\text{c}}\,\eta\,\beta^2 
    + G_{\text{c}} \left(\frac{\beta^2}{\ell} + \ell\,\beta'^2 \right)
    \right]\,dx.
    \label{eq:second_Gat}
\end{equation}
We consider the specific direction $(v_*,\zeta_*,\beta_*)$ given by
\begin{equation}
    v_*(x) = -\frac{2L}{\pi} \cos\left(\frac{\pi\,x}{2\,L}\right),\quad
    \zeta_*(x) = v_*'(x) = \sin\left(\frac{\pi\,x}{2\,L}\right),\quad
    \beta_*(x) = \epsilon\,v_*'(x) = \epsilon\,\sin\left(\frac{\pi\,x}{2\,L}\right),
    \label{eq:par_per}
\end{equation}
where $\epsilon$ is an arbitrarily small parameter. Note that the chosen function $v_*(x)$ satisfies the boundary conditions $v_*(-L) = v_*(L) = 0$. Moreover, in this work we focus exclusively on time-discrete evolutions and are concerned solely with stability of incremental type~\cite{baldelli2021numerical}. As a consequence, the damage perturbation $\beta_*(x)$ is allowed to take negative values. An analysis of time-continuous stability would require additional constraints on $\beta$, but this lies beyond the scope of the present work.

By inserting the perturbations from~\eqref{eq:par_per} into~\eqref{eq:second_Gat}, we obtain
\begin{equation}
    \frac{1}{\sigma_{\text{c}}L}\mathcal{E}''_{\ell}(u,\eta,\alpha)[v_*,\zeta_*,\beta_*][v_*,\zeta_*,\beta_*]= -4\,(1-\alpha)\,\epsilon+\left[2\,\eta+\frac{G_{\text{c}}}{\sigma_{\text{c}} L}\left(\frac{L}{\ell}+\frac{\pi^2}{4}\,\frac{\ell}{L}\right)\right]\,\epsilon^2,
\end{equation}
which is a second-order polynomial in $\epsilon$. In particular, for $0 \leq \alpha < 1$, the coefficient of the linear term is negative, while the coefficient of the quadratic term is always positive. Therefore, it is always possible to choose $\epsilon$ small enough to make the second Gâteaux derivative negative. 
Hence, the homogeneous solution $(u,\eta,\alpha)$ in (\ref{eq:hom_sol_full}) is \textit{always unstable}. This is in contrast with the results known for phase-field modeling of brittle fracture \citep{pham2013stability}, whereby second-order stability of the homogeneous solution depends on the $L/\ell$ ratio. We discuss this and other differences more extensively in Section \ref{comparison_brittle_cohesive}.
%, as one can always choose the test direction given in~\eqref{eq:par_per} such that the second Gâteaux derivative in (\ref{eq:second_Gat}) becomes negative. 

\subsection{The localized solution and the cohesive response}
Let us now consider a non-homogeneous solution. This alternative solution is constructed by assuming that the damage is localized and symmetric with respect to the point \( x = 0 \), monotonically increasing in the region \( (-L, 0) \), and decreasing in the other half, \( (0, L) \). Therefore, the maximum value of damage occurs at \( x = 0 \), i.e., \( \alpha(0) \). Since the stress \( \sigma \) must remain constant along the bar for equilibrium (\ref{eq:equilibrium1D}), the evolution criterion for the eigenstrain (\ref{eq:nl_law_1D})$_1$ requires that
\begin{equation}
    \sigma = a(\alpha(0))\,\sigma_{\text{c}}<a(\alpha(x))\,\sigma_{\text{c}} \quad \text{for} \quad x\in[-L, L] \setminus \{0\}
    \label{eq:sigma_0}
\end{equation}
which implies that
\begin{equation}
    \eta = 0 \quad \text{for} \quad x\in[-L, L]\setminus\{0\}
\end{equation}
and that, on the other hand, $\eta$ is singular at $x=0$. Accordingly, the KKT conditions (\ref{eq:kkt_1D}-\ref{eq:kkt_1D_BC}) require 
\begin{equation}
    \frac{1}{2}\,w'(\alpha)-\,\ell^2\,\alpha''=0\quad \text{in}\quad (-L,L)\setminus\{0\}, 
    \label{eq:dc_ode}
\end{equation}
\begin{equation}
    \jump{u(0)}=\frac{2}{c_w}\,\frac{G_{\text{c}}}{\sigma_{\text{c}}}\,\frac{\jump{\alpha'(0)}}{a'(\alpha(0))}\,\ell
    \label{eq:jump_eq}
\end{equation}
and the boundary conditions $\alpha'(-L)=\alpha'(L)=0$. 
By integrating in $\alpha$ equation (\ref{eq:dc_ode})  we obtain
\begin{equation}
    \vert \alpha'(x)\vert =\frac{1}{\ell}\sqrt{w(\alpha(x))}\quad \text{in}\quad (-L,L)\setminus\{0\}, 
    \label{eq:profile_a}
\end{equation}
that combined with (\ref{eq:jump_eq}) gives
\begin{equation}
    \jump{u(0)}=-\frac{4}{c_w}\,\frac{G_{\text{c}}}{\sigma_{\text{c}}}\frac{\sqrt{w(\alpha(0))}}{a'(\alpha(0))},
    \label{eq:jump_w}
\end{equation}
from which in particular we see that the relationship between the displacement jump $\jump{u(0)}$ and the damage $\alpha(0)$ does not depend on the regularization length $\ell$. Additionally, because of the assumption $a'(1)=0$, we have that the fully broken localized solution, i.e., the localized solution with $\alpha(0)=1$, is obtained only for an infinite magnitude of the displacement jump $\jump{u(0)}$. In this sense, the proposed model is of the Barenblatt type \cite{charlotte2006initiation}.
\par
\begin{figure}
    \centering
    \includegraphics[scale=1]{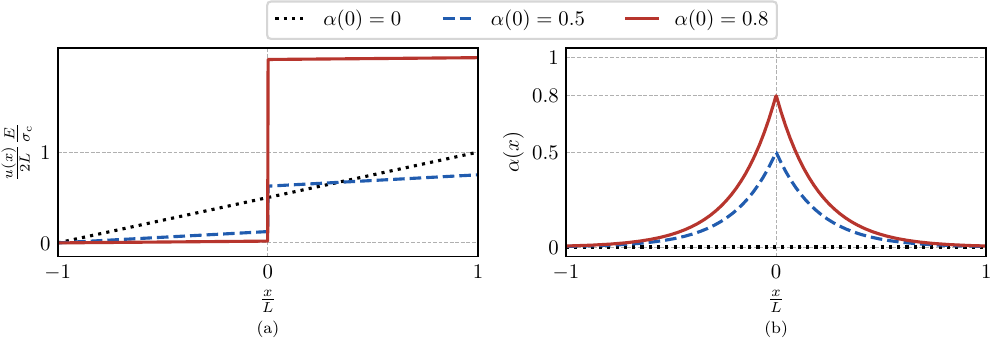}
    \caption{Localized solution for \( \frac{2L}{\ell_\text{ch}} = 2\) and \( \frac{\ell}{2L} = \frac{1}{10} \), shown for different values of the control parameter \( \alpha(0) \): normalized displacement field \( \frac{u(x)}{2L} \frac{E}{\sigma_{\text{c}}} \) (a), and the corresponding damage field \( \alpha(x) \) (b).
} 
    \label{fig:disp_field}
\end{figure}

\begin{figure}
    \centering
    \begin{subfigure}[t]{0.49\textwidth}
        \centering
        \includegraphics[scale=1]{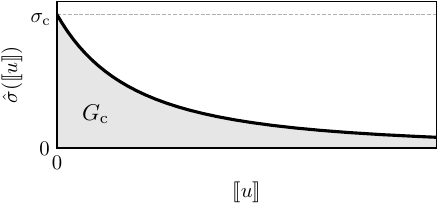}
        \caption{}
    \end{subfigure}
    \begin{subfigure}[t]{0.49\textwidth}
        \centering
        \includegraphics[scale=1]{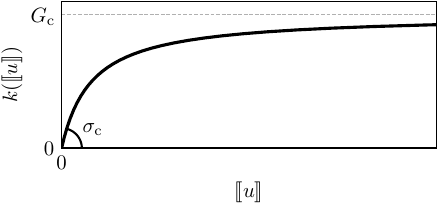}
        \caption{}
    \end{subfigure}
    \caption{Cohesive law $\hat{\sigma}(\jump{u})$ (a) and cohesive surface energy density $k(\jump
    u)$ (b). The area under the cohesive law $\hat{\sigma}(\jump{u})$ corresponds to the  fracture toughness $G_{\text{c}}$ of the material. The initial slope of the cohesive surface energy density $k(\jump{u})$ corresponds to the material strength $\sigma_{\text{c}}$.}
    \label{fig:cohesive_law1D}
\end{figure}

 In order to illustrate an explicit solution, we adopt again the \texttt{AT2} model. Under the simplifying assumption \(L \rightarrow \infty\), the solution to equation~(\ref{eq:dc_ode}) is given by  
\begin{equation}
    \alpha(x) = \alpha(0)\,\exp\left(-\frac{\vert x \vert}{\ell}\right) \quad \text{for} \quad x \in (-L, L),
    \label{eq:opt_prof}
\end{equation}  
as illustrated in Figure~\ref{fig:disp_field}b. The corresponding dissipated energy \(\mathcal{D}_{\ell}(\alpha)\) is  
\begin{equation}
    \mathcal{D}_{\ell}(\alpha) = \int_{-\infty}^{\infty} \frac{G_{\text{c}}}{2} \left( \frac{\alpha^2(x)}{\ell} + \ell\,\alpha'^2(x) \right) dx = G_{\text{c}}\,\alpha^2(0) =:D(\alpha(0)),
    \label{eq:diss_alpha0}
\end{equation}  
which is independent of the regularization length \(\ell\).  
Equation~(\ref{eq:jump_w}) becomes  
\begin{equation}
    \alpha(0) = \frac{\jump{u(0)}}{\frac{G_{\text{c}}}{\sigma_{\text{c}}} + \jump{u(0)}} =:g(\jump{u(0)}).
    \label{eq:alpha_jump}
\end{equation}  
Combining (\ref{eq:sigma_0},\ref{eq:opt_prof},\ref{eq:alpha_jump}), we obtain that $\sigma=\hat{\sigma}(\jump{u(0)})$, where $\hat{\sigma}(\jump{u})$ is the one-dimensional \textit{cohesive law} related to the \texttt{AT2} model (\ref{eq:AT2})
\begin{equation}
    \hat{\sigma}(\jump{u}):=\frac{\sigma_{\text{c}}}{\left(1+\frac{\sigma_{\text{c}}}{G_{\text{c}}}\,\jump{u}\right)^2}\quad \text{with}\quad \jump{u}\geq 0.
    \label{eq:coh_law1D}
\end{equation}
Note that the cohesive law is a convex function of the displacement jump $\jump{u}$, such that $\hat{\sigma}(0)=\sigma_{\text{c}}$, $\lim_{v\rightarrow \infty}\hat{\sigma}(v)=0$, and the area under the $\hat{\sigma}(\jump{u})$ curve is equal to the  fracture toughness $G_{\text{c}}$ of the material (Figure~\ref{fig:cohesive_law1D}).
By integration we obtain the one-dimensional \textit{cohesive surface energy density} $k([u])$:
\begin{equation}
k(\jump{u}):=\int_0^{\jump{u}}\hat{\sigma}(v)\,dv=\frac{G_{\text{c}}\,\jump{u}}{\frac{G_{\text{c}}}{\sigma_{\text{c}}}+\jump{u}}\quad \text{with}\quad \jump{u}\geq 0.
\label{eq:surf_dens_1D}
\end{equation}
The surface energy density $k(\jump{u})$ is a sublinear concave function of the displacement jump $\jump{u}$ such that $k(0)=0$, $k'(0)=\sigma_{\text{c}}$ and $\lim_{v\rightarrow \infty}k(v)=G_{\text{c}}$ (Figure~\ref{fig:cohesive_law1D}). Additionally, in the brittle limit, i.e., when $G_{\text{c}}/\sigma_{\text{c}} \rightarrow 0$, the cohesive energy density yields the Griffith surface energy density. Let us recall that the specific surface energy density derived in (\ref{eq:surf_dens_1D}) depends on the particular choice of the degradation functions \( a(\alpha) \) and \( w(\alpha) \) of the \texttt{AT2} model. On the other hand, by appropriately choosing \( a(\alpha) \) and \( w(\alpha) \), it is possible to recover an arbitrary cohesive law.

\par We observe that both the cohesive law and the surface energy density are independent of the regularization length \( \ell \). In this sense, the localized solution is \(\ell\)-insensitive. Conversely, the homogeneous solution depends explicitly on \( \ell \). This discrepancy between the homogeneous and localized solutions in the context of phase-field modeling of cohesive fracture has already been noted in \cite{zolesi2024stability}, in the context of models exhibiting asymptotic cohesive fracture behavior.

A theoretically important distinction between the models in \cite{zolesi2024stability, lorentz2011convergence} and the one proposed in this work lies in the convergence properties: while the present model also approaches a sharp cohesive model asymptotically, it can additionally be shown to \(\Gamma\)-converge to a sharp cohesive model. In particular, in \cite{Maggiorelli25} it is proven that the \(\Gamma\)-limit of the spatially discretized version of the energy in (\ref{eq:total_en_1D}) as \( \ell \rightarrow 0 \) corresponds to the cohesive energy functional 
\begin{equation}
    \mathcal{E}(u) = \int_{-L}^L \tilde{\psi}(u',0)\,dx + k(\jump{u}) + \sigma_{\text{c}}\,\vert C(u)\vert,
    \label{eq:sharp_model}
\end{equation}
where \( \vert C(u)\vert \) denotes the total variation of the Cantor part of \( u \) \cite{bourdin2008variational,bonacini_iurlano_2024}.

\begin{remark}
    Note that, under the assumptions of our model, the surface energy density does not coincide with the dissipated energy. In fact, it is straightforward to verify that
    \begin{equation}
        k(\jump{u}) \geq D(g(\jump{u}))
    \end{equation}
    and that $k$ equals the dissipated energy $D$ only in the case of a fully developed crack, i.e., when $\jump{u}\rightarrow\infty$ or, equivalently, $\alpha \rightarrow 1$, for which $k = D = G_{\text{c}}$. This mismatch between surface energy density and dissipation is characteristic of cohesive models and it depends on the specific choice of the irreversibility condition, as discussed in Section 5.2 in \cite{bourdin2008variational}. 
    In our case, $k \geq D$ because, as illustrated locally in Figure~\ref{fig:areas}, the portion of the total energy density associated with non-linearities, namely $\pi(\eta, \alpha)$, is not dissipated and can be recovered upon unloading.
\end{remark}

\subsection{The solution of the structural problem}
%With the obtained equations, we can describe the cohesive interface behavior specific of the material. 
With the cohesive law (\ref{eq:coh_law1D}) at hand (valid for the \texttt{AT2} model), we now move on to describe the solution of the structural problem for the bar of length \( 2L \). By inverting equation~(\ref{eq:alpha_jump}), we obtain 
\begin{equation}
    \jump{u(0)} =g^{-1}(\alpha(0))= \frac{\alpha(0)}{1 - \alpha(0)}\,\frac{G_{\text{c}}}{\sigma_{\text{c}}},
\end{equation}
from which we notice that, for a given material, the displacement jump \(\jump{u(0)}\) is controlled by the maximal damage \(\alpha(0)\). 
Thus, the normalized displacement field \(\frac{u(x)}{2L}\,\frac{E}{\sigma_{\text{c}}}\) is given by
\begin{equation}
    \frac{u(x)}{2L}\,\frac{E}{\sigma_{\text{c}}} = \begin{cases}
        \frac{1}{2}(1 - \alpha(0))^2\,\left(\frac{x}{L} + 1\right) & \text{if } -1 \leq \frac{x}{L} < 0,\\[5pt]
        \frac{\alpha(0)}{1 - \alpha(0)}\,\frac{1}{B} + \frac{1}{2}(1 - \alpha(0))^2\,\,\left(\frac{x}{L}+1\right) & \text{if } 0 < \frac{x}{L}\leq 1,
    \end{cases}
\end{equation}
see Figure~\ref{fig:disp_field}a, and is governed by \(\alpha(0)\) and by the \textit{structural brittleness ratio} B defined as
\begin{equation}
    B:=\frac{2L}{\ell_{\text{ch}}}.
    \label{eq:struct_brittleness}
\end{equation}
Accordingly, the normalized prescribed displacement reads
\begin{equation}
    \frac{U_t}{2L}\,\frac{E}{\sigma_{\text{c}}}=\frac{u(L)}{2L}\,\frac{E}{\sigma_{\text{c}}}=\frac{\alpha(0)}{1-\alpha(0)}\,\frac{1}{B}+(1-\alpha(0))^2
    \label{eq:structural_U},
\end{equation}
which again depends exclusively on \(\alpha(0)\) and $B$.

\begin{figure}[H]
    \centering
    \includegraphics[scale=1]{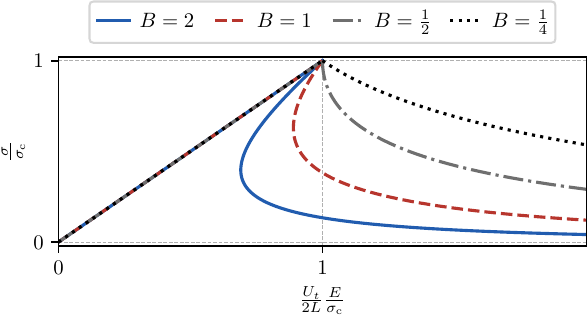}
    \caption{Structural (stress-displacement) response for different values of the structural brittleness ratio $B$. Values of the structural brittleness ratio $B > \frac{1}{2}$ lead to a response exhibiting snap-back.
} 
    \label{fig:structural_snap}
\end{figure}

Combining (\ref{eq:sigma_0}) and (\ref{eq:structural_U}) we can plot the structural (stress-displacement) response in Figure~\ref{fig:structural_snap}. We observe that, as the brittleness ratio $B$ increases, the structural response becomes more brittle. In particular, using equation~(\ref{eq:structural_U}), it is straightforward to derive that for values of the structural brittleness ratio $B > \frac{1}{2}$ the system exhibits a structural snap-back response, otherwise, $U_t$ increases monotonically with the control parameter $\alpha(0)$. Additionally, since the energy dissipated for a fully developed crack, i.e., $\alpha(0)=1$, corresponds to the fracture toughness $G_{\text{c}}$ according to (\ref{eq:diss_alpha0}), one can show that the area under the curve in Figure~\ref{fig:structural_snap} corresponds to the structural brittleness ratio $B$. This further explains the change of the structural response with varying $B$.

In Table \ref{tab:unloading_loc} we visualize the structural response including loading and unloading stages under the assumption of $B>\frac{1}{2}$. If the bar is loaded under displacement control, after reaching the peak point A localization occurs and a cohesive crack forms at $x=0$; as a result, the stress drops and the subsequent branch of the softening response is followed up to point B, which is characterized by a certain value of $\jump{u(0)}$ and $\alpha(0)$. If unloading begins from this point, $\jump{u(0)}$ starts to decrease, leading to a progressive closure of the cohesive crack; however, $\alpha(0)$ is constrained by the irreversibility condition, hence the stress remains constant at the value $a(\alpha(0))\,\sigma_{\text{c}}$ (B$\rightarrow$C). Once $\jump{u(0)}$ has decreased to zero, its value stays at zero and unloading proceeds in a linear elastic fashion (C$\rightarrow$D). Upon reloading, the material first follows a linear elastic path again (D$\rightarrow$E) until the stress reaches the value $a(\alpha(0))\,\sigma_{\text{c}}$. Beyond this point, as loading continues, the cohesive crack gradually reopens at constant damage and stress until the softening curve is encountered again (E$\rightarrow$F). Further loading follows the subsequent branch of the softening curve, with a continued increase in both $\jump{u(0)}$ and $\alpha(0)$.

\subsection{Comparison between phase-field models of brittle and cohesive fracture}
\label{comparison_brittle_cohesive}
In light of the one-dimensional analysis in the previous sections, let us summarize the key differences between predictions of one-dimensional phase-field models for brittle \cite{pham2011gradient} and for cohesive fracture.
\begin{enumerate}
    \item In the phase-field model for brittle fracture, homogeneous damage can be observed in short bars, i.e., when \( 2L/\ell \sim 1 \). In contrast, in the cohesive model proposed here, no homogeneous solution is observable even for short bars. Instead, reaching the critical stress always leads to localization and the formation of a cohesive crack.
    
    \item In the one-dimensional phase-field model for brittle fracture, the regularization length \( \ell \) determines the critical stress that (for sufficiently long bars) leads to localization and the formation of a crack. Thus, $\ell$ has to be interpreted as a material parameter in order for nucleation to be predicted at the correct level of stress. In contrast, in the present model for the one-dimensional setting, \( \ell \) does not affect the critical stress and must only be chosen small enough to accurately resolve the Irwin length, as required by the strain-hardening condition in (\ref{eq:strain_hardening_Irwin}).
    
    \item In the phase-field model for brittle fracture, the unloading path of the homogeneous solution follows the secant stiffness. Conversely, in the cohesive phase-field model, the unloading path takes the form illustrated in Table~\ref{tab:unloading}, which may be debatable. However, stability prevents the homogeneous solution from being observed, hence the issue does not arise in practice.
    \item  For the phase-field model of brittle fracture, the structural unloading response also follows the secant stiffness. In contrast, as shown in Table~\ref{tab:unloading_loc}, the cohesive model displays a different unloading path, including a phase at constant stress (B$\rightarrow$C). For a sufficiently large brittleness ratio (i.e., \( B \gg 1/2 \)), this stress is nearly zero, making the unloading response practically indistinguishable from the secant one. The branch at constant stress in the unloading behavior becomes apparent only when the brittleness parameter \( B \) is low enough for cohesive effects to manifest at the structural level.
    In any case, the observed unloading response is not surprising in this context: the proposed model, being an approximation of the variational sharp cohesive model~(\ref{eq:sharp_model}), inherits both its strengths and its limitations. In particular, a non-secant unloading path naturally arises in the variational sharp cohesive model as a consequence of the specific choice of the irreversibility condition. While irreversibility in the sharp brittle case is a purely geometric notion, in the cohesive case it is more delicate, as it depends on the definition of a suitable memory variable—which is notoriously an open issue, as discussed in Section~5.2 of~\cite{bourdin2000numerical}.
Several alternative formulations have been proposed in the mathematical literature~\cite{dal2007quasi, bourdin2000numerical, cagnetti2011quasistatic, crismale2018cohesive}. In this work, we adopt the irreversibility condition \( \alpha \geq \alpha_p \), which has also been used in the cohesive context in~\cite{bonacini2021cohesive, lorentz2011convergence, zolesi2024stability}. Nevertheless, the selection of an appropriate irreversibility criterion remains a complex topic that warrants further investigation and lies beyond the scope of the present work.

\end{enumerate}

%To conclude this section, let us also summarize the key differences between the predictions of the proposed model and those of previously proposed phase-field models with asymptotic  cohesive response \citep{freddi2010regularized,lorentz2011convergence, Lorentz2012,WU201820,zolesi2024stability}.

%\begin{enumerate}
%    \item Gamma convergence not in \cite{lorentz2011convergence, Lorentz2012,WU201820,zolesi2024stability}. Split 
%    \item Sharp displacement, smeared displacement. 
%    \item Drawbacks of Conti, Iurlano \cite{freddi2010regularized}
%\end{enumerate}

The considerations listed above are valid in the one-dimensional setting. They will partly change in the multi-dimensional case; in particular, we will see in Section \ref{sec:stability3D} that the study of stability of the homogeneous solution becomes more subtle.

\begin{table}
\centering
\begin{tabular}{|>{\centering\arraybackslash}m{4cm}|
                >{\centering\arraybackslash}m{6cm}|
                >{\centering\arraybackslash}m{4cm}|}
\hline
\textbf{\parbox[c][1cm][c]{4cm}{\centering Phase}} & \textbf{\parbox[c][1cm][c]{6cm}{\centering Stress-displacement response}} & \textbf{\parbox[c][1cm][c]{4cm}{\centering Rates of internal variables}} \\
\hline
Initial linear elastic loading &
\vspace{0.5mm}\hspace{1mm}\includegraphics[scale=1, trim=0 0.5cm 0 0.2cm, clip]{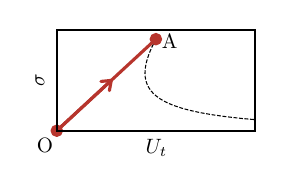}\hspace{1mm}\vspace{0.5mm} &
\begin{equation*}
    \begin{aligned}
        &\dot{\jump{\eta(0)}}=0\\
        &\dot{\alpha}(0)=0
    \end{aligned}
\end{equation*} \\
\hline
Fracture loading &
\vspace{0.5mm}\hspace{1mm}\includegraphics[scale =1, trim=0 0.5cm 0 0.2cm, clip]{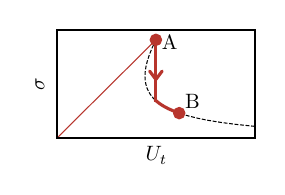}\hspace{1mm}\vspace{0.5mm} &
\begin{equation*}
    \begin{aligned}
        &\dot{\jump{\eta(0)}}>0\\
        &\dot{\alpha}(0)>0
    \end{aligned}
\end{equation*} \\
\hline
Crack-closure unloading &
\vspace{0.5mm}\hspace{1mm}\includegraphics[scale =1, trim=0 0.5cm 0 0.2cm, clip]{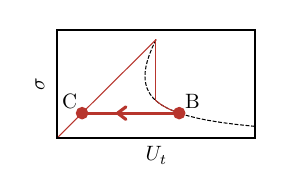}\hspace{1mm}\vspace{0.5mm} &
\begin{equation*}
    \begin{aligned}
        &\dot{\jump{\eta(0)}}<0\\
        &\dot{\alpha}(0)=0
    \end{aligned}
\end{equation*} \\
\hline
Linear elastic unloading &
\vspace{0.5mm}\hspace{1mm}\includegraphics[scale =1, trim=0 0.5cm 0 0.2cm, clip]{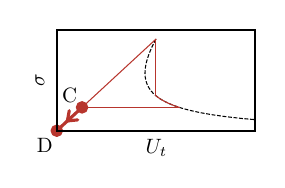}\hspace{1mm}\vspace{0.5mm} &
\begin{equation*}
    \begin{aligned}
        &\dot{\jump{\eta(0)}}=0\\
        &\dot{\alpha}(0)=0
    \end{aligned}
\end{equation*} \\
\hline
Linear elastic reloading &
\vspace{0.5mm}\hspace{1mm}\includegraphics[scale =1, trim=0 0.5cm 0 0.2cm, clip]{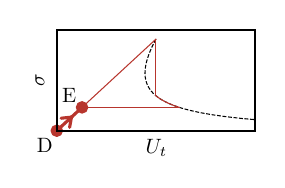}\hspace{1mm}\vspace{0.5mm} &
\begin{equation*}
    \begin{aligned}
        &\dot{\jump{\eta(0)}}=0\\
        &\dot{\alpha}(0)=0
    \end{aligned}
\end{equation*} \\
\hline
Crack-opening reloading &
\vspace{0.5mm}\hspace{1mm}\includegraphics[scale =1, trim=0 0.5cm 0 0.2cm, clip]{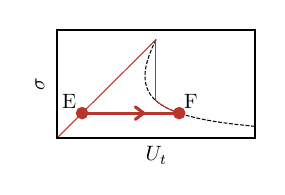}\hspace{1mm}\vspace{0.5mm} &
\begin{equation*}
    \begin{aligned}
        &\dot{\jump{\eta(0)}}>0\\
        &\dot{\alpha}(0)=0
    \end{aligned}
\end{equation*} \\
\hline
Fracture reloading &
\vspace{0.5mm}\hspace{1mm}\includegraphics[scale =1, trim=0 0.5cm 0 0.2cm, clip]{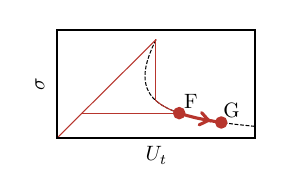}\hspace{1mm}\vspace{0.5mm} &
\begin{equation*}
    \begin{aligned}
        &\dot{\jump{\eta(0)}}>0\\
        &\dot{\alpha}(0)>0
    \end{aligned}
\end{equation*} \\
\hline
\end{tabular}
\caption{Solution of the structural problem: example of a loading--unloading history and corresponding stress--displacement response. The third column reports the signs of the rates of the eigenstrain jump $\dot{\jump{\eta}}$ and of the damage variable $\dot{\alpha}$ at the crack location $x = 0$ (with a little abuse of notation since we are in the time-discrete setting).}
\label{tab:unloading_loc}
\end{table}

\section{Three-dimensional phase-field model of cohesive fracture}
\label{sec:3D}

Now that we have thoroughly analyzed the one-dimensional case and understood the fundamental characteristics of the model and its ability to describe cohesive cracks, we move towards the three-dimensional case and the study of the shape of the elastic domains.

\subsection{Volumetric-deviatoric decomposition}
\label{subsec:voldev}

We consider the isotropic case, where the elasticity tensor $\mathbb{C}$ depends solely on two parameters, such as the Young's modulus $E$ and the Poisson ratio $\nu$, or equivalently the bulk modulus $\kappa$ and the shear modulus $\mu$, or the Lamé parameters $\lambda$, $\mu$. 
By using the volumetric-deviatoric decomposition, in the elastic energy density (\ref{eq:strain_en_d}) we now have
%\begin{equation}     \psi(\boldsymbol{\varepsilon}, \boldsymbol{\eta}, \alpha) = \begin{cases}
 %   \varphi(\boldsymbol{\varepsilon}, \text{tr}(\boldsymbol{\eta}), \Vert \boldsymbol{\eta}_{\text{dev}}\Vert,\alpha),&\quad\text{if}\quad\text{tr}(\boldsymbol{\eta})\geq 0\\
  %  +\infty,&\quad\text{if}\quad\text{tr}(\boldsymbol{\eta})< 0,
%\end{cases}
%    \label{eq:el_en_simp}
%\end{equation}
%where
\begin{equation}
    \psi_e(\boldsymbol{\varepsilon}-\boldsymbol{\eta})=\hat{\psi}_e\left({\text{tr}(\boldsymbol{\varepsilon}})-\text{tr}(\boldsymbol{\eta}), \Vert \boldsymbol{\varepsilon}_{\text{dev}}\Vert-\Vert \boldsymbol{\eta}_{\text{dev}}\Vert\right):=\frac{\kappa}{2} \left( \text{tr}(\boldsymbol{\varepsilon}) - \text{tr}(\boldsymbol{\eta}) \right)^2 + \mu \left( \Vert \boldsymbol{\varepsilon}_{\text{dev}} \Vert - \Vert \boldsymbol{\eta}_{\text{dev}} \Vert \right)^2,
    \label{eq:vol_dev_exp}
\end{equation}
%with the potential $\phi$ defined as 
%\begin{equation}
%    \phi(\text{tr}(\boldsymbol{\eta}), \Vert \boldsymbol{\eta}_{\text{dev}}\Vert):=\pi_0(\boldsymbol{\eta}).
%    \label{eq:k_pot}
%\end{equation}
and
\begin{equation}
    \pi(\boldsymbol{\eta},\alpha)=\hat{\pi}\left(\text{tr}(\boldsymbol{\eta}), \Vert \boldsymbol{\eta}_{\text{dev}}\Vert,\alpha\right),
    \label{eq:k_pot}
\end{equation}
whereas in (\ref{eq:barrier})
\begin{equation}
    \pi_0(\boldsymbol{\eta})=\phi\left(\text{tr}(\boldsymbol{\eta}), \Vert \boldsymbol{\eta}_{\text{dev}}\Vert\right).
    \label{eq:k_pot_0}
\end{equation}
In (\ref{eq:vol_dev_exp}), we exploit the identity of \( \boldsymbol{\eta}_{\text{dev}}/\Vert \boldsymbol{\eta}_{\text{dev}}\Vert \) and \( \boldsymbol{\varepsilon}_{\text{dev}}/\Vert \boldsymbol{\varepsilon}_{\text{dev}}  \Vert \) which arises as a necessary condition of the minimization problem~(\ref{eq:locmin}), see \ref{append:simpler} for further details. Moreover, with~(\ref{eq:k_pot}) we make it explicit that the eigenstrain potential \( \pi(\boldsymbol{\eta}, \alpha) \) is independent of the orientation of the deviatoric part of the eigenstrain \( \boldsymbol{\eta}_{\text{dev}} \) (see again \ref{append:simpler}), which allows us to further simplify the notation.

\par By differentiation, we define the \textit{pressure} \( p \) and the \textit{shear stress} \( \tau \) as
\begin{equation}
    \begin{aligned}
        p(\text{tr}(\boldsymbol{\varepsilon})- \text{tr}(\boldsymbol{\eta})) := \frac{\partial \hat{\psi}_e}{\partial\, \text{tr}(\boldsymbol{\varepsilon})} = \kappa\,\big(\text{tr}(\boldsymbol{\varepsilon}) - \text{tr}(\boldsymbol{\eta})\big), \\ \tau(\Vert\boldsymbol{\varepsilon}_{\text{dev}}\Vert- \Vert\boldsymbol{\eta}_{\text{dev}}\Vert) := \frac{\partial\hat{\psi}_e}{\partial\, \Vert\boldsymbol{\varepsilon}_{\text{dev}}\Vert} = 2\mu\,\big(\Vert\boldsymbol{\varepsilon}_{\text{dev}}\Vert - \Vert\boldsymbol{\eta}_{\text{dev}}\Vert\big),
    \end{aligned}
\end{equation}
so that the stress tensor \( \boldsymbol{\sigma} \) can be expressed as
\begin{equation}
    \boldsymbol{\sigma}%(\boldsymbol{\varepsilon}, \boldsymbol{\eta}) 
    = p\,\boldsymbol{I} + \tau\,\check{\boldsymbol{\varepsilon}}_{\text{dev}},
    \label{eq:stress_split}
\end{equation}
where \( \check{\boldsymbol{\varepsilon}}_{\text{dev}}=\boldsymbol{\varepsilon}_{\text{dev}}/\Vert\boldsymbol{\varepsilon}_{\text{dev}}\Vert \) denotes the director of the deviatoric part of the strain tensor. 
\par In order to characterize the mechanical response given by the phase-field fracture model in the three-dimensional setting, a key quantity is the \textit{residual stress} \( \boldsymbol{\sigma}_R \), which represents the stress at a fully damaged material point, i.e., for \( \alpha = 1 \). This can be defined as
\begin{equation}
    \boldsymbol{\sigma}_R(\boldsymbol{\varepsilon}) := \mathbb{C}(\boldsymbol{\varepsilon} - \tilde{\boldsymbol{\eta}}),
\end{equation}
where \( \tilde{\boldsymbol{\eta}} \) is the eigenstrain at \( \alpha = 1 \), obtained from the constrained minimization problem
\begin{equation}
    \tilde{\boldsymbol{\eta}} = \argmin_{\boldsymbol{\eta},\,\text{tr}(\boldsymbol{\eta}) \geq 0} \psi_e(\boldsymbol{\varepsilon} - \boldsymbol{\eta}).
    \label{eq:structured_defo}
\end{equation}
Note that (\ref{eq:structured_defo}) is identical to the problem used to compute the optimal \textit{structured deformation} associated with the volumetric-deviatoric split~\cite{amor2009regularized} in the context of phase-field modeling of brittle fracture~\cite{freddi2010regularized, freddi2011variational, de2022nucleation}.
Therefore, we can exploit the results obtained therein and the residual stress is
\begin{equation}
    \boldsymbol{\sigma}_R(\boldsymbol{\varepsilon}) = \langle p(\text{tr}(\boldsymbol{\varepsilon})) \rangle_{-}\, \boldsymbol{I} = \kappa \langle \text{tr}(\boldsymbol{\varepsilon}) \rangle_{-} \, \boldsymbol{I},
    \label{eq:res_stress}
\end{equation}
which effectively captures crack-like behavior, as it is insensitive to both deviatoric deformations and positive volumetric strains, as discussed in \cite{vicentini2024energy}. 
\begin{remark}
Note that the minimization problem (\ref{eq:structured_defo}) is constrained by the barrier cone \( \text{tr}(\boldsymbol{\eta}) \geq 0 \), reflecting the assumption that the initial elastic domain \( \mathcal{S}_0 \) is unbounded only along the negative purely volumetric direction. Alternative assumptions lead to a different barrier cone, potentially introducing residual stresses that might not align with physical expectations. The interested reader is referred to \cite{vicentini2024energy} for a discussion on the residual stresses obtained by using domains of different types in the context of phase-field models of brittle fracture with energy decompositions \cite{freddi2010regularized, freddi2011variational,de2022nucleation}. %On the other hand, for any shape of \( \mathcal{S}_0 \), the residual stress remains the one in (\ref{eq:res_stress}) if \( \mathcal{S}_0 \) is properly truncated to satisfy the assumptions on the bounds.
\end{remark}

The strength potential \( \phi \) preserves the properties of \( \pi_0 \) of convexity, positive homogeneity of degree 1, and zero value at the origin. 
Exploiting the expression in (\ref{eq:vol_dev_exp}), the evolution criterion for the eigenstrain (\ref{eq:nl_law}) can be rewritten as
\begin{equation}
    p\,\zeta + \tau\,\xi \leq a(\alpha)\,\phi'(\text{tr}(\boldsymbol{\eta}),\Vert\boldsymbol{\eta}_{\text{dev}}\Vert)[\zeta, \xi]
    %, \quad \forall \zeta , \ \xi: \text{tr}(\boldsymbol{\eta})+\zeta \geq 0, \Vert\boldsymbol{\eta}_{\text{dev}}\Vert+\xi\geq 0.
    \label{eq:criterion_ev}
\end{equation}
for any admissible $\zeta, \xi$. In the case where \( \phi \) is Fréchet differentiable, the Gâteaux derivative is a linear operator, and the evolution criterion (\ref{eq:criterion_ev}) reduces to

\begin{equation}
    \begin{aligned}
        &p \leq a(\alpha)\, \frac{\partial \phi(\text{tr}(\boldsymbol{\eta}), \Vert\boldsymbol{\eta}_{\text{dev}}\Vert)}{\partial \text{tr}(\boldsymbol{\eta})},\quad \text{tr}(\boldsymbol{\eta})\geq 0,\quad \left(-p+a(\alpha)\,\frac{\partial \phi(\text{tr}(\boldsymbol{\eta}), \Vert\boldsymbol{\eta}_{\text{dev}}\Vert)}{\partial \text{tr}(\boldsymbol{\eta})}\right)\,\text{tr}(\boldsymbol{\eta})=0, \\
        &\tau \leq a(\alpha)\, \frac{\partial \phi(\text{tr}(\boldsymbol{\eta}), \Vert\boldsymbol{\eta}_{\text{dev}}\Vert)}{\partial \Vert\boldsymbol{\eta}_{\text{dev}}\Vert}, \quad \Vert\boldsymbol{\eta}_{\text{dev}}\Vert\ge0,\quad \left(-\tau+a(\alpha)\,\frac{\partial \phi(\text{tr}(\boldsymbol{\eta}), \Vert\boldsymbol{\eta}_{\text{dev}}\Vert)}{\partial \Vert\boldsymbol{\eta}_{\text{dev}}\Vert}\right)\,\Vert\boldsymbol{\eta}_{\text{dev}}\Vert=0.
    \end{aligned}
\label{eq:diff_voldev}
\end{equation}

\subsection{Relationship between the strength potential $\phi$ and the nucleation domain}

In Section~\ref{sec:general}, we provided the way of defining \( \pi_0(\boldsymbol{\eta}) \) to obtain a desired initial elastic domain \( \mathcal{S}_0 \), namely, we defined \( \pi_0(\boldsymbol{\eta}) \) as the support function of \( \mathcal{S}_0 \). In this section, we follow the inverse path: starting from a convex, positively 1-homogeneous strength potential \( \phi \) (see the symbol redefinition in (\ref{eq:k_pot_0})), we construct the corresponding initial elastic domain \( \mathcal{S}_0 \). The elastic domains for \( \alpha > 0 \) can easily be derived afterwards by using the assumed homothetic scaling \( \mathcal{S}(\alpha) = a(\alpha)\, \mathcal{S}_0 \). 
\par The boundary of the domain \( \mathcal{S}_0 \) corresponds to the set of stress states at which the eigenstrain starts to evolve from zero. Note that, for the \texttt{AT2} model considered here and under loading from an intact material, the onset of a non-zero eigenstrain \( \boldsymbol{\eta} \) activates the energy release rate \( a'(\alpha)\pi_0(\boldsymbol{\eta}) \), which in turn immediately triggers the evolution of the damage variable \( \alpha \).
Since \( \phi \) is positively 1-homogeneous and vanishes at the origin, we obtain that
\begin{equation}
    \phi'(0,0)[\zeta,\xi] = \phi(\zeta,\xi),
    \label{eq:equi_k}
\end{equation}
so that for the nucleation analysis, i.e. for $\text{tr}(\boldsymbol{\eta})=\Vert\boldsymbol{\eta}_{\text{dev}}\Vert=0$ and $\alpha=0$, the eigenstrain evolution criterion in (\ref{eq:criterion_ev}) simplifies to
\begin{equation}
    p\,\zeta + \tau\,\xi \leq \phi(\zeta, \xi), \quad \forall \zeta \geq 0,\ \xi \geq 0.
    \label{eq:criterion_zx}
\end{equation}
Following the analysis in \cite{charlotte2006initiation}, we distinguish two cases.

\subsubsection{Case of a smooth strength potential $\phi$}
\label{sub:smooth_phi}
The first case occurs when \( \phi \) is Fréchet differentiable at \( (0,0) \), so that \( \phi'(0,0)[\zeta,\xi] \) is a linear function of \( \zeta \) and \( \xi \), yielding
\begin{equation}
    \phi(\zeta, \xi) = p_{\text{c}}\,\zeta + \tau_{\text{c}}\,\xi,
    \label{eq:smooth}
\end{equation}
where we have defined
\begin{equation}
p_c:=\phi(1,0),\quad \tau_c:=\phi(0,1),
\end{equation}
which represent the \textit{critical pressure} and the \textit{shear strength}, respectively. 
From (\ref{eq:smooth}), the eigenstrain evolution criterion (\ref{eq:criterion_zx}) simplifies to
\begin{equation}
    p \leq p_{\text{c}}, \quad \tau \leq \tau_{\text{c}},
    \label{eq:max_p_tau}
\end{equation}
which correspond to \textit{maximum pressure} and \textit{maximum shear} criteria, respectively.

\subsubsection{Case of a non-smooth strength potential $\phi$}
\label{sec:int_curve}
The second case arises when \( \phi \) is non Fréchet differentiable at the origin, and thus cannot be expressed in the form~\eqref{eq:smooth}, although it remains Gâteaux differentiable at that point. For a non-smooth $\phi$, let us consider a perturbation $(\zeta,\xi)$ such that $\zeta = \lambda\,\xi$ with $\lambda \geq 0$ as an arbitrary non-negative real number. Exploiting the positive 1-homogeneity property of $\phi$, the criterion (\ref{eq:criterion_zx}) can be rewritten as
\begin{equation}
    \tau \leq f(p),\quad \text{with} \quad f(p) = \inf_{\lambda \geq 0} \left\{ \phi(\lambda,1) - \lambda\,p \right\},
    \label{eq:intrinsic_f}
\end{equation}
where $f(p)$ is the so-called \textit{intrinsic curve} \cite{charlotte2006initiation}. Accordingly, the corresponding strength surface $\mathcal{F}(\boldsymbol{\sigma})=0$ is given by
\begin{equation}
    \mathcal{F}(\boldsymbol{\sigma}) = \Vert \boldsymbol{\sigma}_{\text{dev}} \Vert - f\left( \frac{\text{tr}(\boldsymbol{\sigma})}{3} \right).
    \label{eq:strength_surface}
\end{equation}
Then, the initial elastic domain \( \mathcal{S}_0 \) follows from  (\ref{eq:elastic_dom}). Note that the obtained domain satisfies all the initial assumptions stated in Section~\ref{sec:general}. Furthermore it can be observed that, in accordance with the normality rule, the perturbation direction \( (\zeta, \xi) = (\lambda^*, 1)\,\xi \), where \( \lambda^* = \operatorname*{arg\,inf}_{\lambda \geq 0} \left\{\phi(\lambda,1) - \lambda\,p\right\} \), is aligned with the gradient of the curve \( \tau - f(p) = 0 \) in the \( (p\text{-}\tau) \) plane, and therefore lies in the normal direction.

Once the strength surface is available, it is possible to implicitly define the \textit{tensile strength} as the quantity $\sigma_{\text{c}}^+$ such that
\begin{equation}
    \mathcal{F}(\sigma_{\text{c}}^+\,\boldsymbol{e}_1\otimes\boldsymbol{e}_1)=0\quad\text{with }\quad \boldsymbol{e}_1=[1,0,0]^\intercal,
    \label{eq:tensile_stength}
\end{equation}
which is the intersection of the strength surface $\mathcal{F}(\boldsymbol{\sigma})=0$ with the positive uniaxial stress axis in the stress space. When the intrinsic curve $f(p)$ is available, the implicit definition in (\ref{eq:tensile_stength}) is equivalent to \begin{equation}
    \sigma_{\text{c}}^+=\sqrt{\frac{3}{2}}\,f\left(\frac{\sigma_{\text{c}}^+}{3}\right).
\end{equation}
Similarly, the \textit{compressive strength} is defined as the quantity $\sigma_{\text{c}}^-$ such that $\mathcal{F}(-\sigma_{\text{c}}^-\,\boldsymbol{e}_1\otimes\boldsymbol{e}_1)=0$.
\par We now illustrate, through specific examples based on particular choices of \( \pi_0(\boldsymbol{\eta}) \), the corresponding evolution criterion for the eigenstrain. In particular, we also show how the strain-hardening condition, the nucleation domain, and the surface energy density specialize in these cases. 

\subsubsection{Example 1: The strength potential $\phi$ as $r$-norm}
\label{sec:norm_pot}
As an illustrative example, we begin with the simple case where $\phi$ is defined as the $r$-norm of the vector composed of the \textit{scaled} trace and deviatoric parts of $\boldsymbol{\eta}$, that is
\begin{equation}
     \phi_r\left(\text{tr}(\boldsymbol{\eta}),\Vert \boldsymbol{\eta}_{\text{dev}}\Vert\right):=  \sqrt[r]{p_\text{c}^r\,\text{tr}^r(\boldsymbol{\eta}) +\tau_\text{c}^r\,\Vert \boldsymbol{\eta}_\text{dev}\Vert^r}\quad\text{with}\quad r\in\mathbb{R}, r\geq1.
\label{eq:pot_phi_n}
\end{equation}
In particular, for $r=1$, we retrieve the smooth case in (\ref{eq:smooth})    
\begin{equation}
\phi_1(\text{tr}(\boldsymbol{\eta}),\Vert \boldsymbol{\eta}_{\text{dev}}\Vert)=p_{\text{c}}\,\text{tr}(\boldsymbol{\eta})+\tau_{\text{c}}\,\Vert \boldsymbol{\eta}_{\text{dev}}\Vert,
\label{eq:norm1}
\end{equation}
for $r=2$ we have the \textit{Euclidean norm}
\begin{equation}
\phi_2(\text{tr}(\boldsymbol{\eta}),\Vert \boldsymbol{\eta}_{\text{dev}}\Vert)=\sqrt{p_{\text{c}}^2\,\text{tr}(\boldsymbol{\eta})^2+\tau_{\text{c}}^2\,\Vert \boldsymbol{\eta}_\text{dev}\Vert ^2},
\label{eq:norm2}
\end{equation}
and the limit for $r\rightarrow \infty$ yields the \textit{max function}
\begin{equation}
\begin{aligned}
	\phi_{\infty}(\text{tr}(\boldsymbol{\eta}),\Vert \boldsymbol{\eta}_{\text{dev}}\Vert)&=\max\left\{p_{\text{c}}\,\text{tr}(\boldsymbol{\eta}),\tau_{\text{c}}\,\Vert \boldsymbol{\eta}_\text{dev}\Vert \right\}=\frac{1}{2}\left(p_{\text{c}}\,\text{tr}(\boldsymbol{\eta})+\tau_{\text{c}}\,\Vert \boldsymbol{\eta}_{\text{dev}}\Vert\right)+\frac{1}{2}\left|p_{\text{c}}\,\text{tr}(\boldsymbol{\eta})-\tau_{\text{c}}\,\Vert \boldsymbol{\eta}_{\text{dev}}\Vert\right      |=\\&= \frac{p_{\text{c}}}{2}\,\text{tr}(\boldsymbol{\eta})+\frac{\tau_{\text{c}}}{2}\,\Vert \boldsymbol{\eta}_{\text{dev}}\Vert+\frac{1}{2}\sqrt{\left(p_{\text{c}}\,\text{tr}(\boldsymbol{\eta})-\tau_{\text{c}}\,\Vert \boldsymbol{\eta}_{\text{dev}}\Vert\right)^2}.
    \end{aligned}
	\label{eq:normInf}
\end{equation}
While $\phi_1$ is Fréchet differentiable,  $\phi_{\infty}$ is not Fréchet differentiable along the line $p_{\text{c}}\,\zeta=\tau_{\text{c}}\,\xi$. For a generic $r\in(1,\infty)$, $\phi_r$ is not Fréchet differentiable in the origin $(\zeta,\xi)=(0,0)$ and Fréchet differentiable everywhere else. On the other hand, all these strength potentials are Gâteaux differentiable.

\begin{figure}[H]
    \centering
    \includegraphics[scale=1]{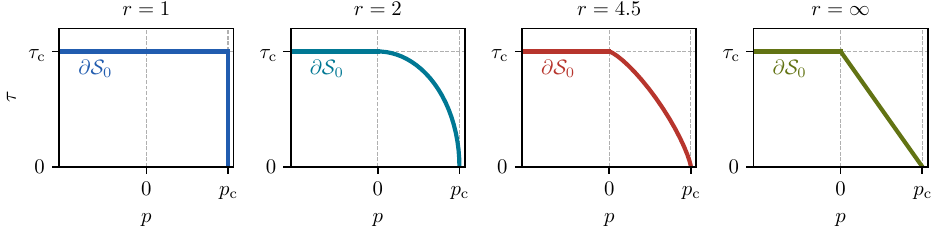}
    \caption{Strength surface $\partial \mathcal{S}_0$  obtained with $\phi_r$ for different values of $r \geq 1$ in the $p-\tau$ plane.  
For $p \geq 0$, as $r$ increases, $\partial\mathcal{S}_0$ changes smoothly from a rectangle ($r = 1$) to a triangle ($r = \infty$), passing through an ellipse ($r = 2$). To illustrate the shape for a generic $r$, we show the Lamé curve for $r = 4.5$. Note that $r$ does not need to be an integer.
} 
    \label{fig:norms}
\end{figure}

\begin{figure}[htbp]
    \centering

    \begin{minipage}[b]{0.45\textwidth}
        \centering
         \subcaption*{\fontsize{10}{12}\selectfont $r = 1$}
        \includegraphics[scale=1,trim=85 55 48 85, clip]{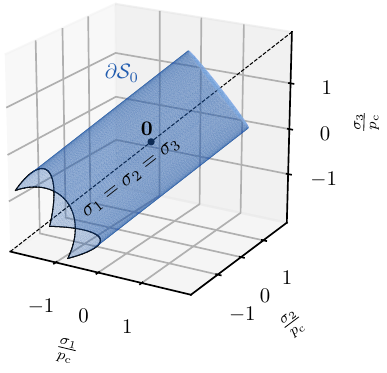}
        \label{fig:subfig1}
    \end{minipage}
    \begin{minipage}[b]{0.45\textwidth}
        \centering
         \subcaption*{\fontsize{10}{12}\selectfont $r = 2$}
        \includegraphics[scale=1,trim=85 55 48 85, clip]{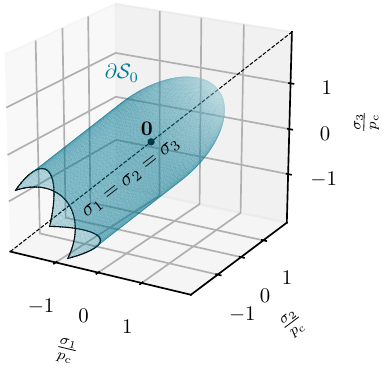}
        \label{fig:subfig2}
    \end{minipage}

    \vspace{0.5cm}

    \begin{minipage}[b]{0.45\textwidth}
        \centering
        \subcaption*{\fontsize{10}{12}\selectfont $r = 4.5$}        
        \includegraphics[scale=1,trim=85 55 48 85, clip]{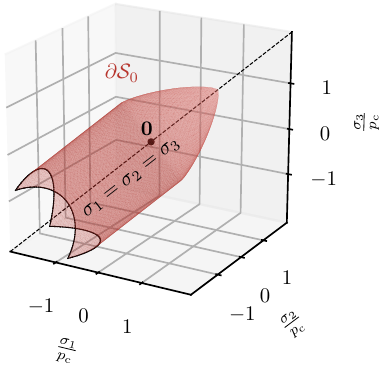}
        \label{fig:subfig3}
    \end{minipage}
    \begin{minipage}[b]{0.45\textwidth}
        \centering
        \subcaption*{\fontsize{10}{12}\selectfont $r = \infty$}        
        \includegraphics[scale=1,trim=85 55 48 85, clip]{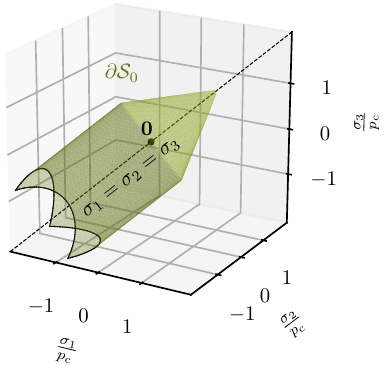}
        \label{fig:subfig4}
    \end{minipage}

    \caption{Strength surface $\partial\mathcal{S}_0$ obtained with $\phi_r$ for different values of $r\geq1$ in the principal stress space. For $p \geq 0$, as $r$ increases, $\partial\mathcal{S}_0$ changes smoothly from a cylinder ($r = 1$) to a cone ($r = \infty$), passing through an ellipsoid ($r = 2$). To illustrate the shape for a generic $r$, we show the strength surface for $r = 4.5$. Note that $r$ does not need to be an integer.}
    \label{fig:3D_domain_r}
\end{figure}

In general, we observe that the strength surface $\partial \mathcal{S}_0$ corresponds to 
\begin{equation}
     \tau=\tau_{\text{c}}\quad \text{if}\quad p<0
\end{equation}
for any value of \( r \geq 1 \), so that it satisfies the assumption of being unbounded along the negative volumetric direction. For \( p \geq 0\), the shape of the strength surface is more interesting, and we now proceed to determine it for different values of \( r \):

\begin{itemize}
\item{For $\phi_1$, as already discussed in Section (\ref{sub:smooth_phi}), the initial elastic domain has a \textit{rectangular} shape in the $p-\tau$ plane bounded by the maximum pressure and the maximum shear criteria (\ref{eq:max_p_tau}) (Figure~\ref{fig:norms}). In the principal stress space this domain has a cylindrical shape truncated at $\text{tr}(\boldsymbol{\sigma})=3\,p_{\text{c}}$ (Figure~\ref{fig:3D_domain_r}).}

\item{For $\phi_r$ with $r\in(1,\infty)$, we show in \ref{app:R_phi_n} that the strength surface can be written as 
\begin{equation}
     \left(\frac{p}{p_{\text{c}}}\right)^\frac{r}{r-1}+\left(\frac{\tau}{\tau_{\text{c}}}\right)^\frac{r}{r-1}=1\quad \text{if}\quad p\geq 0,
    \label{eq:R_param}
\end{equation}
which corresponds to the equation of the so-called \textit{Lamé curve}, or superellipse \cite{gridgeman1970lame} (Figure~\ref{fig:norms}).
Specifically, for $\phi_2$ the curve (\ref{eq:R_param}) in the $p-\tau$ plane for $p\geq 0$  corresponds to the \textit{ellipse} (Figure~\ref{fig:norms}) 
\begin{equation}
     \frac{p^2}{p_{\text{c}}^2}+\frac{\tau^2}{\tau_{\text{c}}^2}=1\quad \text{if}\quad p\geq 0.
     \label{eq:ellipse}
\end{equation}
}
\item{In the case of  $\phi_{\infty}$, the strength surface for $p\geq 0$ becomes the \textit{straight line} (Figure~\ref{fig:norms})
\begin{equation}
     \frac{p}{p_{\text{c}}}+\frac{\tau}{\tau_{\text{c}}}=1\quad \text{if}\quad p\geq 0,
\end{equation}
which in the principal stress space yields a \textit{Drucker-Prager cone} on $\text{tr}(\boldsymbol{\sigma})\geq 0$ (Figure~\ref{fig:3D_domain_r}).}
\end{itemize}

\par
\begin{figure}[H]
    \centering
    \includegraphics[scale=1]{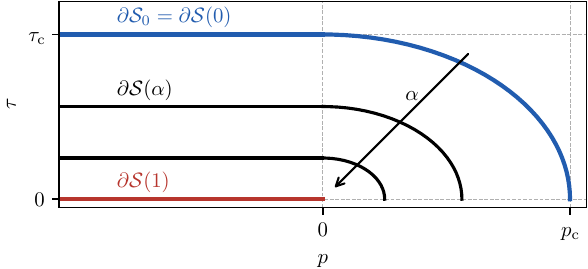}
    \caption{Evolution of the elastic domain boundary $\partial\mathcal{S}(\alpha)$ with increasing $\alpha$ for the strength potential $\phi_2$. As $\alpha$ increases, the domain homothetically contracts in the direction indicated by the arrow. The strength surface $\partial\mathcal{S}_0$ is shown in blue, while the boundary of the final elastic domain $\partial\mathcal{S}(1)$ is shown in red. Intermediate boundaries are colored in black.} 
    \label{fig:homotetic}
\end{figure}

 In Figure~\ref{fig:homotetic}, we show the homothetic contraction of the elastic domain $\mathcal{S}(\alpha)$ with the increase of the phase-field variable $\alpha$, which guarantees stress softening by construction, on the example $r=2$. For the fully damaged material state $\alpha = 1$, the domain collapses to the negative half-line $p \leq 0$ (shown in red in Figure~\ref{fig:homotetic}), corresponding to the admissible residual stresses $\boldsymbol{\sigma}_R$ given by (\ref{eq:res_stress}).

Knowing the strength surface $\partial\mathcal{S}_0$, we can determine the characteristic cohesive length as defined in (\ref{eq:suff_SH}).
Following (\ref{eq:tensile_stength}), we can also determine the tensile strength $\sigma_{\text{c}}^+$ and, similarly, the compressive strength $\sigma_{\text{c}}^-$. In Table~\ref{tab:ell_strength} we present the values of $\ell_{\text{ch}}$, $\sigma_{\text{c}}^+$, and $\sigma_{\text{c}}^-$ for the three norms with $r=1, 2, \infty$. We recall that, for the \texttt{AT2} model, the strain-hardening condition is fulfilled if (\ref{eq:suff_SH}) is satisfied, that is, if $\ell \leq \frac{1}{4} \ell_{\text{ch}}$.

\begin{table}[h!]
\centering
\renewcommand{\arraystretch}{2.8} % Improve vertical spacing
\begin{tabular}{|>{\centering\arraybackslash}m{1.5cm} 
                |>{\centering\arraybackslash}m{4.3cm} 
                |>{\centering\arraybackslash}m{4.3cm} 
                |>{\centering\arraybackslash}m{3.5cm}|}
\hline
\raisebox{0.4em}{$r$} & \raisebox{0.4em}{{$\ell_{\text{ch}}$}} & \raisebox{0.4em}{{$\sigma_{\text{c}}^+$}} & \raisebox{0.4em}{{$\sigma_{\text{c}}^-$}} \\
\hline
\raisebox{0.3em}{1} &
\raisebox{0.4em}{$\dfrac{2\,\mu\,\kappa\,G_{\text{c}}}{2\,\mu\,p_{\text{c}}^2+\kappa\,\tau_{\text{c}}^2}$} &
\raisebox{0.4em}{$\min\left\{3p_{\text{c}},\sqrt{\dfrac{3}{2}}\,\tau_{\text{c}}\right\}$} &
\multirow{3}{*}{$\sqrt{\dfrac{3}{2}}\,\tau_{\text{c}}$} \\
\cline{1-3}
\raisebox{0.2em}{2} &
\multirow{2}{*}{$\min\left\{\frac{\kappa\,G_{\text{c}}}{p_{\text{c}}^2},\frac{2\,\mu\,G_{\text{c}}}{\tau_{\text{c}}^2}\right\}$} &
\raisebox{0.5em}{$\dfrac{3\,p_{\text{c}}\,\tau_{\text{c}}}{\sqrt{6\,p_{\text{c}}^2+\tau_{\text{c}}^2}}$} &
\\
\cline{1-1} \cline{3-3}
\raisebox{0.4em}{$\infty$} & &
\raisebox{0.5em}{$\dfrac{3\,p_{\text{c}}\,\tau_{\text{c}}}{\sqrt{6}\,p_{\text{c}}+\tau_{\text{c}}}$} &
\\
\hline
\end{tabular}
\caption{Characteristic cohesive length $\ell_{\text{ch}}$, tensile strength $\sigma_{\text{c}}^+$ and compressive strength $\sigma_{\text{c}}^-$ for the strength potential $\phi_r$ with $r=1,2,\infty$.}
\label{tab:ell_strength}
\end{table}

\subsubsection{Example 2: Combining $r$-norm and $1$-norm}
Let us consider the $\gamma$-dependent family of strength potentials obtained by combining $r$-norm with $1$-norm
\begin{equation}
	 \phi(\text{tr}(\boldsymbol{\eta}),\Vert \boldsymbol{\eta}_{\text{dev}}\Vert) =(1+\gamma)\,\phi_r(\text{tr}(\boldsymbol{\eta}),\Vert \boldsymbol{\eta}_{\text{dev}}\Vert)-\gamma\,\phi_1(\text{tr}(\boldsymbol{\eta}),\Vert \boldsymbol{\eta}_{\text{dev}}\Vert)\quad\text{with}\quad 0\leq\gamma\leq \frac{1}{\sqrt{2}-1},
     \label{eq:pot_shift}
\end{equation}
where the upper bound on $\gamma$ ensures that the origin of the stress space always remains within the elastic domain, i.e., $\boldsymbol{0} \in \mathcal{S}(\alpha)$ for all $\alpha \in [0,1]$.
\par
\begin{figure}[H]
    \centering
    \hspace{-2.4cm}\includegraphics[scale=1]{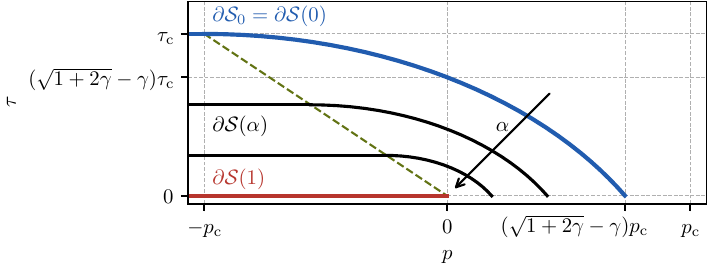}
    \caption{Evolution of the elastic domain boundary $\partial\mathcal{S}(\alpha)$ with increasing $\alpha$ for the strength potential in (\ref{eq:pot_shift}) with $r=2$ and $\gamma=1$. As $\alpha$ increases, the domain homothetically contracts in the direction indicated by the arrow. The strength surface $\partial\mathcal{S}_0$ is shown in blue, while the boundary of the final elastic domain $\partial\mathcal{S}(1)$ is shown in red. Intermediate boundaries are colored in black. To the left of the green dashed line, the boundary of the elastic domain corresponds to the horizontal line $\tau = a(\alpha)\,\tau_{\text{c}}$.} 
    \label{fig:shift}
\end{figure}

 In terms of representation of $\mathcal{S}(\alpha)$ in the $p-\tau$ plane (Figure~\ref{fig:shift}), the effect of multiplying $\phi_r$ by the factor $(1+\gamma)$  is that of homothetically expanding the elastic domain radius by $(1+\gamma)$, while subtracting the norm $\gamma\,\phi_1$ translates the domain by $-\gamma\,[p_{\text{c}},\tau_{\text{c}}]^\intercal$. In Figure~\ref{fig:shift}, we illustrate the elastic domain $\mathcal{S}(\alpha)$ for the specific case $r = 2$ and $\gamma = 1$. The evolution of the elastic domain as the phase-field variable $\alpha$ increases corresponds to a homothetic contraction that preserves the stress-softening behavior. For the fully-broken state $\alpha = 1$, the domain collapses to the negative volumetric half-line, shown in red in Figure~\ref{fig:shift}, corresponding to the region of admissible residual stresses $\boldsymbol{\sigma}_R$ in (\ref{eq:res_stress}). The strength surface $\partial \mathcal{S}_0$ in the principal stress space is illustrated in Figure~\ref{fig:3D_domain_23}.

\subsubsection{Example 3: Parabolic strength surface}
\begin{figure}[htbp]
    \centering

    \begin{minipage}[b]{0.45\textwidth}
        \centering
         \subcaption*{\fontsize{10}{12}\selectfont Example 2}
        \includegraphics[scale=1,trim=85 55 48 85, clip]{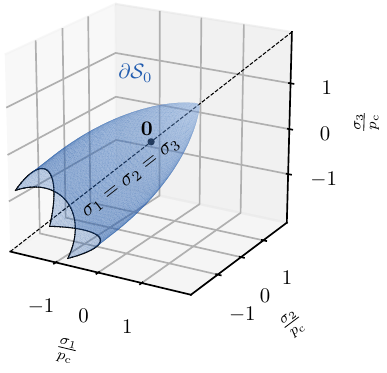}
        \label{fig:subfig1}
    \end{minipage}
    \begin{minipage}[b]{0.45\textwidth}
        \centering
         \subcaption*{\fontsize{10}{12}\selectfont Example 3}
        \includegraphics[scale=1,trim=85 55 48 85, clip]{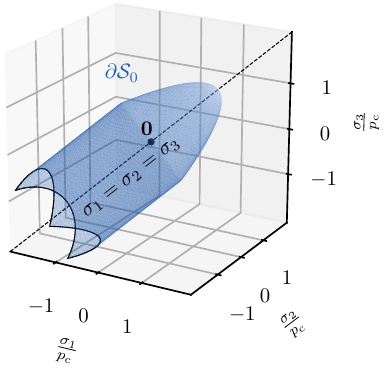}
        \label{fig:subfig2}
    \end{minipage}

    \caption{Strength surface $\partial\mathcal{S}_0$ for Example 2 with $r=2$, $\gamma=1$, $p_{\text{c}}/\tau_{\text{c}}=1$ and Example 3 with $p_{\text{c}}/\tau_{\text{c}}=1$ in the principal stress space.}
    \label{fig:3D_domain_23}
\end{figure}

\begin{figure}[H]
    \centering
    \includegraphics[scale=1]{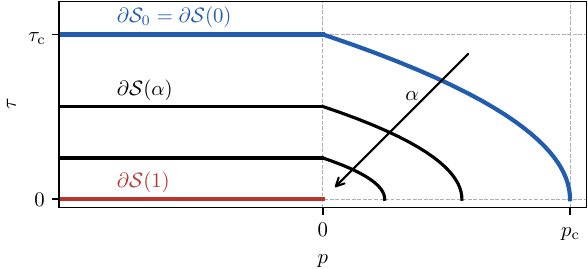}
    \caption{Evolution of the elastic domain boundary $\partial\mathcal{S}(\alpha)$ with increasing $\alpha$ for the strength potential in (\ref{eq:parabola_pot}). As $\alpha$ increases, the domain homothetically contracts in the direction indicated by the arrow. The strength surface $\partial\mathcal{S}_0$ is shown in blue, while the boundary of the final elastic domain $\partial\mathcal{S}(1)$ is shown in red. Intermediate boundaries are colored in black.} 
    \label{fig:parabola}
\end{figure}

So far, we have only presented examples in which  $\phi$ is defined using the $r$-norm or combinations thereof. To highlight the flexibility of the proposed cohesive model in shaping the strength surface $\partial\mathcal{S}_0$, while still ensuring a proper residual stress response, we now illustrate the strength potential associated with a \textit{parabolic} strength surface $\partial\mathcal{S}_0$, given by
\begin{equation}
	\begin{aligned}
		\frac{p}{p_{\text{c}}}+\frac{\tau^2}{\tau_{\text{c}}^2}=1\quad&\text{if}\quad p\geq 0,\\
		\tau=\tau_{\text{c}}\quad&\text{if}\quad p< 0.
        \label{eq:parabola_partial}
	\end{aligned}
\end{equation}
After some calculations (see \ref{app:parabolic}), we obtain the corresponding convex strength potential 
\begin{equation}
	\phi(\text{tr}(\boldsymbol{\eta}),\Vert \boldsymbol{\eta}_{\text{dev}}\Vert)=\begin{cases}
	    p_{\text{c}}\,\text{tr}(\boldsymbol{\eta})+\frac{\tau_{\text{c}}^2}{4\,p_{\text{c}}}\,\frac{\Vert\boldsymbol{\eta}_{\text{dev}}\Vert^2}{\text{tr}(\boldsymbol{\eta})},\quad&\text{if}\quad \Vert\boldsymbol{\eta}_{\text{dev}}\Vert\leq \frac{2\,p_{\text{c}}}{\tau_{\text{c}}}\,\text{tr}(\boldsymbol{\eta}),\\
\tau_{\text{c}}\,\Vert\boldsymbol{\eta}_{\text{dev}}\Vert\quad&\text{if}\quad 0\leq \frac{2\,p_{\text{c}}}{\tau_{\text{c}}}\,\text{tr}(\boldsymbol{\eta})<\Vert\boldsymbol{\eta}_{\text{dev}}\Vert,\\
+\infty\quad&\text{if}\quad\text{tr}(\boldsymbol{\eta})<0.
	\end{cases}
    \label{eq:parabola_pot}
\end{equation}
In Figure~\ref{fig:parabola}, the elastic domain $\mathcal{S}(\alpha)$ is shown along with its stress-softening evolution with the phase-field variable $\alpha$. The strength surface $\partial \mathcal{S}_0$ in the principal stress space is illustrated in Figure~\ref{fig:3D_domain_23}.

\subsection{Stability of the homogeneous solution}
\label{sec:stability3D}
In this section, following the footsteps of \cite{pham2013stability}, we present a stability result under the simplifying assumption of a large domain $\Omega$, and using the strength potential $\phi_2(\text{tr}(\boldsymbol{\eta}), \Vert \boldsymbol{\eta}_{\text{dev}}\Vert)$ defined in (\ref{eq:pot_phi_n}).
\par To this end, we start by assuming that the boundary of the domain $\partial \Omega$ is subjected to a displacement $\boldsymbol{\varepsilon}\,\boldsymbol{x}$, with $\boldsymbol{\varepsilon} \in \mathbb{M}_s^d$. We thus denote by $(\boldsymbol{u}, \eta\,\boldsymbol{N}, \alpha)$ the homogeneous solution that satisfies the first-order stability condition (\ref{eq:first_opt}). In particular, $\boldsymbol{u} = \boldsymbol{\varepsilon} \, \boldsymbol{x}$. The eigenstrain tensor is expressed as $\eta\, \boldsymbol{N}$, where $\eta\geq 0$ is its norm and the tensor $\boldsymbol{N}$ represents its orientation, i.e. $\Vert \boldsymbol{N}\Vert =1$, and fulfills the normality rule (\ref{eq:normality_rule}). The stress associated with this solution is $\boldsymbol{\sigma} = \mathbb{C}(\boldsymbol{\varepsilon} - \eta\,\boldsymbol{N})$. We assume that the damage is $\alpha > 0$, so that $\boldsymbol{\eta} \neq \boldsymbol{0}$; we additionally assume that $\text{tr}(\boldsymbol{\eta}) > 0$, therefore, the total energy is finite and differentiable. Let us take an admissible perturbation $(\boldsymbol{v},\zeta\,\boldsymbol{M},\beta)$ such that $\zeta\geq 0$, $\Vert \boldsymbol{M}\Vert =1$ and $(\boldsymbol{u}+h\,\boldsymbol{v},\boldsymbol{\eta}+h\,\zeta\,\boldsymbol{M},\alpha+h\,\beta)\in\mathcal{Z}_t$, for $h>0$ sufficiently small. Thus, assuming the \texttt{AT2} law and recalling (\ref{eq:nl_law}), the second Gâteaux derivative is
\begin{equation}
\begin{aligned}
    \mathcal{E}''_{\ell}(\boldsymbol{u},\eta\,\boldsymbol{N},\alpha)[\boldsymbol{v},\zeta\,\boldsymbol{M},\beta][\boldsymbol{v},\zeta\,\boldsymbol{M},\beta]=&\int_{\Omega}\Biggl\{\mathbb{C}\left(\boldsymbol{\varepsilon}(\boldsymbol{v})-\zeta\,\boldsymbol{M}\right)\cdot\left(\boldsymbol{\varepsilon}(\boldsymbol{v}\right)-\zeta\,\boldsymbol{M})-\frac{4\,\zeta\,\beta}{1-\alpha}\,\boldsymbol{\sigma}\cdot \boldsymbol{M}+\\
    & + (1-\alpha)^2\,\pi_0''(\eta\,\boldsymbol{N})[\zeta\,\boldsymbol{M}][\zeta\,\boldsymbol{M}]+2\,\eta\,\beta^2\,\pi_0(\boldsymbol{N})+G_{\text{c}}\,\left(\frac{\beta^2}{\ell}+\ell\,\Vert \nabla\beta\Vert ^2\right)\Biggr\}\,d\boldsymbol{x}
\end{aligned}
\label{eq:second_G_n2}
\end{equation}
where the symbol $\boldsymbol{\varepsilon}(\boldsymbol{v})$ denotes the strain tensor associated with the displacement perturbation $\boldsymbol{v}$ and 
\begin{equation}
    \pi_0''(\eta\,\boldsymbol{N})[\zeta\,\boldsymbol{M}][\zeta\,\boldsymbol{M}]=\frac{\zeta^2}{\eta}\,\frac{Q(\boldsymbol{M})}{\pi_0(\boldsymbol{N})},\quad \text{with}\quad Q(\boldsymbol{M}):=\pi_0^2(\boldsymbol{M})-\frac{1}{(1-\alpha)^4}(\boldsymbol{\sigma}\cdot\boldsymbol{M})^2.
    \label{eq:second_pi}
\end{equation}

Let us now study the limit homogeneous solution obtained for $\eta\rightarrow 0^+$ (hence also $\alpha\rightarrow 0^+$). Note that then $Q(\boldsymbol{M})$ is non-negative by the definition of the support function $\pi_0$ in (\ref{eq:support_fun}), and is zero only if $\boldsymbol{M}=\boldsymbol{N}$.
We also see that $\lim_{\eta\rightarrow 0 ^+}\pi_0''(\eta\,\boldsymbol{N})[\zeta\,\boldsymbol{M}][\zeta\,\boldsymbol{M}]=+\infty$ if $\boldsymbol{M}\neq \boldsymbol{N}$. Thus, the interesting case for the stability analysis is given by $\boldsymbol{M}=\boldsymbol{N}$. Taking the limit for $\eta\rightarrow 0^+$, $\alpha\rightarrow 0^+$ and assuming that $\boldsymbol{M}=\boldsymbol{N}$, the second variation in (\ref{eq:second_G_n2}) becomes
\begin{equation}
    \mathcal{E}''_{\ell}(\boldsymbol{u},0^+\boldsymbol{N},0^+)[\boldsymbol{v},\zeta\,\boldsymbol{N},\beta][\boldsymbol{v},\zeta\,\boldsymbol{N},\beta]=\int_{\Omega}\Biggl\{\mathbb{C}\left(\boldsymbol{\varepsilon}(\boldsymbol{v})-\zeta\,\boldsymbol{N}\right)\cdot\left(\boldsymbol{\varepsilon}(\boldsymbol{v}\right)-\zeta\,\boldsymbol{N})-4\,\zeta\,\beta\,\boldsymbol{\sigma}\cdot \boldsymbol{N}+G_{\text{c}}\,\left(\frac{\beta^2}{\ell}+\ell\,\Vert \nabla\beta\Vert ^2\right)\Biggr\}\,d\boldsymbol{x}
\label{eq:second_G_NN}
\end{equation}

\par Following \cite{pham2013stability}, let us further simplify the problem by assuming that the domain $\Omega$ is large. To this end, we first introduce the domain $\check{\Omega}$ scaled homothetically from $\Omega$ with scaling factor $L$, where $L$ is the characteristic dimension of $\Omega$. Accordingly, we 
define the dimensionless position vector $\check{\boldsymbol{x}}=\boldsymbol{x}/L$, the dimensionless gradient operator $\check{\nabla}=L\,\nabla$, and the dimensionless perturbation $\check{\boldsymbol{v}}=\boldsymbol{v}/L$. The latter is admissible when it belongs to $H^1_0(\check{\Omega})$, i.e., to the subspace of $H^1(\check{\Omega})$ of functions that vanish on the boundary $\partial\check{\Omega}$. Moreover, analogously to \cite{pham2013stability}, it is always possible to assume that $\int_{\check{\Omega}}\zeta^2(x)\,dx=1$. Then, we can rewrite the second Gâteaux derivative as

\begin{equation}
\begin{aligned}
    \,\mathcal{E}''_{\ell}(\boldsymbol{u},0^+\boldsymbol{N},0^+)[\check{\boldsymbol{v}},\zeta\,\boldsymbol{N},\beta][\check{\boldsymbol{v}},\zeta\,\boldsymbol{N},\beta]=\left[-\Delta(\check{\boldsymbol{v}},\zeta)+H(\zeta,\beta)+G_{\text{c}}\,\frac{\ell}{L^2}\,\Vert \check{\nabla}\beta\Vert^2\right]\,L^d,
\end{aligned}
\label{eq:second_G_DH}
\end{equation}
where
\begin{equation}
\Delta(\check{\boldsymbol{v}},\zeta):=\int_{\check{\Omega}} \mathbb{C}\boldsymbol{\varepsilon}(\check{\boldsymbol{v}})\cdot\left[2\,\zeta\boldsymbol{N}- \boldsymbol{\varepsilon}(\check{\boldsymbol{v}})\right] d\check{\boldsymbol{x}},\quad
H(\zeta,\beta):=\int_{\check{\Omega}}\left(\zeta^2\mathbb{C}\boldsymbol{N}\cdot \boldsymbol{N}-4\,\boldsymbol{\sigma}\cdot \boldsymbol{N}\,\zeta\,\beta+\frac{G_{\text{c}}}{\ell}\,\beta^2\right)\,d\check{\boldsymbol{x}}.
\label{eq:Delta_H}
\end{equation}
Note that $H$ is non-negative under the strain-hardening condition (see \ref{app:strain_hard}). Therefore, when $\Delta \leq 0$, the Gâteaux derivative in (\ref{eq:second_G_DH}) is positive. Hence, the interesting case for the stability analysis is when $\Delta$ is positive. Assuming $\Delta>0$, we define the Rayleigh ratio $\mathsf{R}_L$ and its limit  for $L\rightarrow \infty$, $\mathsf{R}_\infty$:
\begin{equation}
\mathsf{R}_L(\check{\boldsymbol{v}},\zeta,\beta):=\frac{H(\zeta,\beta)+G_{\text{c}}\,\frac{\ell}{L^2}\,\Vert \check{\nabla}\beta\Vert^2}{\Delta(\check{\boldsymbol{v}},\zeta)},\quad \mathsf{R}_{\infty}(\check{\boldsymbol{v}},\zeta,\beta)=\frac{H(\zeta,\beta)}{\Delta(\check{\boldsymbol{v}},\zeta)}. 
\end{equation}
both non-negative and lower-bounded.
For large domains, the stability condition can thus be expressed as
\begin{equation}
    \tilde{\mathsf{R}}:=\inf_{\check{\boldsymbol{v}}\in H^1_0(\check{\Omega})}\,\inf_{\substack{\zeta \geq 0 \\ \int_{\check{\Omega}} \zeta^2(x)\,dx = 1}}\,\inf_{\beta} \mathsf{R}_{\infty}(\check{\boldsymbol{v}},\zeta,\beta)\geq 1.
    \label{eq:infinfinf}
\end{equation}
Minimizing $H$ with respect to $\beta$ and taking into account the normalization $\int_{\check{\Omega}} \zeta^2(x)\,dx = 1$, we obtain
\begin{equation}
     \tilde{H} = \mathbb{C}\boldsymbol{N}\cdot \boldsymbol{N} - \frac{4\ell}{G_{\text{c}}} \left(\boldsymbol{\sigma} \cdot \boldsymbol{N} \right)^2.
\end{equation}
To attain the infimum in~\eqref{eq:infinfinf}, we now need to compute
\begin{equation}
    \tilde{\Delta} := \sup_{\check{\boldsymbol{v}}\in H^1_0(\check{\Omega})}\, \sup_{\substack{\zeta \geq 0 \\ \int_{\check{\Omega}} \zeta^2(x)\,dx = 1}} \Delta(\check{\boldsymbol{v}},\zeta) = \inf_{\check{\boldsymbol{v}}\in H^1_0(\check{\Omega})}\, \inf_{\substack{\zeta \geq 0 \\ \int_{\check{\Omega}} \zeta^2(x)\,dx = 1}} -\Delta(\check{\boldsymbol{v}},\zeta).
    \label{eq:Delta_prob}
\end{equation}
An important step in this direction is to introduce the subspace $\mathcal{M}(\boldsymbol{n})$ of $\mathbb{M}_s^d$, associated to any unit vector $\boldsymbol{n}\in\mathbb{R}^d$ with $\Vert \boldsymbol{n}\Vert=1$, defined as
\begin{equation}
    \mathcal{M}(\boldsymbol{n}) = \left\{2\, \boldsymbol{n}\otimes_{\text{sym}}\boldsymbol{w}  : \boldsymbol{w} \in \mathbb{R}^d \right\}.
    \label{eq:subsetM}
\end{equation}
To have a physical sense for this subspace, think that the singular part of the eigenstrain tensor $\boldsymbol{\eta}_S\in\mathcal{M}(\boldsymbol{n})$, where $\boldsymbol{n}$ is the normal to the crack set $J(\boldsymbol{z})$, and cannot live outside this subspace according to (\ref{eq:eta_jump}).
 With this definition, problem~\eqref{eq:Delta_prob} can be solved by directly applying the result presented in Lemma 1 of~\cite{pham2013stability}, which reads
\begin{equation}
    \tilde{\Delta} = \max_{\boldsymbol{n} : \Vert \boldsymbol{n} \Vert = 1} \mathbb{C} \tilde{\boldsymbol{\xi}}(\boldsymbol{n}) \cdot \tilde{\boldsymbol{\xi}}(\boldsymbol{n}),
    \label{eq:delta_max}
    \end{equation}
    where $\tilde{\boldsymbol{\xi}}(\boldsymbol{n})$ is obtained by solving the minimization problem
\begin{equation}
    \tilde{\boldsymbol{\xi}}(\boldsymbol{n})=\argmin_{\boldsymbol{\xi} \in \mathcal{M}(\boldsymbol{n})} \mathbb{C}(\boldsymbol{N} - \boldsymbol{\xi})\cdot(\boldsymbol{N} - \boldsymbol{\xi}).
    \label{eq:xi}
\end{equation}
Finally, the Rayleigh ratio $\tilde{\mathsf{R}}$ in (\ref{eq:infinfinf}) is given by
\begin{equation}
    \tilde{\mathsf{R}}=\frac{\tilde{H}}{\tilde{\Delta}}=\frac{\mathbb{C}\boldsymbol{N}\cdot \boldsymbol{N} - \frac{4\ell}{G_{\text{c}}} \left(\boldsymbol{\sigma} \cdot \boldsymbol{N} \right)^2}{\max_{\boldsymbol{n} : \Vert \boldsymbol{n} \Vert = 1} \mathbb{C} \tilde{\boldsymbol{\xi}}(\boldsymbol{n}) \cdot \tilde{\boldsymbol{\xi}}(\boldsymbol{n})},
    \label{eq:Rayleigh}
\end{equation}
Recall that, for a given $\boldsymbol{\sigma}$, $\boldsymbol{N}$ is determined by the normality rule~(\ref{eq:normality_rule}), hence $\tilde{\mathsf{R}}$ depends solely on $\boldsymbol{\sigma}$.
\par Let us now notice that if $\tilde{\boldsymbol{\xi}}(\boldsymbol{n})=\boldsymbol{N}$, then   
\begin{equation}
    \tilde{\mathsf{R}}=\frac{\mathbb{C}\boldsymbol{N}\cdot \boldsymbol{N} - \frac{4\ell}{G_{\text{c}}} \left(\boldsymbol{\sigma} \cdot \boldsymbol{N} \right)^2}{\mathbb{C}\boldsymbol{N}\cdot \boldsymbol{N}}\leq 1,
\end{equation}
meaning that the corresponding homogeneous solution is \textit{unstable} for $\ell>0$. Thanks to Proposition~3 in \cite{pham2013stability}, we know that this occurs if and only if
\begin{enumerate}
    \item the orientation $\boldsymbol{N}$ is a rank-one tensor in $\mathbb{M}_s^d$, or
    \item the orientation $\boldsymbol{N}$ is a rank-two tensor in $\mathbb{M}_s^d$ and its non-zero eigenvalues have opposite signs.
\end{enumerate}
Note that the second case occurs in the case of pure shear strain. Therefore, the homogeneous solution with a purely shear strain state is always unstable for $\ell>0$ (see \ref{app:stability_ps}).

If the orientation of the eigenstrain tensor $\boldsymbol{N}$ does not satisfy one of these conditions, no general conclusions about stability can be drawn. However, further insights may be obtained by analyzing equation~(\ref{eq:Rayleigh}) in more detail. For example, in \ref{app:Rayleigh}, we study the case of volumetric strain for which $\boldsymbol{\sigma} = p_{\text{c}}\,\boldsymbol{I}$ and $\boldsymbol{N} = \frac{1}{\sqrt{3}}\,\boldsymbol{I}$. For this solution, we obtain the Rayleigh ratio
\begin{equation}
    \boldsymbol{\sigma} = p_{\text{c}}\,\boldsymbol{I}: \quad \tilde{\mathsf{R}} = 12\,\frac{1 - \nu}{1 + \nu}\,\left(1 - 4\,\ell\frac{p_{\text{c}}^2}{\kappa\,G_{\text{c}}}\right).
    \label{eq:Rayleigh_vol}
\end{equation}
From Table~\ref{tab:ell_strength} we know that, if $\frac{\kappa\,G_{\text{c}}}{p_{\text{c}}^2}\leq \frac{2\,\mu\,G_{\text{c}}}{\tau_{\text{c}}^2}$ then $\ell_{\text{ch}}=\frac{\kappa\,G_{\text{c}}}{p_{\text{c}}^2}$. Thus, in this case it is always possible to choose $\ell$ sufficiently small to satisfy the strain-hardening condition $\ell\leq \frac{1}{4}\,\ell_{\text{ch}}$ but sufficiently large to make the homogeneous solution unstable. This is similar to the observation in~\cite{zolesi2024stability} that, in asymptotically cohesive phase-field models, reducing the length ratio $\ell/L$ promotes instability of the homogeneous solution, while decreasing the ratio $\ell/\ell_{\text{ch}}$ has the opposite effect. These models admit a localized solution that is independent of $\ell$, however, in the general multi-axial case, the homogeneous solution and its stability depends on $\ell$. Clearly, choosing $\ell$ not too much smaller than $\ell_{\text{ch}}$ is also an advantage in terms of  computational cost, as a very small $\ell$ would require an accordingly fine mesh.

\subsection{Cohesive law in the three-dimensional setting}
\label{sec:cohesive_law_3D}
\begin{figure}[H]
\centering
\includegraphics[scale =1]{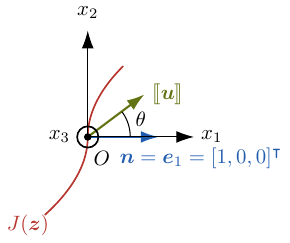}
\caption{Schematic representation of the jump set $J(\boldsymbol{z})$ on the plane passing through the point $O$, where both the outer unit normal $\boldsymbol{n}$ (in blue) and the displacement jump $\jump{\boldsymbol{u}}$ at $O$ lie. The reference frame $\{x_1,x_2,x_3\}$ is centered in $O$, it is oriented such that the axis $x_3$ is perpendicular to the plane and the axis $x_1$ is aligned with $\boldsymbol{n}$. The angle between $\boldsymbol{n}$ and $\jump{\boldsymbol{u}}$ is denoted with $\theta\in[-\pi/2,\pi/2]$.}
\label{fig:2d_crack}
\end{figure}

Given a crack set \( J(\boldsymbol{z}) \), let us consider a point \( O \in J(\boldsymbol{z}) \). We denote by \( \boldsymbol{n} \) the outward unit normal to the crack set, and by \( \jump{\boldsymbol{u}} \) the displacement jump at this point. We then consider the plane passing through \( O \) that contains both \( \boldsymbol{n} \) and \( \jump{\boldsymbol{u}} \). On this plane, we define a coordinate system \( \{x_1, x_2, x_3\} \) centered at \( O \), as illustrated in Figure~\ref{fig:2d_crack}. The angle \( \theta \) between \( \boldsymbol{n} \) and \( \jump{\boldsymbol{u}} \) is denoted with $\theta\in[-\pi/2,\pi/2]$.

In the chosen reference frame, the components of the displacement jump $\jump{\boldsymbol{u}} = \left[\delta_n, \delta_t, 0\right]^\intercal$ can be expressed as
\begin{equation}
\delta_n = \delta \cos(\theta), \quad 
\delta_t = \delta\sin(\theta)\quad\text{with } \theta \in \left[-\frac{\pi}{2}, \frac{\pi}{2}\right],
\end{equation}
where we defined $\delta=\big\Vert\jump{\boldsymbol{u}}\big\Vert $. We interpret $\delta_n$ as the \textit{normal (opening) relative displacement}  and $\delta_t$ as the \textit{tangential relative displacement} across the crack surface. It follows that
\begin{equation}
\jump{\boldsymbol{u}} \otimes_{\text{sym}} \boldsymbol{n} = \delta
\begin{bmatrix} 
\cos \theta & \frac{1}{2} \sin \theta & 0\\ 
\frac{1}{2} \sin \theta & 0 & 0\\
0 & 0 & 0
\end{bmatrix}  \quad \text{in } O.
\label{eq:rank2}
\end{equation}
From this expression, we obtain in particular:
\begin{equation}
\label{voldev}
\text{tr}(\jump{\boldsymbol{u}} \otimes_{\text{sym}} \boldsymbol{n}) = \delta_n = \delta\,\cos(\theta) , \quad 
\big\Vert(\jump{\boldsymbol{u}} \otimes_{\text{sym}} \boldsymbol{n})_{\text{dev}}\big\Vert = \delta\sqrt{\frac{1}{2} + \frac{\cos^2(\theta)}{6}} ,
\quad \theta \in \left[-\frac{\pi}{2}, \frac{\pi}{2}\right].
\end{equation}

We observe that whenever a displacement jump occurs, i.e., $\delta > 0$, the symmetric tensor $\jump{\boldsymbol{u}} \otimes_{\text{sym}} \boldsymbol{n}$ necessarily has a non-zero deviatoric part, regardless of the angle $\theta \in \left[-\frac{\pi}{2}, \frac{\pi}{2}\right]$. In other words, it is impossible to have a purely volumetric sharp crack as the fracture process inherently entails a change in shape. This contrasts with the case of homogeneous damage, where shape preservation is admissible.
\par If $\boldsymbol{\eta}$ belongs to the subspace $\mathcal{M}(\boldsymbol{n})$, introduced in Eq.~(\ref{eq:subsetM}), it can be written in the form given by Eq.~(\ref{eq:rank2}). Consequently, we refer to $\mathcal{M}(\boldsymbol{n})$ as the space of \textit{jump-compatible} eigenstrains. If $\boldsymbol{\eta}$ is \textit{jump-compatible}, i.e., $\boldsymbol{\eta} \in \mathcal{M}(\boldsymbol{n})$, no homogeneous solution occurs, and damage localization and cohesive crack formation automatically arise, as discussed in the following remark.
\begin{remark}
   Consider the homogeneous state characterized by a jump-compatible eigenstrain tensor $\boldsymbol{\eta}$ whose matrix representation takes the form given in equation~(\ref{eq:rank2}). This tensor has eigenvalues
\begin{equation}
    \lambda_1 = \frac{1 + \cos(\theta)}{2}, \quad
    \lambda_2 = \frac{-1 + \cos(\theta)}{2}, \quad
    \lambda_3 = 0,
\end{equation}
and is therefore either a rank-1 tensor or a rank-2 tensor with non-zero eigenvalues of opposite signs. According to the stability results discussed in Section~\ref{sec:stability3D}, such a homogeneous state is unstable.
In other words, when the eigenstrain tensor is clearly associated with a crack, the system will prefer localization and crack formation over the development of diffuse damage. On the other hand, when this is not the case, especially when the volumetric part of the eigenstrain dominates over the deviatoric part, the formation of diffuse damage cannot be excluded a priori and the conclusion on whether localization occurs or not requires a more detailed analysis.
\end{remark}
Assuming that the characteristic length of the domain $\Omega$ is much larger than the regularization length $\ell$, and that the radii of curvature of the jump set $J(\boldsymbol{z})$ are also much larger than $\ell$, the phase field $\alpha$ along the direction $\boldsymbol{n}$, i.e., $\alpha(x_1 \boldsymbol{n})$, is described by the localization profile obtained in the one-dimensional setting. Consequently, for the directional derivative along $\boldsymbol{n}$, we can use the result in (\ref{eq:profile_a}) and we obtain:
\begin{equation}
\left\Vert\nabla \alpha(x_1 \boldsymbol{n}) \cdot \boldsymbol{n} \right\Vert = 
\left\Vert \frac{\partial \alpha(\boldsymbol{x})}{\partial x_1} \right\Vert_{\boldsymbol{x} = x_1 \boldsymbol{n}} = 
\frac{1}{\ell} \sqrt{w(\alpha(x_1 \boldsymbol{n}))} \quad \text{for } x_1 \neq 0,
\end{equation}
which implies:
\begin{equation}
\jump{\nabla \alpha} \cdot \boldsymbol{n} = -\frac{2}{\ell} \sqrt{w(\alpha)} \quad \text{in } O.
\end{equation}
Assuming $\delta > 0$, the KKT conditions (\ref{eq:KKT_damage2}) in $O$ require that
\begin{equation}
\label{eq:jump_law_0}
a'(\alpha)\, \pi_0\left(\jump{\boldsymbol{u}} \otimes_{\text{sym}} \boldsymbol{n} \right) = 2 \frac{G_{\text{c}}\ell}{c_w}  \jump{\nabla \alpha} \cdot \boldsymbol{n} = -4 \frac{G_{\text{c}}}{c_w} \sqrt{w(\alpha)} \quad \text{in } O,
\end{equation}
where, recalling  (\ref{eq:k_pot}) and (\ref{voldev}),
\begin{equation}
\pi_0\left(\jump{\boldsymbol{u}} \otimes_{\text{sym}} \boldsymbol{n} \right)= \phi\left(\delta\cos(\theta), \delta\sqrt{\frac{1}{2} + \frac{\cos^2(\theta)}{6}} \right)=\delta\,\sigma_{\text{c}}^*(\theta) 
\label{eq:jump_law}
\end{equation}
with
\begin{equation}
\sigma_{\text{c}}^*(\theta) := \phi\left(\cos(\theta), \sqrt{\frac{1}{2} + \frac{\cos^2(\theta)}{6}} \right)
\end{equation}
as the angle-dependent \textit{equivalent cohesive strength}.
From (\ref{eq:jump_law_0}) and (\ref{eq:jump_law}), we obtain the relation between the magnitude of the displacement jump and the value of the phase-field variable $\alpha$ in O:
\begin{equation}
\delta =  -\frac{4}{c_w} \frac{G_{\text{c}}}{\sigma_{\text{c}}^*(\theta)} \frac{\sqrt{w(\alpha)}}{a'(\alpha)}=:g^{-1}_{\theta}(\alpha) \quad \text{in } O
\label{eq:jump_alpha}
\end{equation}
which is the analogous of (\ref{eq:jump_w}) for the three-dimensional case.

Letting $\boldsymbol{\sigma}$ denote the stress tensor at point $O$, we define the \textit{traction} vector $\boldsymbol{T}$ as
\begin{equation}
\boldsymbol{T} := \boldsymbol{\sigma} \, \boldsymbol{n} = [T_1, T_2, T_3]^\intercal.
\end{equation}
The infinitesimal work done by the traction on the infinitesimal displacement jump $d\jump{\boldsymbol{u}}$ can be expressed in terms of the stress tensor and the infinitesimal increment of the symmetric product $d\left(\jump{\boldsymbol{u}} \otimes_{\text{sym}} \boldsymbol{n}\right)$ as:
\begin{equation}
\boldsymbol{T} \cdot d\jump{\boldsymbol{u}} = \boldsymbol{\sigma} \cdot  d\left(\jump{\boldsymbol{u}} \otimes_{\text{sym}} \boldsymbol{n} \right).
\end{equation}
The eigenstrain evolution criterion (\ref{eq:eta_ev}) applied to $\boldsymbol{\eta}=\boldsymbol{\eta}_S$ (hence on $J(z)$) yields
\begin{equation}
\boldsymbol{\sigma} \cdot \boldsymbol{\eta}_S = \pi(\boldsymbol{\eta}_S,\alpha) 
\quad \text{with} \quad \boldsymbol{\eta}_S = \jump{\boldsymbol{u}} \otimes_{\text{sym}} \boldsymbol{n} \, \delta_{J(\boldsymbol{z})},
\end{equation}
from which we deduce
\begin{equation}
\boldsymbol{T} \cdot d\jump{\boldsymbol{u}} = \hat{\sigma}_{\theta}\left( \delta \right) \, d\delta, \quad 
\text{with} \quad \hat{\sigma}_{\theta}\left( \delta \right) := 
a\left( g_{\theta}\left( \delta \right) \right) \, \sigma_{\text{c}}^*(\theta),
\label{eq:cohesive_law_2D}
\end{equation}
where $\hat{\sigma}_{\theta}$ defines the \textit{cohesive law} in the direction $\theta$. By integration, we obtain the \textit{cohesive surface energy density} $k$
\begin{equation}
k\left(\jump{\boldsymbol{u}}\right) = 
\hat{k}_{\theta}\left( \delta \right) := 
\int_0^{\delta} \hat{\sigma}_{\theta}(v) \, dv.
\end{equation}
Using the \texttt{AT2} model, the cohesive law and the cohesive surface energy density are
\begin{equation}
\hat{\sigma}_{\theta}\left( \delta \right)=\frac{\sigma_{\text{c}}^*(\theta)}{\left(1+\frac{\sigma_{\text{c}}^*(\theta)}{G_{\text{c}}}\,\delta\right)^2},\quad \hat{k}_{\theta}\left( \delta  \right)=\frac{G_{\text{c}}\,\delta}{\frac{G_{\text{c}}}{\sigma_{\text{c}}^*(\theta)}+\delta},
\end{equation}
which closely resemble (\ref{eq:coh_law1D}) and(\ref{eq:surf_dens_1D}) obtained in the one-dimensional case, but now include the dependence on $\theta$ through the equivalent cohesive strength $\sigma_{\text{c}}^*(\theta)$ in place of $\sigma_{\text{c}}$.
\par
\begin{figure}[H]
    \centering
    \includegraphics[scale=1]{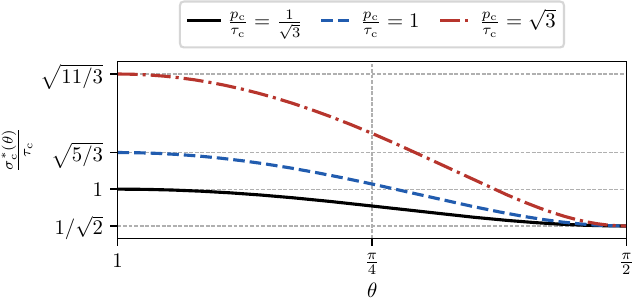}
    \caption{Equivalent cohesive strength $\sigma_{\text{c}}^*(\theta)$ as a function of $\theta$ for different values of the strength ratio $\frac{p_{\text{c}}}{\tau_{\text{c}}}=\frac{1}{\sqrt{3}}, 1,\sqrt{3}$.} 
    \label{fig:stress_eq}
\end{figure}

\begin{figure}[H]
    \centering
    \includegraphics[scale=1]{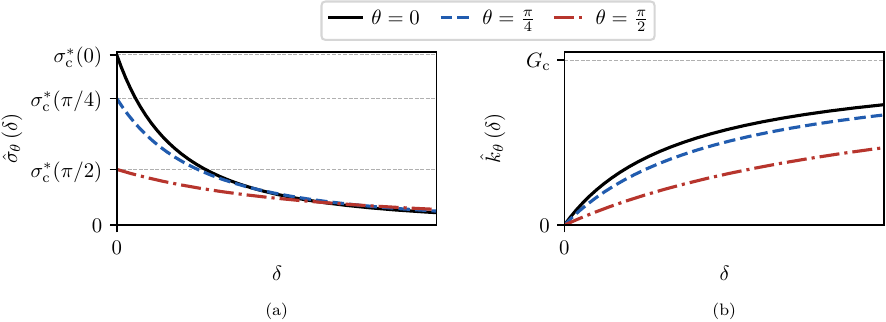}
    \caption{Cohesive law $\hat{\sigma}_\theta(\delta)$ (a) and cohesive surface energy density $\hat{k}_{\theta}(\delta)$ (b) for $\frac{p_\text{c}}{\tau_{\text{c}}}=2$ and different values of the angle $\theta=0, \frac{\pi}{4},\frac{\pi}{2}$.} 
    \label{fig:cohesive_law_theta}
\end{figure}

 For illustrative purposes, let us consider $\phi_r$ in (\ref{eq:pot_phi_n}) for $r=2$, for which the equivalent stress $\sigma_{\text{c}}^*(\theta)$ (Figure~\ref{fig:stress_eq}) is given by
\begin{equation}
	\sigma_{\text{c}}^*(\theta)=\sqrt{p_{\text{c}}^2\,\text{cos}^2(\theta)+\tau_{\text{c}}^2\,\left(\frac{1}{2}+\frac{\text{cos}^2(\theta)}{6}\right)}.
    \label{eq:sigma_star}
\end{equation}
Note that, according to Eq.~\eqref{eq:sigma_star}, we have $\sigma_C^*(0) > \sigma_C^*(\pi/2)$, as a consequence of the previously observed fact that any jump-compatible $\boldsymbol{\eta}$ necessarily has a non-zero deviatoric part. We then plot in Figure~\ref{fig:cohesive_law_theta} the corresponding cohesive law and cohesive energy density as functions of $\delta$, for different values of the angle $\theta$. As in the one-dimensional setting, the cohesive laws are of the Barenblatt type: they are convex, starting from the peak value $\sigma_C^*(\theta)$ and decreasing monotonically to zero. Similarly, the cohesive surface energy densities are concave; they vanish at $\delta = 0$, have initial slope given by $\sigma_C^*(\theta)$, and asymptotically tend to $G_c$.

\section{Computational aspects and numerical examples}
\label{sec:numerics}
After having explored the behavior of the proposed model in the one- and three-dimensional cases, we now discuss its numerical implementation. We also present some numerical examples that aim at validating the analytical results.

\subsection{Discretization and solution strategy}
For the numerical implementation, we employ a standard approach with the finite element (FE) method. For the strength potential, we implement the three cases $\phi_1$, $\phi_2$, and $\phi_\infty$ presented in Section~\ref{sec:norm_pot}. Among these, $\phi_2$ and $\phi_\infty$ are not Fréchet-differentiable at the origin $(\text{tr}(\boldsymbol{\eta}), \Vert \boldsymbol{\eta}_{\text{dev}}\Vert) = (0, 0)$. To deal with non-differentiability in the numerical context, we employ a regularization by means of a small numerical constant $0 < \epsilon \ll 1$, that is
\begin{subequations}
    \begin{equation}
        \phi_2(\text{tr}(\boldsymbol{\eta}), \|\boldsymbol{\eta}_{\text{dev}}\|) \approx \phi_2^\epsilon(\text{tr}(\boldsymbol{\eta}), \|\boldsymbol{\eta}_{\text{dev}}\|)=\sqrt{p_{\text{c}}^2\,\text{tr}(\boldsymbol{\eta})^2+\tau_{\text{c}}^2\,\|\boldsymbol{\eta}_{\text{dev}}\|^2 + \epsilon} \qquad\text{and}
        \label{eq:norms_reg}
    \end{equation}
    \begin{equation}
        \phi_{\infty}(\text{tr}(\boldsymbol{\eta}),\|\boldsymbol{\eta}_{\text{dev}}\|) \approx \phi_{\infty}^\epsilon(\text{tr}(\boldsymbol{\eta}),\|\boldsymbol{\eta}_{\text{dev}}\|)=\frac{1}{2}\left(p_{\text{c}}\text{tr}(\boldsymbol{\eta})+\tau_{\text{c}}\,\|\boldsymbol{\eta}_{\text{dev}}\|\right)+\frac{1}{2}\sqrt{\left(p_{\text{c}}\,\text{tr}(\boldsymbol{\eta})-\tau_{\text{c}}\,\|\boldsymbol{\eta}_{\text{dev}}\|\right)^2+\epsilon}
    \end{equation}
\end{subequations}
for the Euclidean norm \eqref{eq:norm2}, and the max-function \eqref{eq:normInf}, respectively.
The value of $\epsilon$ should be large enough to avoid ill-posedness of the algebraic system or convergence issues in the numerical context, and small enough to avoid numerical artifacts in the solution due to the non-vanishing values of $\phi^\epsilon_2(0,0)$ and $\phi^\epsilon_\infty(0,0)$.
We found $\epsilon = (3 \cdot 10^{-16})^2$ and $\epsilon = 10^{-17}$ to be robust choices for $\phi_2^\epsilon$ and $\phi_{\infty}^\epsilon$, respectively.

We discretize the displacement field $\boldsymbol{u}$ and the phase field $\alpha$ with standard linear FEs. Being the strain constant within the elements, a natural choice is to discretize the eigenstrain $\boldsymbol{\eta}$ with a piecewise constant ansatz, i.e. a single degree of freedom (DOF) per element. In the multi-axial setting, we exploit the volumetric-deviatoric decomposition introduced in Subsec.~\ref{subsec:voldev}, so that the problem in terms of $\boldsymbol{\eta}$ depends exclusively on $\text{tr}(\boldsymbol{\eta})$ and $\|\boldsymbol{\eta}_{\text{dev}}\|$.
This approach allows us to work with two scalar quantities, $\text{tr}(\boldsymbol{\eta})$ and $\|\boldsymbol{\eta}_{\text{dev}}\|$, which we discretize separately  with a single DOF per element each.  
% To properly account for the fact that the second scalar variable represents a norm which must be non-negative, the constraint $\operatorname{tr}(\boldsymbol{\eta}) \geq 0$ must be complemented with the additional condition $\|\boldsymbol{\eta}_{\text{dev}}\| \geq 0$.
We use \texttt{FEniCSx} \cite{baratta_2023,scroggs_2022a,scroggs_2022b}, where we exploit the automatic differentiation capabilities of \texttt{UFL} \cite{alnaes_2014} to automatically derive from the regularized energy functional the residuals and their linearizations giving rise to the coupled non-linear FE equation system.

For the solution, we exploit the separate convexity of the energy functional with respect to  $(\boldsymbol{u}, \text{tr}(\boldsymbol{\eta}), \|\text{dev}(\boldsymbol{\eta})\|)$ and $\alpha$ and employ an \textit{alternate minimization} (also known as \textit{staggered}) solution strategy as outlined in Algorithm \ref{alg:altmin}.
In other words, we minimize first with respect to $(\boldsymbol{u}, \text{tr}(\boldsymbol{\eta}), \|\text{dev}(\boldsymbol{\eta})\|)$  while keeping $\alpha$ fixed, and then minimize with respect to $\alpha$ while keeping $(\boldsymbol{u}, \text{tr}(\boldsymbol{\eta}), \|\text{dev}(\boldsymbol{\eta})\|)$ fixed.
This procedure is repeated until convergence is reached, which we monitor based on the norm of the re-evaluated residual of the $(\boldsymbol{u}, \text{tr}(\boldsymbol{\eta}), \|\text{dev}(\boldsymbol{\eta})\|)$ problem with the updated $\alpha$.
Since both sub-problems are convex, this solution approach guarantees an energy descent and ultimately convergence of the scheme \cite{bourdin2007numerical}.
For both separate minimzation problems, we adopt the iterative Newton-Raphson scheme implemented in \texttt{PETSc} \cite{balay_2024,balay_1997,dalcin_2011} using the variational inequality solvers outlined in \cite{benson_2006} to handle the constraints on $\boldsymbol{\eta}$ and $\alpha$. The constraints on $\boldsymbol{\eta}$ are the non-negativity of the non-linear strain $\eta$ in the one-dimensional setting, and the non-negativity of both $\text{tr}(\boldsymbol{\eta})$ and $\Vert\boldsymbol{\eta}_{\text{dev}}\Vert$ in the three-dimensional setting. The constraint on $\alpha$ is given by the irreversibility condition $\alpha \geq \alpha_p$, where the lower bound $\alpha_p$ is continuously updated to the converged solution from the previous load step.
Due to the strong non-linearities of both problems, we augment the Newton-Raphson scheme with a line-search algorithm as detailed in \cite{heinzmann_robust_2025}.
Our implementation is publicly available at \url{https://github.com/jonas-heinzmann/phase_field_cohesive_fracture} where
further details can be found.

\normalem %otherwise, the conditional statements will be underlined
\begin{algorithm}
    \caption{Alternate minimization strategy at a generic time step $n$. We choose $\mathtt{TOL}_{\mathbf{R}} = 10^{-8}$.}\label{alg:altmin}
    \DontPrintSemicolon
    \KwData{$\alpha_{p}=\alpha_{n-1}$}
    \KwResult{$\boldsymbol{u}_n$, $\boldsymbol{\eta}_n$, $\alpha_n$}
    $i = 0$, $\boldsymbol{\alpha}_n^0 = \alpha_{p}$ \Comment*[r]{initialize alternate minimization}
    \While{$\|\mathbf{R}_{\boldsymbol{u}\boldsymbol{\eta}}^i \|_2 > \mathtt{TOL}_{\mathbf{R}}$}{
        $i \gets i + 1$ \Comment*[r]{increment staggered iteration number}
        $(\boldsymbol{u}_n^i, \boldsymbol{\eta}_n^i) = \underset{(\boldsymbol{u}, \boldsymbol{\eta}), \text{tr}(\boldsymbol{\eta}) \geq 0}{\argmin} \, \mathcal{E}_\ell (\boldsymbol{u}, \boldsymbol{\eta}, \alpha_n^{i-1})$ \Comment*[r]{solve coupled $(\boldsymbol{u}, \boldsymbol{\eta})$ problem with Newton-Raphson}
        $\alpha_n^i = \underset{\alpha \geq \alpha_{p}}{\argmin} \, \mathcal{E}_\ell (\boldsymbol{u}_n^i, \boldsymbol{\eta}_n^i, \alpha)$ \Comment*[r]{solve $\alpha$ problem with Newton-Raphson}
        $\mathbf{R}_{\boldsymbol{u}\boldsymbol{\eta}}^i = \mathbf{R}_{\boldsymbol{u}\boldsymbol{\eta}}^i (\boldsymbol{u}_n^i, \boldsymbol{\eta}_n^i, \alpha_n^i)$ \Comment*[r]{recompute residual for convergence check}
    }
\end{algorithm}
\ULforem

%\par Lastly, we point out that in this work we opted for the mixed \((\boldsymbol{u}, \boldsymbol{\eta})\) formulation, which enables the use of a global gradient-based solution algorithm for the coupled sub-problem, while also allowing efficient handling of the non-negativity constraint on the eigenstrain through bound-constrained optimization solvers. On the other hand, employing a piecewise constant ansatz per element for \( \eta \) (or for \( \text{tr}(\boldsymbol{\eta}) \) and \( \|\boldsymbol{\eta}_{\text{dev}}\| \)) corresponds to using a single Gauss point per element, at which the eigenstrain \( \eta \) is updated. Additionally, given that the eigenstrain law~\eqref{eq:nl_law} depends only on local quantities, this naturally suggests, as an alternative, the use of a return-mapping algorithm. However, this has not yet been explored and is considered a topic for future work focused on numerical aspects, remaining out of the scope of the present study.

\subsection{One-dimensional case}
We start by evaluating the numerically obtained solution for the one-dimensional case discussed in Section~\ref{sec:1D} with the setup in Figure~\ref{fig:bar}.
We adopt  $L=1.0$~mm, $E=10^4$~MPa, $G_{\text{c}}=10^{-3}$~N/mm, $\ell=0.05$~mm as well as $\sigma_{c}=5$~MPa (giving a structural brittleness ratio of $B=5$), and (unless otherwise specified) we spatially discretize the domain with $\ell/h \approx 5$, where $h$ is the element size.
The prescribed displacement up to a maximum value of $U_t=0.005$~mm is incrementally applied by $500$ uniform steps.

\begin{figure}[H]
    \centering
    \includegraphics{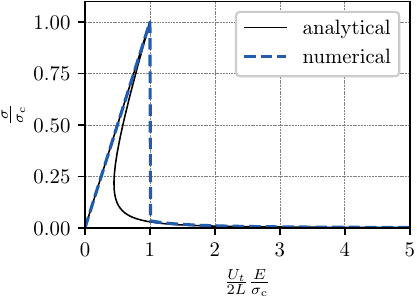}
    \caption{Numerically obtained structural response of the bar in Figure~\ref{fig:bar} for $B=5$.}
     \label{fig:1D_fu}
\end{figure}

We first look at the structural response. As shown in Figure~\ref{fig:1D_fu}, the dimensionless stress-displacement response obtained numerically is in excellent agreement with the theoretical solution \eqref{eq:structural_U} (see also Figure~\ref{fig:structural_snap}), with the difference that the structural snap-back cannot be captured in the numerical setting under displacement control.
We observe that the numerical post-peak response is slightly stiffer than theoretically predicted, which will be explained later. 

\begin{figure}[H]
    \centering
    \includegraphics{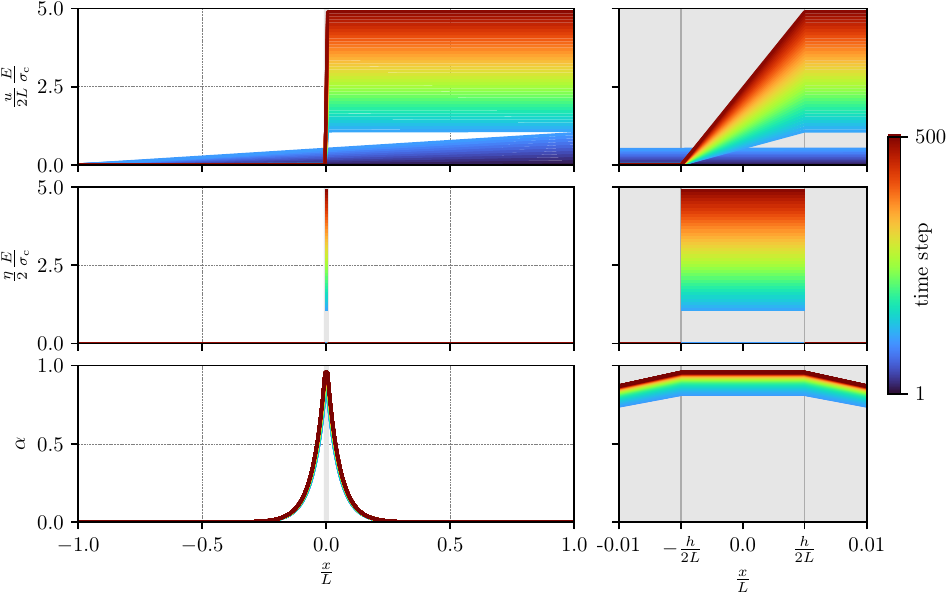}
    \caption{Numerically obtained solution fields $(u, \eta, \alpha)$ for the bar loaded in tension. The plots on the left show the solution over the whole domain; those on the right are enlarged views within a small region around the crack position, highlighted as gray area on the left. Note that the shown value for $\eta$ corresponds to the numerically obtained value $\eta_e^h$ multiplied by the element size $h$.}
    \label{fig:1D_fields}
\end{figure}

\begin{figure}[H]
\hspace{-1.5cm}
    \centering
    \begin{subfigure}[t]{0.43\textwidth}
        \centering
\includegraphics{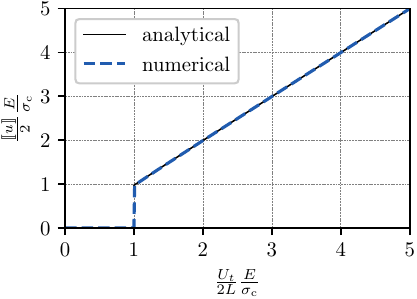}
        \caption{}
                \label{fig:1D_jump}
    \end{subfigure}
    \begin{subfigure}[t]{0.43\textwidth}
        \centering
        \includegraphics{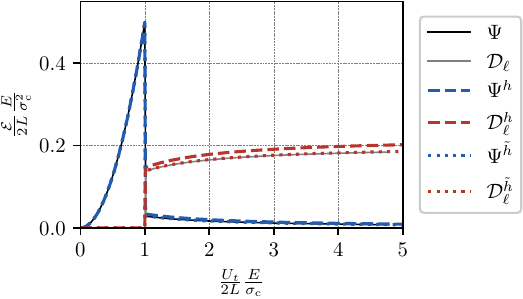}
        \caption{}
        \label{fig:1D_energies}
    \end{subfigure}
    \caption{Comparison between the theoretical jump $\jump{u}$ and the numerical quantity $\eta_e^h\,h$ as functions of the prescribed displacement $U_t$ (a). Comparison between the theoretical  energy contributions and those obtained numerically with  $\ell/h\approx 5$ (dash-dotted curves), and the same discretization, but where the element in which $\eta$ is allowed to take non-zero values has instead a size of $\ell/\tilde{h}\approx 25$ (dotted curves) (b).}
\end{figure}

%The numerically obtained displacement jump is also in very good agreement with the theoretical solution, as shown in Figure~\ref{fig:1D_jump}.
%Due to the structural snap-back not captured numerically, the displacement jump suddenly appears after the boundary of the elastic domain has been reached.
In Figure~\ref{fig:1D_fields}, we present the numerically obtained $(u, \eta, \alpha)$ fields for all loading increments.
During the elastic phase, the displacement is linear (hence strain and stress are constant in the domain), while the phase field and the eigenstrain are both zero.
When the critical stress $\sigma_\text{c}$ is reached, the eigenstrain $\eta$ becomes non-zero. In theory, $\eta$ should localize at a single point; however, due to the piecewise constant FE ansatz, the numerical solution for $\eta$ localizes over an entire 'cracked' element with a value $\eta_e^h\approx\jump{u}/h $, where $\jump{u}$ is the theoretical jump for the prescribed displacement $U_t$ (Figure~\ref{fig:1D_jump}).  In the absence of additional constraints, the specific element where localization occurs is dictated by numerical noise, as a crack could, in theory, nucleate anywhere along the bar. Without loss of generality, we constrain $\eta$ to evolve only in the central element of the domain, enabling a direct comparison with the analytical solution.

Similarly, the displacement cannot exhibit a jump at one point due to the use of linear elements. As a result, the numerical displacement field approximates a jump through a steep gradient across the `cracked' element. This leads the numerically approximated phase field to exhibit an approximately constant value within the `cracked' element, around which its profile is centered. The above approximations of the displacement jump and of the maximum phase-field value are also encountered at fully cracked conditions with the discretized phase-field model of brittle fracture when using linear FEs. As a result, the same expression for the approximate dissipated energy derived in Section~8.1 of~\cite{bourdin2008variational} also applies here. For a given \( \alpha(0) \), the discrepancy between the numerical and theoretical dissipated energies $\mathcal{D}_\ell^h$ and  $\mathcal{D}_\ell $ is given by
\begin{equation}
    \mathcal{D}_\ell^h - \mathcal{D}_\ell = \frac{G_{\text{c}} w(\alpha(0))}{c_w} \frac{h}{\ell}
    \label{eq:D_ell_overestim}
\end{equation}
which shows that the numerical overestimation of the dissipated energy scales linearly with \( h \).
However, with the present model the displacement jump is not only associated with \( \alpha(0) = 1 \) but also appears for lower values of \( \alpha(0) \). Consequently, the discrepancy in the dissipated energy persists and accumulates throughout the evolution of the localized solution, as illustrated in Figure~\ref{fig:1D_energies}.
It can be seen that the discretization almost correctly captures the fracture energy $\mathcal{D}_\ell(\alpha)$ right after nucleation, while the discrepancy grows larger afterwards.
On the contrary, the elastic energy of the system $\Psi(\boldsymbol{u},\boldsymbol{\eta},\alpha)$ is overestimated right after nucleation, while the discrepancy decreases afterwards (Figure~\ref{fig:1D_energies}). 

%This is in fact related to the numerical post-peak response being slightly stiffer than the theoretical solution as evident in Figure~\ref{fig:1D_fu}, and stems from our choice to approximate a discontinuous displacement field with a continuous numerical approximation.

\subsection{Plane-strain case: strength surface}
\label{sec:2D_plate} 
\begin{figure}[H]
    \centering
    \includegraphics{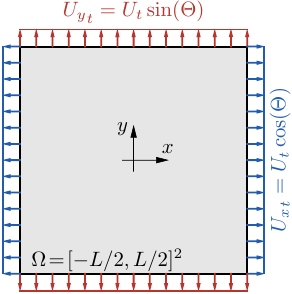}
    \caption{Test setup to numerically obtain the strength surface.}
    \label{fig:2D_elastic_domain_setup}
\end{figure}

Next, we move on to the multi-axial case, assuming plane-strain conditions. First, we aim to numerically reproduce the theoretical strength surface. To this end, we consider the initially intact square domain of edge length $L=1$~mm illustrated in Figure~\ref{fig:2D_elastic_domain_setup}. 
%The center of the geometry coincides with the origin of the $x$- and $y$-axes, while the out-of plane direction is the $z$-axis.
At the boundary, displacements are prescribed in opposite directions on opposing edges. In particular, the displacement in the \(x\)-direction is given by \({U_x}_t = U_t \cos(\Theta)\), and in the \(y\)-direction by \({U_y}_t = U_t \sin(\Theta)\), where \(\Theta\) is a fixed angle and \(U_t\) is uniformly incremented from zero up to the maximum value \(L/4\) over 2500 steps.  

Initially, the solution remains homogeneous, with strain components \(\varepsilon_{xx} = 2\,{U_x}_t / L\), \(\varepsilon_{yy} = 2\,{U_y}_t / L\), and \(\varepsilon_{xy} = 0\), so that  
\begin{equation}
    \frac{\varepsilon_{yy}}{\varepsilon_{xx}} = \tan \Theta.
\end{equation}

Accordingly, through \(\Theta\), we can directly control the strain ratio and thus indirectly explore the stress space. For each value of \(\Theta\), we numerically identify the time step at which the phase field becomes non-zero for the first time and record the corresponding stress value. We present these results in both the \(p\)-\(\tau\) and \(\sigma_{xx}\)-\(\sigma_{yy}\) diagrams. By running multiple simulations for different values of \(\Theta\) and plotting the identified points, we reconstruct the strength surface \(\partial \mathcal{S}_0\). We sample 72 angles, \(\Theta \in [0, 2\pi]\).

To prevent spurious crack nucleation at the boundary, we impose $\alpha = 0$ along the edges of the sample.
The geometry is discretized with a structured mesh of triangular elements with edge length $h$ such that $\ell/h \approx 5$.
For the material parameters, we consider $E=100$~MPa, $\nu=0.3$, $G_{\text{c}} = 0.2$~MPa~mm, $\ell=0.025$~mm.
For $r=2$ and $r=\infty$, we set $p_c = \tau_c = 12.4$~MPa, while for $r=1$ we set $p_c = \tau_c = 8.9$~MPa, which correspond to the respective limits of $\ell / \ell_{\text{ch}} = 1/4$ (see Table~\ref{tab:ell_strength}). This will be of relevance later on.

\begin{figure}[p]
    \centering
    \includegraphics{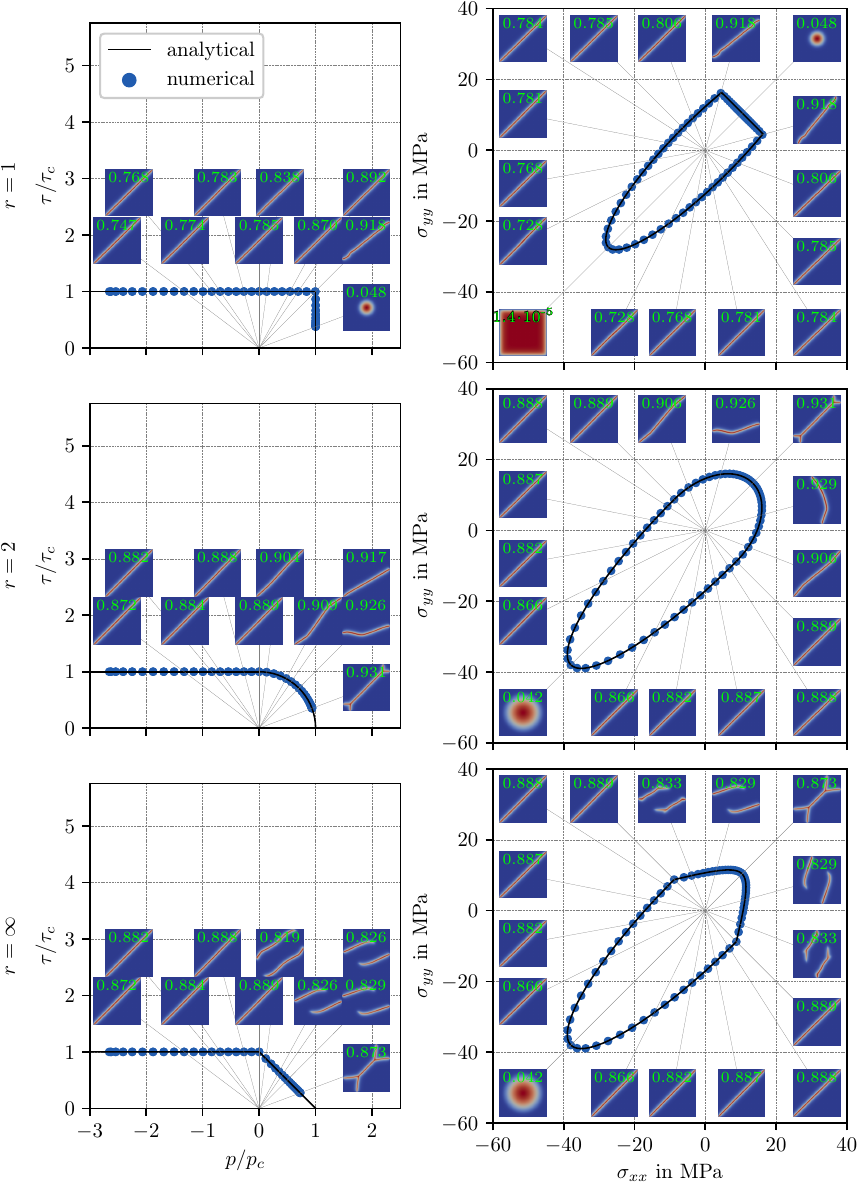}\hspace{0.25cm}\raisebox{8.5cm}{\includegraphics{colorbar_alpha_max.pdf}}
    \caption{Strength surfaces in the $(p,\tau)$ (left) and $(\sigma_{xx},\sigma_{yy})$ (right) planes for different strength potentials $\phi_r$ with $r=1$, $r=2$, and $r=\infty$.
    Since the phase fields in the square inserts have varying maximum value, their maximum is annotated in green.}
    \label{fig:2D_elastic_domains}
\end{figure}

The obtained strength surfaces for each $\Theta$ are depicted in Figure~\ref{fig:2D_elastic_domains} for $\phi_1$, $\phi_2$ and $\phi_{\infty}$ in both the $p$-$\tau$ and $\sigma_{xx}$-$\sigma_{yy}$ planes.
%For each row in Figure~\ref{fig:2D_elastic_domains} corresponding to the different norms $\phi_r$, the plot on the left shows the elastic boundary in the $p$-$\tau$ plane, while the plot on the right shows the $\sigma_{xx}$-$\sigma_{yy}$ plane.
The black lines corresponds to the theoretical curves, cf. \ref{app:R_phi_n}. While nucleation in theory occurs when the damage variable \(\alpha\) becomes strictly positive, in the present numerical setting it is defined through the condition \(\alpha > 10^{-6}\).
The blue points in Figure~\ref{fig:2D_elastic_domains} correspond to the time at which \(\max_{\boldsymbol{x} \in \Omega} \alpha(\boldsymbol{x}) > 10^{-6}\) is observed first. For selected values of \(\Theta\), we also display the corresponding \(\alpha\)-field at that instant. With this, the test does not only validate the numerically obtained strength surface, but also provides valuable insights on  the crack patterns at nucleation under different loading conditions. Similarly to the one-dimensional case (Figure~\ref{fig:structural_snap}), discontinuous damage evolution  over time can occur due to structural snap-back events, leading to initial values of damage significantly larger than \(10^{-6}\). The extent of the snap-back depends on the specific test, hence, different tests may exhibit different maximum damage values.

The \(\alpha\)-fields are plotted using a common colorbar, whereby \(\alpha = 0\) corresponds to blue and the maximum value appears in red. However, since the maximum value of  \(\alpha\) varies across the plots, we annotate each plot with its maximum value as a small green number to ensure a correct interpretation of the color scale.

The theoretical strength surface is well matched by the numerical results.  
In the \(\sigma_{xx}\)-\(\sigma_{yy}\) plane, the elastic domains are symmetric with respect to the bisector of the first and third quadrants.  
Due to the plane-strain conditions assumed in this test, no purely volumetric deformation state is accessible. As a result, the elastic domain is closed and damage onset occurs for all values of \(\Theta\).
In the region where \(p < 0\), i.e. below the line \(\sigma_{yy} = -\sigma_{xx}\), the elastic domain coincides with the same ellipse for \(r = 1\), \(r = 2\), and \(r = \infty\).  
Above that line, the shape of the boundary depends on the norm: we observe a straight line for \(r = 1\), an ellipse for \(r = 2\), and a hyperbola for \(r = \infty\), corresponding to sections of the respective strength surfaces in the principal stress space.
\par

As for the damage patterns, the fracture orientation tends to be in agreement with the normality rule. This is evident for \( p < 0 \), where we consistently obtain mainly shear cracks inclined at \( 45^{\circ} \). For \( p > 0 \), the obtained pattern depends on the specific choice of the strength domain. 
Along the axis \( \sigma_{yy} = \sigma_{xx} \), for \( p < 0 \), the results show diffuse damage patterns. This result appears physically reasonable, as under compressive volumetric loading it is more realistic to expect a diffuse rather than a localized damage pattern. Moreover, this result is  analogous to what was obtained in the brittle case under compression using the volumetric-deviatoric split~\citep{amor2009regularized}, see Figure~12 in~\citep{vicentini2024energy}.

\begin{figure}[H]
    \centering
    \includegraphics{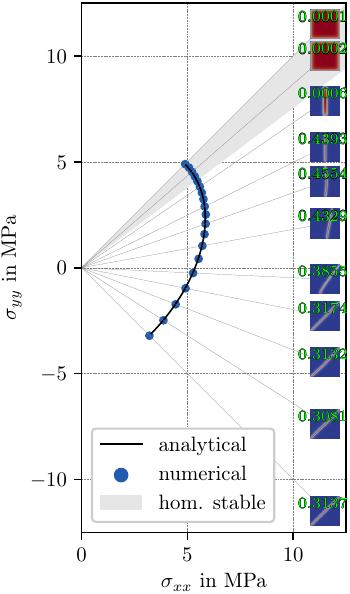}\raisebox{3.625cm}{\hspace{0.2cm}\includegraphics{colorbar_alpha_max.pdf}}
    \caption{Strength surface in the $(\sigma_{xx},\sigma_{yy})$ plane with $\phi_2$ for $\ell/\ell_{\text{ch}}=1/30$ and $\Theta \in [-\pi/4, \pi/4]$ to study the stability of the homogeneous solution.}
    \label{fig:2D_stability}
\end{figure}

It is evident that the damage patterns  presented in Figure~\ref{fig:2D_elastic_domains} for \( r = 2 \) and \( p > 0 \) are all localized, i.e. no homogeneous solution is observed. As discussed in Section~\ref{sec:stability3D}, a crucial role in determining the stability of the homogeneous solution is played by the  \( \ell / \ell_{\text{ch}} \) ratio. While being small enough to fulfill the strain-hardening condition, this ratio must be sufficiently large to render the homogeneous solution unstable. As previously mentioned, the results in Figure~\ref{fig:2D_elastic_domains} for \( r = 2 \) are obtained with \( \ell / \ell_{\text{ch}} = 1/4 \). As shown in  Figure~\ref{fig:stability_plane_strain}, for \( \ell / \ell_{\text{ch}} = 1/4 \) the homogeneous solution is expected to be unstable and localization naturally emerges. On the other hand, from Figure~\ref{fig:stability_plane_strain} we also observe that for smaller ratios the homogeneous solution is expected to become stable starting from a certain angle \( \Theta^* \) in the \( \varepsilon_{xx} \)-\( \varepsilon_{yy} \) plane. 
In particular, for \( \ell / \ell_{\text{ch}} = 1/30 \) we have
\( \Theta^* \approx \pi/7 \). To validate this result numerically, we consider the same setup and evaluation procedure as before, again with \(r=2 \), changing only the value of \( p_c = \tau_c = 2 \sqrt{1/30} \sqrt{\mu G_{\text{c}} / 2 \ell} = 4.5 \)~MPa in order to obtain \( \ell / \ell_{\text{ch}} = 1/30 \). We repeat the numerical tests for \( \Theta \in [-\pi/4, \pi/4] \) and retrieve the strength surface in the \( \sigma_{xx} \)-\( \sigma_{yy} \) diagram, see Figure~\ref{fig:2D_stability}. The corresponding \( \alpha \)-field solutions for illustrative angles are also plotted in Figure~\ref{fig:2D_stability}. Within the same diagram, the theoretical region where the homogeneous solution is expected to be stable is highlighted in light gray. In particular, the angle \( \Theta^* \) in the \( \varepsilon_{xx} \)-\( \varepsilon_{yy} \) plane corresponds to an angle of approximately \( 37^\circ \) in the \( \sigma_{xx} \)-\( \sigma_{yy} \) plane. As seen in the phase-field plots, the numerical results confirm this prediction: within the light gray region, the phase field exhibits what can be regarded as a homogeneous solution, given the imposed Dirichlet boundary condition \( \alpha = 0 \) along the boundaries. 
Near the boundary of the light gray region, the localized damage displays a wider profile.

\subsection{Plane-strain case: cohesive law}
\begin{figure}[H]
    \centering
    \begin{subfigure}[t]{0.45\textwidth}
        \centering
        \includegraphics{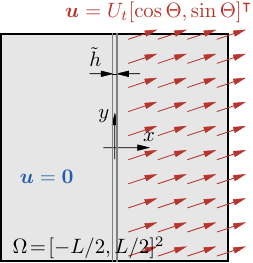}
        \caption{}
        \label{fig:2D_cohesive_law_setup}
    \end{subfigure}
    \begin{subfigure}[t]{0.45\textwidth}
        \centering
        \includegraphics[width=3.9cm]{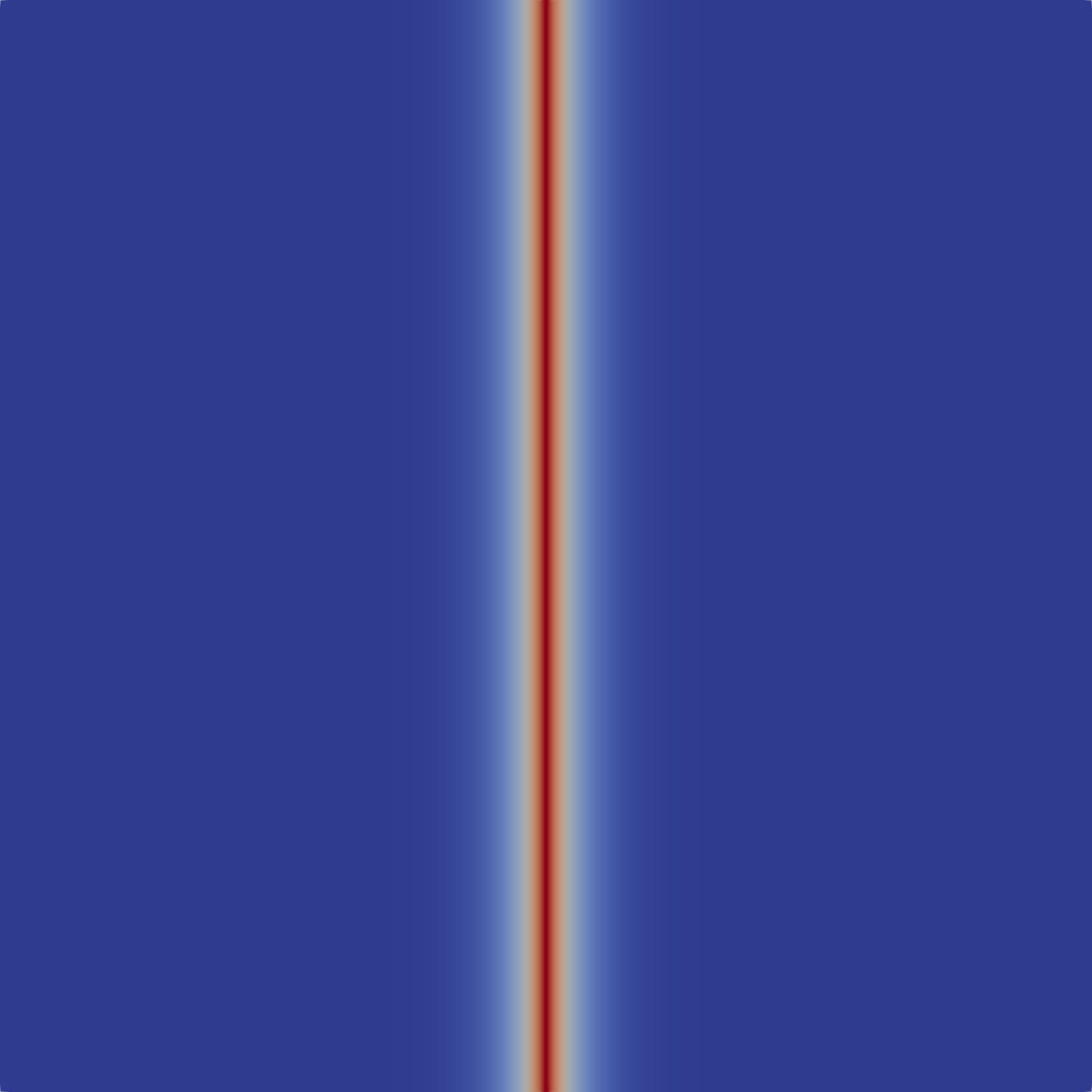}\hspace{0.1cm}\raisebox{0.1cm}{\includegraphics{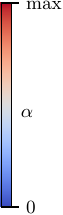}}
        \caption{}
        \label{fig:2D_cohesive_law_alpha}
    \end{subfigure}
    \caption{Setup for the cohesive law test in plane strain (a) and representative phase field (b).}
    \label{fig:2D_cohesive_law_setup_fields}
\end{figure}

In the following numerical example, we validate the cohesive law in a plane-strain setting.  
To this end, we employ the quadratic block setup with edge length $L=1$~mm illustrated in Figure~\ref{fig:2D_cohesive_law_setup}.  
The discretization of the square domain features a central column of elements with a narrow horizontal size \( \tilde{h} = \ell / 25 \) to minimize the numerical overestimation of $G_{\text{c}}$, whereas outside this column we use a structured triangular mesh with \( h \approx \ell / 5 \).

To replicate the configuration used in the analytical derivation of Equation~\eqref{eq:cohesive_law_2D}, where the crack normal is \( \boldsymbol{n} = \boldsymbol{e}_1 \), we impose a zero rigid displacement \( \boldsymbol{u} = \boldsymbol{0} \) on the portion of the domain to the left of the narrow column. On the right-hand side, we prescribe a rigid displacement \( \boldsymbol{u} = U_t [\cos \theta, \sin \theta]^\intercal \), where \( \theta \in [0, \pi/2] \) denotes a fixed loading angle. As a natural consequence, the eigenstrain \( \boldsymbol{\eta} \) localizes within the central narrow column and remains zero elsewhere in the domain.

\begin{figure}[H]
    \centering
    \includegraphics[scale=0.75]{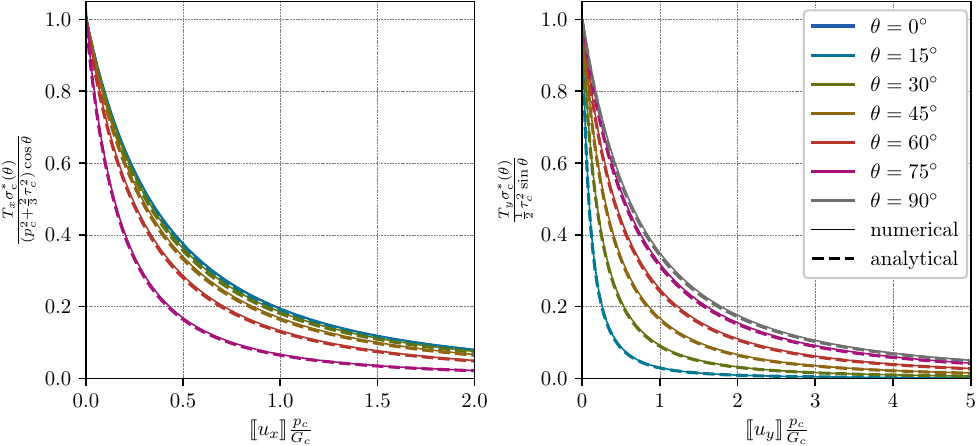}
    \caption{Comparison of the theoretically predicted and numerically obtained traction forces for $\ell/\ell_{\text{ch}}=1/8$ with $\phi_2$.}
    \label{fig:2D_cohesive_law}
\end{figure}

For the material parameters, we consider $E=100$~MPa, $\nu=0.3$, $G_{\text{c}} = 0.2$~MPa~mm, $\ell=0.025$~mm, as well as $p_c = \tau_c = 2 \sqrt{1/8} \sqrt{\mu G_{\text{c}} / 2\ell} = 8.8$~MPa, yielding $\ell/\ell_{\text{ch}}=1/8$. We test different angles, namely $\theta \in \{0^\circ, 15^\circ, 30^\circ, 45^\circ, 60^\circ, 75^\circ, 90^\circ\}$. For each angle, we decompose the traction into horizontal ($T_x$) and vertical ($T_y$) components, and we plot the results for the different angles in Figure~\ref{fig:2D_cohesive_law}.
In this figure we verify that the numerical results  overlap with the theoretical ones. As expected, the cohesive laws exhibit a convex, Barenblatt-type decrease in the traction values. We confirm, as anticipated, that with increasing angle the horizontal traction component $T_x$ decreases in magnitude, while the vertical traction component $T_y$ increases.

\section{Conclusions}
\label{sec:conclusions}

In this paper, we propose a new variational phase-field model of cohesive fracture that enables the flexible tuning of a desired convex strength surface. The total energy functional is similar to that of a gradient damage model coupled with plasticity \cite{alessi2014gradient}, but without a degradation function acting on the elastic energy. The place of the plastic strain is taken by a reversible eigenstrain variable $\boldsymbol{\eta}$, associated with a potential $\pi_0$ that directly depends on the initial elastic domain $\mathcal{S}_0$. This yields a functional structure analogous to that of the brittle phase-field model, with a modified elastic energy density. In particular, the non-linear elastic response governed by $\boldsymbol{\eta}$ reflects the energy relaxation  characteristic of variational sharp cohesive fracture models.

In the paper, we first present an overview of the model and derive its main properties. We then examine its core features in a simplified one-dimensional setting, providing analytical solutions. We introduce the orthogonal volumetric–deviatoric decomposition to facilitate the extension to three dimensions. We also present examples of  $\pi_0$ and derive the corresponding strength surfaces $\partial\mathcal{S}_0$, as well as the associated three-dimensional cohesive laws. Finally, we outline the strategy for the numerical FE implementation and illustrate simulation results. 

Throughout the paper, we comment on the differences between our model and various types of variational gradient damage models, including the classical phase-field model for brittle fracture \cite{bourdin2000numerical} and its versions with energy decomposition~\cite{vicentini2024energy}, asymptotically cohesive models (e.g.,~\cite{lorentz2011convergence,WU201820,zolesi2024stability}), and the model in~\cite{bonacini2021cohesive} which, in addition to being asymptotically cohesive, is proved to be $\Gamma$-convergent to the variational sharp cohesive model. 
%These models typically assume an elliptical initial elastic domain and symmetric response in tension and compression \cite{vicentini2024energy}. Modifying the domain shape via energy decomposition~\cite{de2022nucleation} leads often to spurious residual stresses, except for the recently proposed star-convex model~\cite{vicentini2024energy}. As a result, some authors have deviated from the variational setting~\cite{kumar2020revisiting}, at the cost of giving up the advantages and the rich set of mathematical results available for energy-based fracture models \cite{bourdin2008variational}.
The key differences and contributions of our approach are summarized below:

\begin{itemize}
  \item In the proposed model, the strength surface can be freely prescribed in shape and size as the boundary of the initial elastic domain \(\mathcal{S}_0\), which is explicitly embedded in the model by defining \(\pi_0\) as the support function of \(\mathcal{S}_0\). 
  
    \item We adopt quite general geometric assumptions for $\mathcal{S}_0$: convexity, axisymmetry with respect to the hydrostatic axis, and unboundedness exclusively in the hydrostatic direction under compressive (negative volumetric) stresses. The assumption of axisymmetry can be relaxed. However, the requirement that the domain be unbounded only along the hydrostatic axis under negative volumetric stresses is crucial to ensure the `crack-like' residual stress condition~\eqref{eq:res_stress} (see \citep{chambolle2018approximation} and earlier studies on damage models, e.g. \citep{RICHARD20101203}) when homothetic scaling of $\mathcal{S}(\alpha)$ is assumed. This scaling is a common simplifying assumption~\cite{ortiz1985constitutive}, though it may also be relaxed. As a result, the proposed approach inherently satisfies the non-interpenetration condition without requiring any additional energy decomposition—a step that is necessary in classical gradient models~\cite{amor2009regularized, chambolle2018approximation}.

  \item The proposed phase-field model predicts sharp cohesive cracks. This distinguishes it from previous cohesive gradient damage models~\cite{lorentz2011convergence, WU201820, zolesi2024stability, bonacini_iurlano_2024}, where the displacement jump is smoothed out by the phase-field variable. Similar sharp cracks have also been observed by~\cite{alessi2014gradient} in the context of gradient damage models coupled with plasticity. Having an explicit crack opening is advantageous for some applications (e.g. hydraulic fracture), for which other phase-field approaches require a post-processing step to recover the crack opening displacement from the smeared displacement field~\citep{YOSHIOKA2020113210}. In this sense, the proposed model can be interpreted as a sharp cohesive formulation, where the phase-field variable acts as an indicator of the regularized crack set and affects its evolution.

    \item  The presence of a displacement jump allows for a straightforward identification of the cohesive law and of the surface energy density in the three-dimensional setting and enables a direct comparison with sharp-interface models~\cite{charlotte2006initiation}. In contrast, the literature on cohesive gradient damage models usually limits itself to the study of cohesive laws in the one-dimensional setting \cite{lorentz2011convergence}. 

  \item Unlike with previous gradient damage models \cite{pham2011gradient, zolesi2024stability}, with the present model the homogeneous solution for the one-dimensional bar in tension is always unstable. This makes the model inherently insensitive to the regularization length in the one-dimensional setting. In three dimensions, especially under high volumetric stress, stability becomes more subtle: as already observed in~\cite{zolesi2024stability}, the localized fracture response remains insensitive to the regularization length, while the homogeneous state and its stability depend on the ratio of the regularization length to the cohesive length. This ratio must be small enough to ensure strain hardening and, if the nucleation of cracks is desired, large enough to guarantee the instability of the homogeneous states. On the other hand, for the specific case of the strength potential in Eq.~(\ref{eq:norm2}), we show that a \textit{jump-compatible} eigenstrain always leads to damage localization and the formation of cohesive cracks.

  \item The proposed model \(\Gamma\)-converges to a sharp cohesive fracture formulation \citep{Maggiorelli25}. Similarly, the gradient damage model coupled with plasticity \citep{alessi2014gradient} converges to the sharp cohesive fracture model with plasticity~\cite{dal2016fracture}. In contrast, for asymptotically cohesive models \cite{lorentz2011convergence, zolesi2024stability}, the \(\Gamma\)-limit remains an open question \cite{lorentz2011convergence}. The model proposed in~\cite{bonacini_iurlano_2024} also \(\Gamma\)-converges to a sharp cohesive model. However, its formulation is more restrictive due to the underlying gradient-damage structure. As a consequence, extensions to multi-axial scenarios typically rely on classical energy decomposition strategies~\cite{lammen2023finite}, and studies investigating the flexibility in the choice of the elastic domain shape are still lacking.

    \item  The present model can be implemented using standard discretization and staggered solution schemes, relying on a simple regularization of the non-differentiable strength potential \(\pi_0\). When a continuous ansatz based on linear finite elements is used to approximate the discontinuous displacement field, the numerical overestimation of the dissipated energy scales linearly with the element size. This is the same scaling that applies in the brittle context to the overestimation of \(G_c\)~\cite{bourdin2008variational}.
    
\end{itemize}

%The evolution of the elastic domain \( \mathcal{S}(\alpha) \) has been assumed homothetic, as commonly done in previous works \cite{ortiz1985constitutive}. This is a sufficient, though not necessary, condition to satisfy the stress-softening requirement. However, this assumption can be relaxed by considering more complex evolutions of the domain.

\section{Acknowledgments}
This research has received funding from the Swiss National Science Foundation through Grant No. 200021-219407 `Phase-field modeling of fracture and fatigue: from rigorous theory to fast predictive simulations', which we gratefully acknowledge. We also thank Prof. Corrado Maurini for having suggested not to multiply the elastic strain energy density by the degradation function.

\section{Code Availability}
Our FE implementation used for the numerical results in Sec.~\ref{sec:numerics} is publicly available at \url{https://github.com/jonas-heinzmann/phase_field_cohesive_fracture}.

\appendix

\section{Normality rule}
\label{append:hill}
Given the eigenstrain tensor $\boldsymbol{\eta}$, consider the Gâteaux derivative of $\pi$ at $(\boldsymbol{\eta}, \alpha)$ in the specific direction $(\boldsymbol{\eta}, 0)$. Exploiting the fact that $\pi_0$ is positively homogeneous of degree 1, for $h>0$, we obtain that
\begin{equation}
    \pi'(\boldsymbol{\eta},\alpha)[\boldsymbol{\eta},0]=\lim_{h \to 0} \frac{\pi((1+h)\,\boldsymbol{\eta},\alpha) - \pi(\boldsymbol{\eta},\alpha)}{h}=\pi(\boldsymbol{\eta},\alpha).
    \label{eq:gat_der}
\end{equation}
Using (\ref{eq:gat_der}) and the definition of support function in (\ref{eq:support_fun}), we obtain
\begin{equation}
    \boldsymbol{\sigma}^*\cdot \boldsymbol{\eta}\leq \pi(\boldsymbol{\eta},\alpha) =\pi'(\boldsymbol{\eta},\alpha)[\boldsymbol{\eta},0],\quad\forall \boldsymbol{\sigma}^*\in\mathcal{S}(\alpha),
    \label{eq:adm_stress}
\end{equation}
where $\boldsymbol{\sigma}^*\in\mathcal{S}(\alpha)$ represents an arbitrary \textit{admissible stress}.
\par Given an eigenstrain tensor $\boldsymbol{\eta} \neq \boldsymbol{0}$, now consider the Gâteaux derivative of $\pi$ at $(\boldsymbol{\eta}, \alpha)$ in the direction $(-\boldsymbol{\eta}, 0)$, namely $\pi'(\boldsymbol{\eta}, \alpha)[-\boldsymbol{\eta}, 0]$. For a sufficiently small $h > 0$, one can exploit the fact that $\pi_0$ is positively homogeneous of degree 1, yielding:
\begin{equation}
    \pi'(\boldsymbol{\eta}, \alpha)[-\boldsymbol{\eta}, 0] = \lim_{h \to 0} \frac{\pi((1-h)\,\boldsymbol{\eta}) - \pi(\boldsymbol{\eta},\alpha)}{h} = -\pi(\boldsymbol{\eta},\alpha) = -\pi'(\boldsymbol{\eta}, \alpha)[\boldsymbol{\eta}, 0].
    \label{eq:flip}
\end{equation}
Note that this result does not hold when $\boldsymbol{\eta} = \boldsymbol{0}$, as $\pi_0$ is not homogeneous of degree 1, but only \textit{positively} homogeneous of degree 1.

According to \eqref{eq:flip}, the Gâteaux derivative of $\pi$ at $(\boldsymbol{\eta}, \alpha)$ in the direction $(\boldsymbol{\eta}, 0)$ is opposite to that in the direction $(-\boldsymbol{\eta}, 0)$, for $\boldsymbol{\eta} \neq \boldsymbol{0}$. Exploiting this result and considering the non-linear law in \eqref{eq:nl_law} for the particular admissible perturbations $\boldsymbol{\zeta} = \pm \boldsymbol{\eta}$, we obtain:
\begin{equation}
    \text{if} \quad \boldsymbol{\eta} \neq \boldsymbol{0}, \quad \text{then} \quad \boldsymbol{\sigma}(\boldsymbol{\varepsilon}, \boldsymbol{\eta}) \cdot \boldsymbol{\eta} = \pi'(\boldsymbol{\eta}, \alpha)[\boldsymbol{\eta}, 0].
\end{equation}
Substituting this expression into \eqref{eq:adm_stress} yields
\begin{equation}
    \text{if} \quad \boldsymbol{\eta} \neq \boldsymbol{0}, \quad \text{then} \quad \boldsymbol{\sigma}^* \cdot \boldsymbol{\eta} \leq \boldsymbol{\sigma}(\boldsymbol{\varepsilon}, \boldsymbol{\eta}) \cdot \boldsymbol{\eta}, \quad \forall\, \boldsymbol{\sigma}^* \in \mathcal{S}(\alpha).
\end{equation}
As a consequence, we obtain
\begin{equation}
    \boldsymbol{\sigma}(\boldsymbol{\varepsilon}, \boldsymbol{\eta}) \in \mathcal{S}(\alpha), \quad \left(\boldsymbol{\sigma}(\boldsymbol{\varepsilon}, \boldsymbol{\eta}) - \boldsymbol{\sigma}^*\right) \cdot \boldsymbol{\eta} \geq 0, \quad \forall\, \boldsymbol{\sigma}^* \in \mathcal{S}(\alpha),
\end{equation}
which we refer to as the \textit{normality rule}.

\section{A sufficient condition for strain hardening}
\label{app:strain_hard}

Given a tensor $\boldsymbol{\zeta} \in \mathbb{M}_s^d$ such that $\text{tr}(\boldsymbol{\zeta}) \geq 0$, and under the assumption that the elastic domain $\mathcal{S}(\alpha)$ is unbounded only in the negative direction of purely volumetric stress, the definition of the support function (\ref{eq:support_fun}) ensures the existence of a tensor $\boldsymbol{\sigma}_\text{c} \in \partial \mathcal{S}_0$ such that
\begin{equation}
\boldsymbol{\sigma}_\text{c} \cdot \boldsymbol{\zeta} := \pi_0(\boldsymbol{\zeta}).
\label{eq:3D_strength}
\end{equation}
Since $\pi_0$, being the support function of a convex set, is subadditive and positively homogeneous of degree 1, for any $0 < h < 1$ we have
\begin{equation}
\pi_0(\boldsymbol{\eta} + h\,\boldsymbol{\zeta}) \leq \pi_0(\boldsymbol{\eta}) + h\,\pi_0(\boldsymbol{\zeta}).
\label{eq:pi_conv}
\end{equation}
Rearranging the terms in (\ref{eq:pi_conv}), dividing by $h$, and taking the limit as $h \to 0$, we obtain
\begin{equation}
\pi_0'(\boldsymbol{\eta})[\boldsymbol{\zeta}] \leq \pi_0(\boldsymbol{\zeta}) = \boldsymbol{\sigma}_\text{c} \cdot \boldsymbol{\zeta}.
\label{B3}
\end{equation}

We are now in the position to define a condition that guarantees the strain-hardening requirement (\ref{eq:strain_hardening}). For convenience, we introduce the function
\begin{equation}
\Phi(\boldsymbol{\eta}, \alpha) := W_{\ell}(\boldsymbol{\varepsilon}, \boldsymbol{\eta}, \alpha, 0).
\end{equation}
In order to satisfy the strain-hardening condition in (\ref{eq:strain_hardening}), we require that the second Gâteaux derivative of $\Phi$ at $(\boldsymbol{\eta},\alpha)$ with $\text{tr}(\boldsymbol{\eta}) \geq 0$ evaluated in the direction of the admissible perturbation $(\boldsymbol{\zeta}, \beta)$, be non-negative:
\begin{equation} 
\Phi''(\boldsymbol{\eta}, \alpha)[\boldsymbol{\zeta}, \beta][\boldsymbol{\zeta}, \beta] \geq 0,
\end{equation}
where
\begin{equation}
\begin{aligned}
\Phi''(\boldsymbol{\eta}, \alpha)[\boldsymbol{\zeta}, \beta][\boldsymbol{\zeta}, \beta] &= \mathbb{C}\,\boldsymbol{\zeta} \cdot \boldsymbol{\zeta} + 2\,a'(\alpha)\,\pi_0'(\boldsymbol{\eta})[\boldsymbol{\zeta}]\,\beta \\
&\quad + a''(\alpha)\,\pi_0(\boldsymbol{\eta})\,\beta^2 + a(\alpha)\,\pi_0''(\boldsymbol{\eta})[\boldsymbol{\zeta}][\boldsymbol{\zeta}] + w''(\alpha)\,\frac{G_{\text{c}}}{c_w\,\ell}\,\beta^2.
\end{aligned}
\end{equation}
By the definition of $\pi_0$ in (\ref{eq:support_fun}) and using its convexity and (\ref{B3}), we obtain the following lower bound:
\begin{equation}
\begin{aligned}
\Phi''(\boldsymbol{\eta}, \alpha)[\boldsymbol{\zeta}, \beta][\boldsymbol{\zeta}, \beta] &\geq \mathbb{C}\,\boldsymbol{\zeta} \cdot \boldsymbol{\zeta} + 2\,a'(\alpha)\,\boldsymbol{\sigma}_\text{c} \cdot \boldsymbol{\zeta}\,\beta + w''(\alpha)\,\frac{G_{\text{c}}}{c_w\,\ell}\,\beta^2 \\
&= \mathbb{C}\left(\boldsymbol{\zeta} + a'(\alpha)\,\mathbb{S}\,\boldsymbol{\sigma}_\text{c}\,\beta\right) \cdot \left(\boldsymbol{\zeta} + a'(\alpha)\,\mathbb{S}\,\boldsymbol{\sigma}_\text{c}\,\beta\right) \\
&\quad + \left(w''(\alpha)\,\frac{G_{\text{c}}}{c_w\,\ell} - a'^2(\alpha)\,\mathbb{S}\,\boldsymbol{\sigma}_\text{c} \cdot \boldsymbol{\sigma}_\text{c}\right)\,\beta^2.
\end{aligned}
\end{equation}
Let us now assume to use the \texttt{AT2} model. Minimizing with respect to $\boldsymbol{\zeta}$ we obtain
\begin{equation}
\min_{\boldsymbol{\zeta} \in \mathbb{M}_s^d} \Phi''(\boldsymbol{\eta}, \alpha)[\boldsymbol{\zeta}, \beta][\boldsymbol{\zeta}, \beta] = \left(\frac{G_{\text{c}}}{\ell} - 4\,\mathbb{S}\,\boldsymbol{\sigma}_\text{c} \cdot \boldsymbol{\sigma}_\text{c}\right)\,\beta^2.
\end{equation}
Thus, a sufficient condition for strain hardening with the \texttt{AT2} model is
\begin{equation}
\ell \leq \min_{\boldsymbol{\sigma}_\text{c}\in\partial\mathcal{S}_0}\frac{G_{\text{c}}}{4\,\mathbb{S}\,\boldsymbol{\sigma}_\text{c} \cdot \boldsymbol{\sigma}_\text{c}}.
\label{eq:st_cond}
\end{equation}

\section{Orientation of the deviatoric eigenstrain with respect to the deviatoric strain}
\label{append:simpler}
\par By exploiting the volumetric-deviatoric decomposition of the tensors $\boldsymbol{\varepsilon}$ and $\boldsymbol{\eta}$, the elastic energy density (\ref{eq:strain_en_d}) can be reformulated as
\begin{equation}
\psi(\boldsymbol{\varepsilon}, \boldsymbol{\eta},\alpha) = \frac{\kappa}{2} \left(\text{tr}(\boldsymbol{\varepsilon}) - \text{tr}(\boldsymbol{\eta})\right)^2 + \mu\,\Vert \boldsymbol{\varepsilon}_{\text{dev}} - \boldsymbol{\eta}_{\text{dev}} \Vert^2 + \pi(\boldsymbol{\eta},\alpha).
    \label{eq:simplifying}
\end{equation}
Throughout this work, we assume that the initial elastic domain \( \mathcal{S}_0 \) is axisymmetric about the purely volumetric direction. Owing to this symmetry, it is sufficient to characterize the set \( \mathcal{S}(\alpha) \) using only two scalar quantities: the trace of the stress, \( \text{tr}(\boldsymbol{\sigma}) \), and the \textit{modulus} of its deviatoric component, \( \Vert \boldsymbol{\sigma}_{\text{dev}} \Vert \). This eliminates any dependence on the orientation, \( \check{\boldsymbol{\sigma}}_{\text{dev}} = \boldsymbol{\sigma}_{\text{dev}} / \Vert \boldsymbol{\sigma}_{\text{dev}} \Vert \). Consequently, the strength surface \( \mathcal{F}(\boldsymbol{\sigma}) \) can be expressed as
\begin{equation}
    \mathcal{F}(\boldsymbol{\sigma}) = \Vert \boldsymbol{\sigma}_{\text{dev}} \Vert - f\left(\frac{\text{tr}(\boldsymbol{\sigma})}{3}\right),
\end{equation}
where the function \( f\left(\cdot\right) \) is the intrinsic curve. The  division by $3$ in its argument is due to the fact that the intrinsic curve is conveniently expressed as a function of the pressure $p:=\frac{1}{3}\text{tr}(\boldsymbol{\sigma})$ (see Section \ref{sec:int_curve}).

\par As a result, the function \( \pi_0(\boldsymbol{\eta}) \) also depends solely on the two scalar quantities \( \text{tr}(\boldsymbol{\eta}) \) and \( \Vert \boldsymbol{\eta}_{\text{dev}} \Vert \), again removing any dependence on the orientation \( \check{\boldsymbol{\eta}}_{\text{dev}} = \boldsymbol{\eta}_{\text{dev}} / \Vert \boldsymbol{\eta}_{\text{dev}} \Vert \). Indeed, by the definition of support function, we have
\begin{equation}
    \pi_0(\boldsymbol{\eta}) = \sup_{\text{tr}(\boldsymbol{\sigma}),\, \check{\boldsymbol{\sigma}}_{\text{dev}}} \left\{ \frac{1}{3} \text{tr}(\boldsymbol{\sigma})\, \text{tr}(\boldsymbol{\eta}) + f\left(\frac{\text{tr}(\boldsymbol{\sigma})}{3}\right)\, \Vert \boldsymbol{\eta}_{\text{dev}} \Vert\, \check{\boldsymbol{\sigma}}_{\text{dev}} \cdot \check{\boldsymbol{\eta}}_{\text{dev}} \right\} = \sup_{\text{tr}(\boldsymbol{\sigma})} \left\{ \frac{1}{3} \text{tr}(\boldsymbol{\sigma})\, \text{tr}(\boldsymbol{\eta}) + f\left(\frac{\text{tr}(\boldsymbol{\sigma})}{3}\right)\, \Vert \boldsymbol{\eta}_{\text{dev}} \Vert \right\}.
\end{equation}
Therefore, in light of (\ref{eq:barrier}), the function \( \pi(\boldsymbol{\eta},\alpha) \) also becomes independent of the deviatoric orientation \( \check{\boldsymbol{\eta}}_{\text{dev}} \), and the following lower bound holds:
\begin{equation}
    \begin{aligned}
    \Vert \boldsymbol{\varepsilon}_{\text{dev}} - \boldsymbol{\eta}_{\text{dev}} \Vert^2 + \pi(\boldsymbol{\eta},\alpha)
    &= \Vert \boldsymbol{\varepsilon}_{\text{dev}} \Vert^2 + \Vert \boldsymbol{\eta}_{\text{dev}} \Vert^2 - 2\, \boldsymbol{\varepsilon}_{\text{dev}} \cdot \boldsymbol{\eta}_{\text{dev}} + \pi(\boldsymbol{\eta},\alpha) \\
    &\geq \left( \Vert \boldsymbol{\varepsilon}_{\text{dev}} \Vert - \Vert \boldsymbol{\eta}_{\text{dev}} \Vert \right)^2 + \pi(\boldsymbol{\eta},\alpha).
    \end{aligned}
\end{equation}
This inequality implies that the minimum of the elastic energy density \( \psi(\boldsymbol{\varepsilon}, \boldsymbol{\eta}, \alpha) \) in (\ref{eq:simplifying}) is achieved when \( \boldsymbol{\eta}_{\text{dev}} \) and \( \boldsymbol{\varepsilon}_{\text{dev}} \) have the same orientation, i.e., when $ \check{\boldsymbol{\eta}}_{\text{dev}}=\boldsymbol{\varepsilon}_{\text{dev}}/\Vert\boldsymbol{\varepsilon}_{\text{dev}}\Vert$. Since we are ultimately interested in the configuration that minimizes the total energy functional, assuming that \( \boldsymbol{\eta}_{\text{dev}} \) and \( \boldsymbol{\varepsilon}_{\text{dev}} \) have the same orientation prior to the minimization does not affect the result and simply serves to simplify the formulation.

\section{Strength surface for the $r$-norm strength potential}
\label{app:R_phi_n}
Let us consider $\phi_r$ in (\ref{eq:pot_phi_n}) for $r \in (1,\infty)$. For $\text{tr}(\boldsymbol{\eta}) > 0$ and $\Vert \boldsymbol{\eta}_{\text{dev}} \Vert > 0$, $\phi_r$ is Fréchet differentiable, and the eigenstrain evolution criterion (\ref{eq:diff_voldev}) yields the following set of equations
\begin{equation}
\begin{aligned}
    &p = a(\alpha)\,\frac{p_{\text{c}}^r\,\text{tr}^{r-1}(\boldsymbol{\eta})}{\left(p_{\text{c}}^r\,\text{tr}^r(\boldsymbol{\eta}) + \tau_{\text{c}}^r\,\Vert \boldsymbol{\eta}_{\text{dev}} \Vert^r \right)^{\frac{r-1}{r}}},\\
    &\tau = a(\alpha)\,\frac{\tau_{\text{c}}^r\,\Vert \boldsymbol{\eta}_{\text{dev}} \Vert^{r-1}}{\left(p_{\text{c}}^r\,\text{tr}^r(\boldsymbol{\eta}) + \tau_{\text{c}}^r\,\Vert \boldsymbol{\eta}_{\text{dev}} \Vert^r \right)^{\frac{r-1}{r}}}.
\end{aligned}
\label{eq:diff_n}
\end{equation}
Since both $\text{tr}(\boldsymbol{\eta})$ and $\Vert \boldsymbol{\eta}_{\text{dev}} \Vert$ are positive, there exist $\rho > 0$ and $\omega \in [0, \pi/2]$ such that
\begin{equation}
    p_{\text{c}}\,\text{tr}(\boldsymbol{\eta}) = \rho\,\cos^{\frac{2}{r}}(\omega),\quad \tau_{\text{c}}\,\Vert \boldsymbol{\eta}_{\text{dev}} \Vert = \rho\,\sin^{\frac{2}{r}}(\omega).
\end{equation}
Substituting into (\ref{eq:diff_n}), we obtain
\begin{equation}
    p = a(\alpha)\,p_{\text{c}}\,\cos^{\frac{2(r-1)}{r}}(\omega),\quad \tau = a(\alpha)\,\tau_{\text{c}}\,\sin^{\frac{2(r-1)}{r}}(\omega).
\end{equation}
Finally, in the limit for $\rho \to 0$, $\alpha = 0$, hence we have
\begin{equation}
    p=p_{\text{c}}\,\cos^{\frac{2(r-1)}{r}}(\omega)\quad\text{and}\quad \tau=\tau_{\text{c}}\,\sin^{\frac{2(r-1)}{r}}(\omega),\quad \omega\in[0,\pi/2]\quad \text{if}\quad \text{tr}(\boldsymbol{\sigma})\geq 0
\end{equation}
from which we recover (\ref{eq:R_param}).

\section{Strength potential for the parabolic strength surface}
\label{app:parabolic}
First, using equation (\ref{eq:strength_surface}), the intrinsic function \( f(p) \) for the parabolic strength surface proposed in (\ref{eq:parabola_partial}) can be written as  
\begin{equation}
    f(p) = \tau_{\text{c}} \sqrt{1 - \frac{p}{p_{\text{c}}}}, \quad \text{for } p \geq 0.
\end{equation}
Next, let us denote by \( (\zeta, \xi) \geq (0, 0) \) the pair \( (\text{tr}(\boldsymbol{\eta}), \Vert\boldsymbol{\eta}_{\text{dev}}\Vert) \), so that
\begin{equation}
    \boldsymbol{\eta} = \frac{\zeta}{3} \, \boldsymbol{I} + \xi \, \check{\boldsymbol{\varepsilon}}_{\text{dev}}.
\end{equation}
Recalling equation (\ref{eq:stress_split}), and using the definition of the support function in (\ref{eq:support_fun}), we find that for values of \( \boldsymbol{\eta} \) such that \( p \geq 0 \), the following holds:
\begin{equation}
    \pi_0(\boldsymbol{\eta}) = \sup_{0 \leq p \leq p_{\text{c}}} \mathcal{H}(p), \quad \text{with} \quad \mathcal{H}(p) = p \, \zeta + f(p) \, \xi.
    \label{eq:H_exp}
\end{equation}
The function \( \mathcal{H}(p) \) is concave and attains its global maximum at
\begin{equation}
    \bar{p} = \arg\max_p \mathcal{H}(p) = p_{\text{c}} - \frac{\tau_{\text{c}}^2}{4 p_{\text{c}}} \, \frac{\xi^2}{\zeta^2}.
    \label{eq:p_bar}
\end{equation}
From (\ref{eq:p_bar}) we observe that \( \bar{p} \leq p_{\text{c}} \) automatically. On the other hand, \( \bar{p} \geq 0 \) only if \( \xi \leq \frac{2 p_{\text{c}}}{\tau_{\text{c}}} \, \zeta \). Substituting this into (\ref{eq:H_exp}), we obtain
\begin{equation}
    \pi_0(\boldsymbol{\eta}) = p_{\text{c}} \, \zeta + \frac{\tau_{\text{c}}^2}{4 p_{\text{c}}} \, \frac{\xi^2}{\zeta}, \quad \text{for} \quad \xi \leq \frac{2 p_{\text{c}}}{\tau_{\text{c}}} \, \zeta.
    \label{eq:supp_part1}
\end{equation}
Taking into account this part of  (\ref{eq:supp_part1}) and the fact that for \( p < 0 \) the strength surface corresponds to the horizontal line \( \tau = \tau_{\text{c}} \), we obtain  (\ref{eq:parabola_pot}).

\section{Rayleigh ratio for the homogeneous solution with volumetric stress state}
\label{app:Rayleigh}
Let us consider the homogeneous solution $(\boldsymbol{u}, \boldsymbol{\eta}, \alpha) = \left(\frac{p_{\text{c}}}{3\,\kappa}\,\boldsymbol{x}, 0^+\,\boldsymbol{I}, 0^+\right)$, characterized by the volumetric stress $\boldsymbol{\sigma} = p_{\text{c}}\,\boldsymbol{I}$. Since, for the chosen strength potential $\phi_2$, the elastic domain $\mathcal{S}(\alpha)$ for $p \geq 0$ is the centered ellipse in (\ref{eq:ellipse}) within the $p$--$\tau$ space, the normality rule (\ref{eq:normality_rule}) requires that $\boldsymbol{N} = \frac{1}{\sqrt{3}}\,\boldsymbol{I}$.

Following \cite{pham2011gradient}, in order to find $\boldsymbol{\xi}(\boldsymbol{n})$ in (\ref{eq:xi}), we introduce the $\boldsymbol{n}$-dependent vector $\boldsymbol{v}(\boldsymbol{n}) \in \mathbb{R}^d$ such that
\begin{equation}
    \boldsymbol{\xi}(\boldsymbol{n}) = 2\,\boldsymbol{n} \otimes_{\text{sym}} \boldsymbol{v}(\boldsymbol{n}).
    \label{eq:xi_v}
\end{equation}
Accordingly, the minimization problem (\ref{eq:xi}) can be reformulated in terms of $\boldsymbol{v}(\boldsymbol{n})$ as
\begin{equation}
    \boldsymbol{v}(\boldsymbol{n}) = \argmin_{\boldsymbol{v} \in \mathbb{R}^d} \left[-\frac{1}{\sqrt{3}}\,\left(3\,\lambda + 2\,\mu\right)\,\boldsymbol{v} \cdot \boldsymbol{n} + (\lambda + \mu)(\boldsymbol{v} \cdot \boldsymbol{n})^2 + \mu\,\boldsymbol{v} \cdot \boldsymbol{v}\right]
\end{equation}
from which we derive
\begin{equation}
    \boldsymbol{v}(\boldsymbol{n}) = \frac{1}{2\,\sqrt{3}}\,\frac{3\,\lambda + 2\,\mu}{\lambda + 2\,\mu}\,\boldsymbol{n}.
\end{equation}
Consequently, the term $\mathbb{C} \boldsymbol{\xi}(\boldsymbol{n}) \cdot \boldsymbol{\xi}(\boldsymbol{n})$ in (\ref{eq:delta_max}) becomes
\begin{equation}
    \mathbb{C} \boldsymbol{\xi}(\boldsymbol{n}) \cdot \boldsymbol{\xi}(\boldsymbol{n}) = \frac{(3\,\lambda + 2\,\mu)^2}{12\,(\lambda + 2\,\mu)},
\end{equation}
which is $\boldsymbol{n}$-independent. Thus, $\tilde{\Delta} = \frac{(3\,\lambda + 2\,\mu)^2}{12\,(\lambda + 2\,\mu)}$, and we finally obtain
\begin{equation}
    \boldsymbol{\sigma} = p_{\text{c}}\,\boldsymbol{I}: \quad\tilde{\mathsf{R}} = 12\,\frac{\lambda + 2\,\mu}{(3\,\lambda + 2\,\mu)^2}\,\left(3\,\kappa - \frac{12\,\ell\,p_{\text{c}}^2}{G_{\text{c}}}\right)
\end{equation}
which, when rearranged, yields (\ref{eq:Rayleigh_vol}).

\section{Rayleigh ratio for the plane-strain homogeneous solution under bi-axial strain state}
\label{app:stability_ps}
Adopting the plane-strain assumption, we are interested in analyzing the stability of the homogeneous solution 
\((\boldsymbol{u}, \boldsymbol{\eta}, \alpha) = \left(\boldsymbol{\varepsilon}_C\boldsymbol{x}, 0^+\,\boldsymbol{N}, 0^+\right)\), 
characterized by the bi-axial strain state \(\boldsymbol{\varepsilon}_C\) defined as
\begin{equation}
    \boldsymbol{\varepsilon}_C(\Theta) =
    \begin{bmatrix}
    \cos \Theta & 0 & 0 \\
    0 & \sin \Theta & 0 \\
    0 & 0 & 0
    \end{bmatrix} \, \epsilon_C(\Theta), 
    \quad \text{with} \quad -\frac{\pi}{4} \leq \Theta \leq \frac{\pi}{4}.
\end{equation}
Accordingly, we have
\begin{equation}
    \operatorname{tr}(\boldsymbol{\varepsilon}_C(\Theta)) = \left(\cos \Theta + \sin\Theta \right)\, \epsilon_C(\Theta),
    \quad 
    \left\| {\boldsymbol{\varepsilon}_C(\Theta)} \right\|_{\text{dev}} = \sqrt{\frac{2 - \sin(2\Theta)}{3}} \, \epsilon_C(\Theta).
\end{equation}
The associated stress \(\boldsymbol{\sigma} = \mathbb{C}\boldsymbol{\varepsilon}_C(\Theta)\) lies on the strength surface \(\partial\mathcal{S}_0\). Using the strength potential 
$\pi_0(\boldsymbol{\eta}) = \phi_2(\operatorname{tr}(\boldsymbol{\eta}), \left\| \boldsymbol{\eta}_{\text{dev}} \right\|)$
defined in~(\ref{eq:pot_phi_n}), we obtain
\begin{equation}
    \epsilon_C(\Theta) =
    \left(
    \frac{\kappa^2}{p_{\text{c}}^2}\left(1 + \sin(2\Theta)\right) + 
    \frac{4\mu^2}{3\tau_{\text{c}}^2}\left(2 - \sin(2\Theta)\right)
    \right)^{-\frac{1}{2}}.
\end{equation}
The eigenstrain tensor \(\boldsymbol{\eta} = \Vert \boldsymbol{\eta} \Vert\, \boldsymbol{N}\) is characterized by an orientation tensor \(\boldsymbol{N}\) that follows the normality rule (\ref{eq:normality_rule}). Accordingly,
\begin{equation}
\boldsymbol{N} = \frac{\boldsymbol{N}^*}{\Vert \boldsymbol{N}^* \Vert}, \quad \text{with} \quad 
\boldsymbol{N}^* := \frac{2}{3}\, \frac{\kappa}{p_{\text{c}}^2} \, \operatorname{tr}(\boldsymbol{\varepsilon}_C(\Theta))\, \boldsymbol{I} + \frac{4\,\mu}{\tau_{\text{c}}^2} \, {\boldsymbol{\varepsilon}_C(\Theta)}_{\text{dev}}.
\end{equation}
In order to account for the plane-strain assumption, let us first introduce the vectors $\boldsymbol{n} = [n_1, n_2, 0]^\intercal$, with $n_1^2 + n_2^2 = 1$, and $\boldsymbol{w} = [w_1, w_2, 0]^\intercal$, both with a zero out-of-plane component. We also define the associated rank-2 tensor  
\begin{equation}
\boldsymbol{\xi}^*(\boldsymbol{w}, \boldsymbol{n}) = 2\,\boldsymbol{n} \otimes_{\text{sym}} \boldsymbol{w} = 
\begin{bmatrix}
2\,n_1\,w_1 & n_1\,w_2 + n_2\,w_1 & 0 \\
n_1\,w_2 + n_2\,w_1 & 2\,n_2\,w_2 & 0 \\
0 & 0 & 0
\end{bmatrix}.
\end{equation}
Then, the minimization problem~\eqref{eq:xi} can be reformulated as the problem of finding $\boldsymbol{w}(\boldsymbol{n}) = [w_1(\boldsymbol{n}), w_2(\boldsymbol{n}), 0]^\intercal$ such that
\begin{equation}
\nabla g_{\boldsymbol{n}}(w_1(\boldsymbol{n}), w_2(\boldsymbol{n})) = \boldsymbol{0}, \quad \text{with} \quad g_{\boldsymbol{n}}(w_1, w_2) := \mathbb{C}(\boldsymbol{N} - \boldsymbol{\xi}^*(\boldsymbol{w}, \boldsymbol{n})) \cdot (\boldsymbol{N} - \boldsymbol{\xi}^*(\boldsymbol{w}, \boldsymbol{n})).
\end{equation}
This formulation is justified, first, by the fact that under the plane-strain assumption the displacement perturbation has no out-of-plane component, and second, because the function $g_{\boldsymbol{n}}(w_1, w_2)$ is convex in $\boldsymbol{w}$.
At this point, we introduce the function
\begin{equation}
    \delta(n_1) := \mathcal{L}\left(n_1, \sqrt{1 - n_1^2}\right), \quad \text{with} \quad 0 \leq n_1 \leq 1,
\end{equation}
where
\begin{equation}
    \mathcal{L}(n_1, n_2) := \mathbb{C} \boldsymbol{\xi}\left(\boldsymbol{w}(\boldsymbol{n}), \boldsymbol{n}\right) \cdot \boldsymbol{\xi}\left(\boldsymbol{w}(\boldsymbol{n}), \boldsymbol{n}\right).
\end{equation}
In this way, the maximization problem in~\eqref{eq:delta_max} reduces to finding $\tilde{n}_1$ such that 
\begin{equation}
    \tilde{n}_1=\argmax_{0 \leq n_1 \leq 1} \delta(n_1).
\end{equation}
To solve this problem, we observe that $\delta(n_1)$ is a fourth-order polynomial in $n_1$ with three real stationary points. Furthermore, $\delta(n_1)$ is an even function, which implies that one of the stationary points corresponds to $n_1 = 0$. This stationary point is a local minimum; in fact, one can verify that $\delta''(0) \geq 0$ for $-1 \leq \nu \leq 1/2$ and $-\pi/4 \leq \Theta \leq \pi/4$. 
Therefore, the other two real stationary points, $n_1 = \pm n_1^*$, correspond to the locations where the maximum value of $\delta(n_1)$ is attained. Taking into account the constraint $0\leq n_1\leq 1$, we obtain
\begin{equation}
    \tilde{n}_1=\min (n_1^*,1),\quad \text{with}\quad n_1^* :=\frac{1}{3}\,\sqrt{\frac{ 
\frac{(1 + \nu)^2}{(1 - 2\nu)}\frac{\tau_{\text{c}}^2}{p_{\text{c}}^2}  (\cos^2\Theta - \sin^2\Theta)
+ 3 \left((1-2\,\nu)\cos^2\Theta - \frac{3}{2}  \sin(2\Theta) + 1+\nu\right)
}{
1 - \sin(2\Theta)
}}.
\end{equation} 
Accordingly, the Rayleigh ratio in (\ref{eq:Rayleigh}) becomes
\begin{equation}  \boldsymbol{\sigma}=\mathbb{C}\boldsymbol{\varepsilon}_C(\Theta):\quad \tilde{\mathsf{R}}=\mathsf{R}_{\Theta}\left(\nu,\frac{p_{\text{c}}}{\tau_{\text{c}}},\frac{\ell}{\ell_{\text{ch}}}\right):=\frac{\mathbb{C}\boldsymbol{N}\cdot \boldsymbol{N} - \frac{4\ell}{G_{\text{c}}} \left(\mathbb{C}\boldsymbol{\varepsilon}_C(\Theta) \cdot \boldsymbol{N} \right)^2}{\delta(\tilde{n}_1)}.
\end{equation}
Note that for a given angle $\Theta\in[-\pi/4,\pi/4]$, the Rayleigh ratio exclusively depends on the Poisson's ratio $\nu$, the strength ratio $\frac{p_{\text{c}}}{\tau_{\text{c}}}$ and the length ratio $\frac{\ell}{\ell_{\text{ch}}}$, where $\ell_{\text{ch}}$ is defined as in Table~\ref{tab:ell_strength}.
\par In the case of pure shear, i.e., $\Theta = -\pi/4$, we obtain
\begin{equation}
    \boldsymbol{\sigma}=\mathbb{C}\boldsymbol{\varepsilon}_C(-\pi/4):\quad \tilde{\mathsf{R}} = 1 -4\, \frac{\ell}{2\,\mu\,G_{\text{c}}/\tau_{\text{c}}^2},
\end{equation}
hence, for $\ell > 0$, the corresponding homogeneous solution is always unstable. This is in agreement with the fact that, for pure shear, $\boldsymbol{N}$ is a rank-2 tensor with opposite eigenvalues (see Section~\ref{sec:stability3D}).
\par
\begin{figure}[H]
    \centering
    \includegraphics[scale=1]{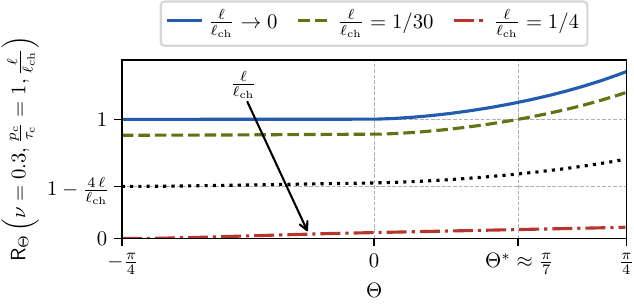}
    \caption{Rayleigh ratio for the plane-strain homogeneous solution under bi-axial strain state for  $\nu=0.3$, $\frac{p_{\text{c}}}{\tau_{\text{c}}}=1$ and different values of $\frac{\ell}{\ell_{\text{ch}}}$:  $\frac{\ell}{\ell_{\text{ch}}}=10 ^{-5}$ (solid blue line), $\frac{\ell}{\ell_{\text{ch}}}=1/30$ (dashed green line), $\frac{\ell}{\ell_{\text{ch}}}=\frac{9}{64}$ (dotted black line) and $\frac{\ell}{\ell_{\text{ch}}}=\frac{1}{4}$ (dashdotted red line). For these parameters $\ell_{\text{ch}}=\frac{2\,\mu\,G_{\text{c}}}{\tau_{\text{c}}^2}$ (see Table~\ref{tab:ell_strength}). For the length ratio $\frac{\ell}{\ell_{\text{ch}}}=\frac{1}{30}$, the tested homogeneous solution is unstable only for loading angles $\Theta \leq \Theta^*\approx \frac{\pi}{7}$.
} 
    \label{fig:stability_plane_strain}
\end{figure}

 Figure~\ref{fig:stability_plane_strain} shows the behavior of the Rayleigh ratio for various values of $\Theta$ and $\ell/\ell_{\text{ch}}$, with $\nu = 0.3$ and $p_{\text{c}}/\tau_{\text{c}} = 1$, corresponding to the parameters used in the numerical section (see Section~\ref{sec:2D_plate}). As discussed in Section~\ref{sec:stability3D}, in order to obtain unstable homogeneous solutions, it is advisable to choose values of $\ell/\ell_{\text{ch}}$ that are not too small, while still satisfying the strain-hardening condition $\ell \leq \frac{1}{4}\ell_{\text{ch}}$.

\bibliographystyle{elsarticle-num}
\bibliography{References}

%===================================================================================================
\end{document}